\documentclass{aa}
\usepackage[utf8]{inputenc}
\usepackage{natbib}
\usepackage[flushleft]{threeparttable}
\usepackage{pdflscape}
\usepackage{tikz}
\usepackage{graphicx}
\usetikzlibrary{positioning}
\bibpunct{(}{)}{;}{a}{}{,} 

\title{Empirical mass-loss rates and clumping properties of O-type stars in the LMC}
\author{C. Hawcroft\inst{\ref{inst1},\ref{inst2}} \and L. Mahy\inst{\ref{inst3}} \and H. Sana\inst{\ref{inst1}} \and J.O. Sundqvist\inst{\ref{inst1}} \and M. Abdul-Masih\inst{\ref{inst4},\ref{inst5},\ref{inst6}} \and S. A. Brands\inst{\ref{inst7}} \and L. Decin\inst{\ref{inst1}} \and A. de Koter\inst{\ref{inst7}, \ref{inst1}} \and J. Puls\inst{\ref{inst8}}}

\institute{Institute of Astronomy, KU Leuven, Celestijnenlaan 200D, 3001, Leuven, Belgium \label{inst1} \and Space Telescope Science Institute, 3700 San Martin Drive, Baltimore, MD 21218, USA \\ email: chawcroft@stsci.edu \label{inst2} \and Royal Observatory of Belgium, Avenue Circulaire/Ringlaan 3, B-1180 Brussels, Belgium \label{inst3} \and European Southern Observatory, Alonso de Cordova 3107, Vitacura, Casilla 19001, Santiago de Chile, Chile\label{inst4} \and Instituto de Astrofísica de Canarias, C. Vía Láctea, s/n, 38205 La Laguna, Santa Cruz de Tenerife, Spain\label{inst5} \and Universidad de La Laguna, Dpto. Astrofísica, Av. Astrofísico Fran- cisco Sánchez, 38206 La Laguna, Santa Cruz de Tenerife, Spain\label{inst6} \and Astronomical Institute Anton Pannekoek, Amsterdam University, Science Park 904, 1098 XH Amsterdam, The Netherlands \label{inst7} \and LMU M{\"u}nchen, Universit{\"a}tssternwarte, Scheinerstr. 1, 81679 M{\"u}nchen, Germany\label{inst8}}

\abstract{The nature of mass-loss in massive stars is one of the most important and difficult to constrain processes in the evolution of massive stars. The largest observational uncertainties are related to the influence of metallicity and wind structure with optically thick clumps.
}{We aim to constrain the wind parameters of sample of 18 O-type stars in the LMC, through analysis with stellar atmosphere and wind models including the effects of optically thick clumping. This will allow us to determine the most accurate spectroscopic mass-loss and wind structure properties of massive stars at sub-solar metallicity to date. This will allow us to gain insight into the impact of metallicity on massive stellar winds.
}{Combining high signal to noise (S/N) ratio observations in the ultraviolet and optical wavelength ranges gives us access to diagnostics of multiple different ongoing physical processes in the stellar wind. We produce synthetic spectra using the stellar atmosphere modelling code FASTWIND, and reproduce the observed spectra using a genetic algorithm based fitting technique to optimise the input parameters. 
}{We empirically constrain 15 physical parameters associated with the stellar and wind properties of O-type stars from the dwarf, giant and supergiant luminosity classes. These include temperature, surface gravity, surface abundances, rotation, macroturbulence and wind parameters. 
}{We find, on average, mass-loss rates a factor of 4-5 lower than those predicted by \cite{Vink2001}, in good agreement with predictions from \cite{Bjorklund2021}, and the best agreement with the predictions from \cite{Krticka2018}. In the 'weak-wind' regime we find mass-loss rates orders of magnitude below any theoretical predictions. We find a positive correlation of clumping factors with effective temperature with an average $f_{\rm{cl}} = 14 \pm 8$ for the full sample. It is clear that there is a difference in the porosity of the wind in velocity space, and interclump density, above and below a temperature of roughly 38 kK. Above 38 kK an average $46 \pm 24 \%$ of the wind velocity span is covered by clumps and the interclump density is 10-30\% of the mean wind. Below an effective temperature of roughly 38 kK there must be additional light leakage for supergiants. For dwarf stars at low temperatures there is a statistical preference for very low clump velocity spans, however it is unclear if this can be physically motivated as there are no clearly observable wind signatures in UV diagnostics.
}

\keywords{stars: early-type -- stars: massive -- stars: mass-loss -- stars: atmospheres -- stars: winds, outflows -- stars: fundamental parameters} 

\begin{document}

\maketitle

\section{Introduction}

The mass-loss rates of hot OB type stars ($M_{\rm{ini}} > 8M_{\odot}$) are one of the paramount parameters in our understanding of massive stars, with a substantial influence on the course of the stellar evolutionary pathway, and yet they remain one of the most uncertain elements in stellar modelling (see reviews by \citealp{Puls2008}, \citealp{Langer2012}, \citealp{Smith2014}, \citealp{Vink2021}). The uncertainty is highlighted not only in stellar evolution but also in the statistical predictions of end-of-life products, including exotic and energetic objects such as stellar-mass black-holes and gravitational wave events. Stellar mass loss is also a major consideration in the evolution of the interstellar medium as it contributes significant amounts of mass, energy and momentum \citep{Weaver1977}. This mass loss is associated with strong radiatively driven winds, which are powered by momentum transfer through scattering between the UV radiation field and metal resonance line transitions in the outer atmosphere \citep{Castor1975}. This process is known to be very effective as Doppler shifted material can be further accelerated by light of a higher wavelength than that of the rest wavelength of the transition \citep{Sobolev1960}. If even small perturbations are present at the onset of the outflows this so-called line-deshadowing leads to highly unstable winds (\citealp{MacGregor1979}, \citealp{Carlberg1980}, \citealp{Owocki1984}, \citealp{Eversberg1998}, \citealp{Lepine2008}). This instability gives rise to strong shocks which result in a multi-component wind structure comprised of dense, slow moving clumps of material where line opacities can increase and an inter-clump medium occupied by under-dense high-velocity material, which allows greater photon escape (\citealp{Owocki1988}, \citealp{Feldmeier1995}, \citealp{Dessart2003}, \citealp{Sundqvist2010}, \citealp{Sundqvist2013}, \citealp{Driessen2019}). 
The uncertainty in mass loss was identified thanks to a discrepancy between empirically derived mass-loss rates from different physical diagnostics. In spectroscopic studies, synthetic spectra are produced from a radiative transfer solution of unified stellar atmosphere and wind models which take into account the effects of non-local-thermodynamic-equilibrium (e.g. CMFGEN, \citealp{Hillier1998}; PoWR, \citealp{Grafener2002} or FASTWIND, \citealp{Puls2005}). Generally, spectroscopic analyses assumed the wind to be a smooth outflow. This resulted in the mass-loss rates derived using optical recombination line profiles (which depend on density as $\rho^{2}$) to be much larger than those determined using UV resonance lines (which depend linearly on $\rho$). However, as the effects of wind structure were included, in a simplified form accounting for enhanced density in clumps, the discrepancy was reduced. Some issues still remained, mainly that the absorption of light emitted by high velocity metal ions such as phosphorus was over-predicted (\citealp{Pauldrach1994}, \citealp{Hillier2003}, \citealp{Fullerton2006}, \citealp{Bouret2012}). This appears to be solved by the consideration of the velocity span of clumps, which is limited at high velocity, resulting in reduced absorption of blue-shifted light from these metal ion transitions (e.g. \citealp{Oskinova2007}, \citealp{Sundqvist2010}, \citealp{Sundqvist2011a}, \citealp{Surlan2013}, \citealp{Hawcroft2021}). 

As empirical mass-loss rates generally began to converge with the inclusion of clumping corrections there still remained a discrepancy when comparing to theoretical predictions. The magnitude of which varies in different physical regimes, but generally a factor 3 was found for O-type stars (see e.g. \citealp{Bouret2012, Hawcroft2021,Brands2022}). This could not be attributed to wind clumping as clumping is thought to have little impact on theoretical predictions of mass loss \citep{Muijres2011}. The problem may come from physical assumptions in such model predictions as significantly different rates are predicted using different numerical techniques (\citealp{Vink2001}, \citealp{Krticka2018}, \citealp{Bjorklund2021}, \citealp{VinkSander2021}, \citealp{Bjorklund2022}).  

In the realm of typical O-stars (5.2 < log$(L/L_{\odot})$ < 6.2), thanks to this inclusion of wind structure in spectroscopic models and new hydrodynamically consistent theoretical predictions, the mass-loss rates in theory and observations appear to be converging. \cite{Hawcroft2021} find mass-loss rates between a factor of 2-3 lower than the rates predicted by \cite{Vink2001} but agree well with the predictions by \cite{Bjorklund2021} for a sample of Galactic O-supergiants. \cite{Brands2022} cover a larger range of O stars in the Large Magellanic Cloud (LMC), with a large representation of main sequence O dwarfs. These authors find good agreement with \cite{Krticka2018} and \cite{Bjorklund2021} up until high luminosities (log$(L/L_{\odot})$ > 6.2) where there is better agreement with the predictions of \cite{VinkSander2021}. Although the predictions must be extrapolated to these luminosities as the model grids in these works extend only up until log$(L/L_{\odot})$ = 6.0. Higher luminosity predictions are available from \cite{Vink2018} but there are concerns regarding the validity of a steady-state approach in this regime (\citealp{Jiang2018}, \citealp{Schultz2020}). Uncertainty persists at high luminosity as stars approach the Eddington limit (\citealp{Grafener2008}, \citealp{Vink2011}). Although perhaps the largest remaining discrepancy is at low luminosities log$(L/L_{\odot})$ < 5.2. Mass-loss rates of late O dwarfs and giants are found to be orders of magnitudes lower than predicted (\citealp{Bouret2003}, \citealp{Martins2004}, \citealp{Martins2005b}, \citealp{Marcolino2009}, \citealp{deAlmeida2019}, \citealp{Rubio-Diez2022}, \citealp{Brands2022}). 

Another major effect in our understanding of stellar winds is the metallicity dependence. As the chemical composition of the star changes, specifically as the abundance of key wind driving metals varies, so should the strength of the wind. The aforementioned theoretical rates all predict a metallicity mass-loss relation, of the order $\sim{} Z^{0.67}$ to $\sim{} Z^{0.95}$  (\citealp{Vink2001}, \citealp{Krticka2018} and \citealp{Bjorklund2021}). This has been difficult to test empirically due to lack of simultaneous UV and optical coverage at low metallicity. \cite{Mokiem2007} use optical diagnostics to estimate the empirical $\dot{M}(Z)$ dependence, finding $\dot{M} \propto Z^{0.83\pm0.16}$. \cite{Marcolino2022} also provide an empirical relation of $\dot{M} \propto Z^{\sim 0.5 - 0.8}$ for bright stars (log$(L/L_{\odot})$ > 5.4), with a weaker dependence at lower luminosity.
It is unclear whether the strength of the line-deshadowing instability (LDI) and therefore the wind structure is likely to change with metallicity, or with other stellar parameters. While \cite{Brands2022} investigated wind properties in the LMC, it is difficult to draw conclusions as their sample follows a close positive relation between luminosity and temperature, meaning it is difficult to disentangle which parameter is leading the relation with wind properties. The only other attempts to constrain wind structure are studies of variability in emission-line profiles, through which it is possible to see the direct influence of clumps moving through the wind and shaping the profile over time. Using this method \cite{Lepine2008} found no evidence of a change in clumping properties between Galactic O supergiants and WR stars. \cite{Marchenko2007} studied 3 WR stars from the SMC and again are unable to find evidence for changing clumping properties. \cite{Driessen2022} investigate the metallicity dependence of clumping properties theoretically using 2D LDI simulations of O-stars at fixed luminosity, and conclude that such a relation exists, although it is fairly weak $f_{\rm{cl}} \propto Z^{0.15}$. These authors also note that this provides a moderate correction to the aforementioned empirical metallicity mass loss relation of \cite{Mokiem2007} which uses H$\alpha$ diagnostics, resulting in $\dot{M} \propto Z^{0.76}$. Finally, \cite{Driessen2022} find generally lower clumping factors relative to 1D simulations, as the added spatial dimension allows for some smoothing of the over-densities. It is thus likely that the clumping factors will decrease even further in 3D LDI simulations. 

Another physical effect of low metallicity is faster rotation, due to the reduced mass-loss at low metallicity resulting in less removal of angular momentum. This is thought to have implications for stellar evolution and internal mixing and has been studied rather more than the mass-loss (e.g. \citealp{Yoon2008}, \citealp{Hunter2009}, \citealp{Brott2011b}, \citealp{RiveroGonzalez2012b}, \citealp{RiveroGonzalez2012a}, \citealp{Bouret2013}, \citealp{Georgy2013}, \citealp{Ramirez-Agudelo13}, \citealp{Ramirez-Agudelo15}, \citealp{Grin2017}, \citealp{Keszthelyi2017}, \citealp{Groh2019}, \citealp{Bouret2021}, \citealp{Murphy2021}, \citealp{Eggenberger2021}).

In this study we focus on a sample of O stars, covering the luminosity range 5.2 < log$(L/L_{\odot})$ < 6.0, in the LMC, i.e., at half the solar metallicity. For this we have obtained UV spectroscopic observations of 18 O stars in the LMC (GO: 15629, PI: Mahy) and utilise archival optical coverage. This allows us to investigate stellar and wind parameters. For this analysis we simultaneously determine a number of stellar properties, with a focus on investigating the mass-loss rates, clumping factors and wind structure parameters with high accuracy across the upper part of the Hertzsprung-Russell diagram (HRD).

The paper progresses with a description of the sample and observations in Sect. \ref{sec: sample}. This is followed by a description of our modelling technique in Sect. \ref{sec: methods}. We present stellar and wind parameters in Sect. \ref{sec: results} . In Sect. \ref{sec: discussion} we discuss our findings, with a focus on wind parameters. The paper concludes in Sect. \ref{sec: conclusions}.

\begin{table}
\centering
\caption{Overview of the capability of the instruments used to obtain spectra for this sample. All optical spectra were taken with the GIRAFFE spectrograph on FLAMES.}
\label{tbl:Instruments}

\begin{tabular}{cccc}
\hline
\hline

Regime & Instrument	& Wavelength \AA & Resolving Power \\
\hline
FUV-UV	&	HST COS	&	1110-2150	&	3\,000	\\
UV	&	HST STIS	&	1145-1710	&	45\,000	\\
Optical	&	LR02	&	3960-4564	&	7\,000	\\
Optical	&	LR03	&	4499-5071	&	8\,500	\\
Optical	&	HR15N &	6442-6817	&	16\,000	\\
\hline
\\              
\end{tabular}
\end{table}

\section{Sample \& Observations} \label{sec: sample}

The sample of 18 LMC stars analysed in this work is distributed throughout the O-star spectral sub-types from early to late including: one supergiant, three bright giants, four giants and 10 dwarfs or sub-giants. The targets are selected from the VFTS catalogue (VLT-FLAMES Tarantula Survey, \citealp{Evans2011}). We also utilise the optical spectra obtained as part of VFTS in our fitting. The VFTS spectra were obtained using the Medusa mode of the ESO FLAMES instrument on the Very Large Telescope (VLT) combined with the GIRAFFE spectrograph. The LR02, LR03 and HR15N setups were used, delivering spectroscopy of key stellar and wind diagnostic spectral lines from 4000 to 4500 \AA \,at resolving power $\sim 7000$, from 4500 to 5000 \AA \,at resolving power $\sim 8500$ and at a resolving power $\sim 16000$ in the range 6400 to 6800 \AA. The VFTS observations are further detailed in \cite{Evans2011}.

In order to capture essential wind diagnostics at UV wavelengths, a series of follow-up observations were carried out with the Hubble Space Telescope (HST). The majority of the observations were taken using the Cosmic Origins Spectrograph (COS), with six stars observed using the Space Telescope Imagining Spectrograph (STIS) due to the presence of bright stars surrounding the target. Table \ref{tbl:Instruments} shows the instruments used to obtain the observations for this work and their spectroscopic capabilities. Compared to \cite{Brands2022}, the UV spectra we have obtained are of a higher resolution by roughly a factor of three, therefore increasing the S/N per equivalent resolution element. This means we have higher S/N for low luminosity (log$(L/L_{\odot})$ < 5.2) stars, which allows us to constrain wind parameters for these types of stars for the first time. We have a smaller sample than \cite{Brands2022} ($\sim 20$ compared to $\sim 50$), but a more even ratio of giants and bright giants to dwarfs, while the sample of \cite{Brands2022} is primarily comprised of dwarf stars. We also note the complementary optical spectra are quite different between our work and \cite{Brands2022}. These authors use optical spectra from HST-STIS as the focus is on the central core of the R136 cluster, which was excluded from the VFTS observing campaign due to crowding and can only be resolved with HST. As a result, the optical spectra used in \cite{Brands2022} are of lower resolving power and, on average, S/N $\sim$ 20 which is a factor of 10 lower than those available in this study from VFTS (see \cite{Brands2022} for further details on the data used in their study). Typically S/N $\sim$ 100-200 is required to constrain individual stellar parameters from optical spectra of O stars, with higher S/N required for fast rotators ($v \sin i$ > 200 $\rm{km}\, \rm{s^{-1}}$, \citealp{Grin2017}).

The sample contains four stars (VFTS143, VFTS184, VFTS422, VFTS608) with significant radial velocity variations and amplitudes large enough to fulfill the binarity criteria of \cite{Sana2013}, although the nature of this variability is unconstrained. We therefore cannot make any systematic analysis of the influence of tentative binarity on global wind physics with this dataset but comment on whether this radial velocity variability has any impact on the fits for the relevant star in Appendix \ref{appendix}. We also take care to select the stars such that the intrinsic spectral variability is small. All targets are far from the Luminous Blue Variable (LBV) regime, therefore the non-contemporaneity of the UV and optical observations will not affect the results and conclusions of the current study.

\section{Methods}\label{sec: methods}

In order to determine stellar and wind parameters we compare the observed spectra with synthetic spectra, produced using the code FASTWIND (v10.3, \citealp{Santolaya-Rey1997a}, \citealp{Puls2005}, \citealp{RiveroGonzalez2011}, \citealp{Carneiro2016}, \citealp{Sundqvist2018}), with the parameters of interest given as inputs. FASTWIND is a stellar atmosphere and wind modelling software which calculates model atmospheres including the effects of line-blocking/blanketing for background ions while giving a full co-moving frame transport treatment to explicit elements. Differentiating between these two groups of elements allows the code to treat relevant spectral diagnostics with high precision while maintaining a performance speed which allows computing thousands of models per object entirely feasible. For this study, line transitions from H, He, C, N, O, P and Si are treated as explicit elements. FASTWIND is uniquely suited for this analysis as it has the functionality to include clumps of arbitrary optical thickness and account for the effect of a highly structured wind on non-local thermodynamic equilibrium (NLTE) occupation numbers and the subsequent projection onto synthetic spectra by computing an effective opacity for the two-component outflow (consisting of dense clumps and an under-dense inter-clump medium). For further details of the clumping parameterisation in FASTWIND we refer to \cite{Sundqvist2018}, \cite{Hawcroft2021} and \cite{Brands2022}. 

The speed of the atmosphere modelling code is a high priority in our analysis as we employ a computationally expensive fitting algorithm to simultaneously optimise all the stellar parameters. Our fitting technique is to use a genetic algorithm (GA), which is based on the principles of genetic evolution and adapted from the pikaia code (\citealp{Charbonneau1995}, \citealp{Mokiem2005}). Such a technique has been used successfully to determine stellar and wind parameters for various samples of massive stars (\citealp{Mokiem2006}, \citealp{Mokiem2007}, \citealp{Tramper2011}, \citealp{Tramper2014}, \citealp{RamirezAgudelo2017}, \citealp{Abdul-Masih2019}, \citealp{Abdul-Masih2021}, \citealp{Hawcroft2021}, \citealp{Johnston2021}, \citealp{Fabry2021}, \citealp{Brands2022}, \citealp{Shenar2022}). The version used in this study is the same as presented in \cite{Hawcroft2021}. 

The GA works by computing a large initial population of FASTWIND models spanning the full range of the parameter space\footnote{The range of values considered can vary and is tailored for each object to ensure the solution sits well within the range covered by the parameter space.}, and assessing the quality of their reproduction of the spectra using a $\chi^{2}$-based fitness metric in which all line profiles are weighted equally (see equation 7 in \citealp{Hawcroft2021}). These models are then selected in pairs and their parameters combined to create another generation of models, with selection preference given to models with higher fitness. This process is iterated until the fit to the data is no longer being improved significantly with each new generation. Various random mutations are applied to the model parameters at each generation to enhance the parameter space exploration.

Such a fitting technique is a marked improvement on the established 'by-eye' fitting method as there is now a consistent statistical evaluation criterion we can use to assess fit quality. The process is also automated, allowing a more thorough exploration of the parameter space. We note that equally high quality fits can be found using the 'by-eye' approach, \cite{Markova2020} found good agreement between GA and by-eye fitting approaches. In both cases appropriate spectral regions must be manually selected and prepared carefully (normalisation, contamination). In an additional pre-processing step, we remove clear instances of interstellar contamination so these contributions are not reproduced by the automated fitting routine. This is most common in the case of nebular emission in the cores of optical hydrogen recombination lines. The final spectra used for the fits are shown throughout Appendix \ref{sec:app-lmc-fits} and regions which have been removed are clearly visible.

We optimise a number of parameters using the GA and leave others fixed following certain assumptions. Thirteen of these parameters are inputs to the FASTWIND code, including effective temperature and surface gravity. Four abundance parameters are used to determine the surface abundances of helium, nitrogen, carbon and oxygen. The remaining FASTWIND inputs are wind parameters, including mass-loss rate $\dot{M}$, wind acceleration $\beta$, terminal wind speed $v_{\infty}$ and four clumping parameters. The clumping parameters are: the clumping factor $f_{\rm{cl}}$, velocity filling factor $f_{\rm{vel}}$, interclump density $f_{\rm{ic}}$ and the clumping onset velocity $v_{\rm{cl}}$ (which can also be expressed as the clumping onset radius $R_{\rm{cl}}$, and is defined further in Section 3.1 of \citealp{Hawcroft2021}). The final two free parameters are rotational and macroturbulent broadening which are applied to the synthetic spectra produced by FASTWIND in post-processing before the fitness to observations is assessed. All models are computed with LMC metallicity $Z=0.5Z_{\odot}$, meaning that all fixed abundances are scaled to half of solar where $Z_{\odot}$ is defined as in \cite{Asplund2009}. Taking distance to the LMC of 50 kpc (\citealp{Gibson2000}, \citealp{Pietrzynski2019}), Ks-band photometry (\citealp{Kato2007}, \citealp{Evans2011}) and an average K-band extinction (\citealp{RamirezAgudelo2017}) we estimate an absolute K-band magnitude which works as a calibration for stellar radius in combination with an input effective temperature.

We adopt a microturbulent velocity of 10 $\rm{km}\, \rm{s^{-1}}$ in the NLTE computation of FASTWIND, and during the formal integral we allow the microturbulence to increase linearly with wind speed as 0.1$v$ which results in an overall prescription of fixed microturbulence at the photosphere, which increases to a maximum in the outer wind at 0.1$v_{\infty}$. This approach is also taken in \cite{Abdul-Masih2021} and \cite{Hawcroft2021}. Similarly, the clumping parameters begin increasing at the input onset velocity and increase linearly until they reach their maximum at twice the onset velocity. The maximum values of the clumping parameters are therefore the free parameters, and these best-fit maximum clumping parameters are listed in Tables \ref{tbl: Wind Parameters} \& \ref{tbl:Parameters}.

We do not include the effects of X-rays in our models. Therefore we do not attempt to fit wind line profiles which are sensitive to the X-ray luminosity; these include UV metal lines such as N\,{\sc v}\,$\lambda \lambda$1239-1243. 

We focus on a number of key line diagnostics, around which we optimise the fit. In the UV we generally include  He\,{\sc ii}\,$\lambda$1640, C\,{\sc iv}\,$\lambda$1169, C\,{\sc iii}\,$\lambda$1176, C\,{\sc iv}\,$\lambda \lambda$1548-1551, C\,{\sc iii}\,$\lambda$1620, N\,{\sc iv}\,$\lambda$1718,  O\,{\sc iv}\,$\lambda \lambda$1340-1344, O\,{\sc v}\,$\lambda$1371, Si\,{\sc iv}\,$\lambda \lambda$1394, 1403 and P\,{\sc v}\,$\lambda \lambda$1118-1128. In the optical we have the Balmer series from H$\alpha$ to H$\delta$, He\,{\sc i}\,$\lambda$4009, He\,{\sc i}+\,{\sc ii}\,$\lambda \lambda$4025-4026, He\,{\sc i}\,$\lambda$4471, He\,{\sc i}\,$\lambda$4922, He\,{\sc ii}\,$\lambda$4200, He\,{\sc ii}\,$\lambda$4541, C\,{\sc iii}\,$\lambda \lambda$4068-70, N\,{\sc iii}\,$\lambda \lambda$4510-4514-4518, N\,{\sc iii}\,$\lambda \lambda$-4634-4640-4642 and N\,{\sc iv}\,$\lambda$4058. We note that some lines may be excluded or others included depending on spectral type, the full line list for each object can be seen in Appendix \ref{sec:app-lmc-fits}.

\section{Results}\label{sec: results}

\onecolumn

\begin{figure*}[t!]
	\centering
	\includegraphics[scale=0.35]{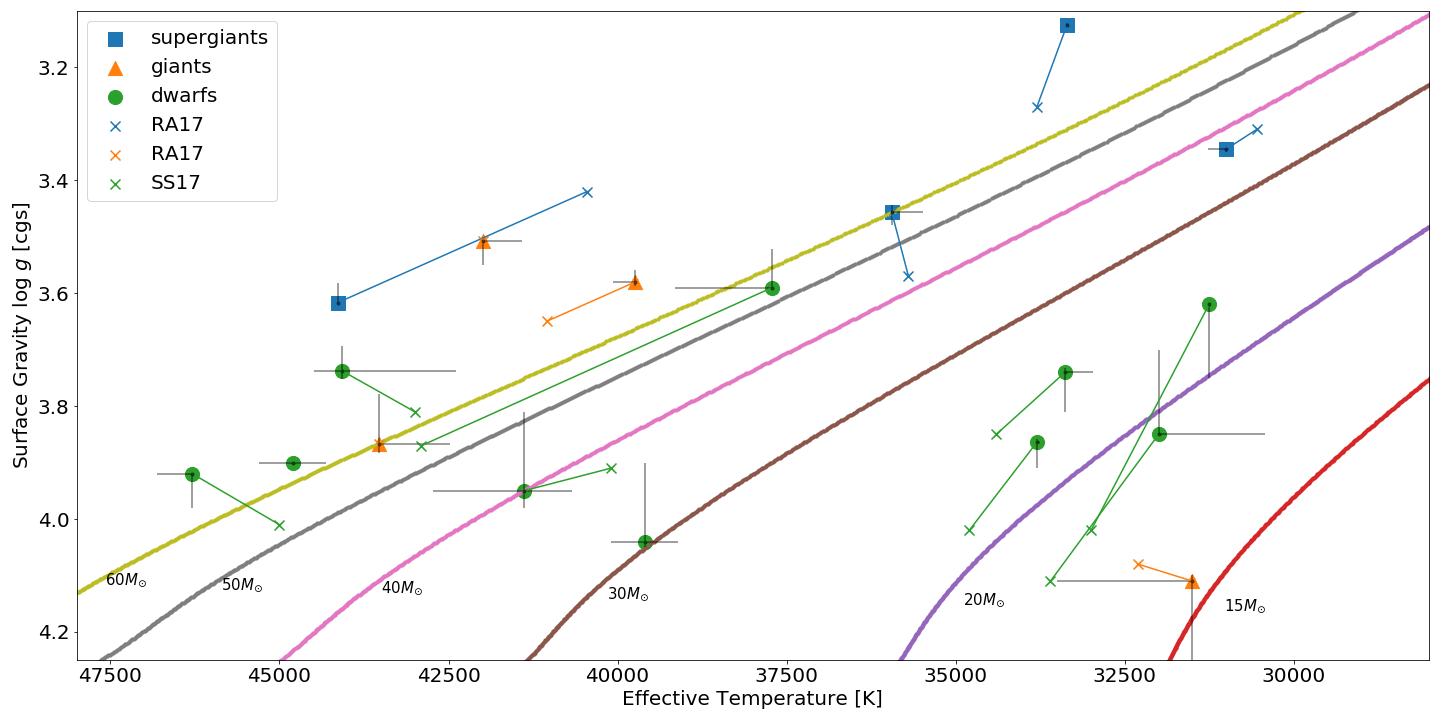}
	\caption{Kiel diagram showing best fit effective temperatures and surface gravities for the sample. Shapes represent the best-fit parameters from the GA, blue squares are the supergiants, orange triangles the giants and green circles the dwarf stars. Crosses are literature measurements of the same parameters from \citealp{RamirezAgudelo2017} (RA17) for (super)giants or \citealp{SabinSanjulian2017} (SS17) for dwarfs. Lines are drawn between points to guide the eye between measurements made for the same star from this work and literature studies. Overplotted against evolutionary tracks from \cite{Brott2011} at LMC metallicity $Z=0.5Z_{\odot}$ and moderate rotation $v \sin i \sim$ 150-200 $\rm{km}\, \rm{s^{-1}}$. Each evolutionary track is annotated with the input initial stellar mass.} 
	\label{fig: Kiel-diagram}
\end{figure*}

\begin{table}
	\tiny
	\centering
	\caption{Best-fit wind parameters from GA fitting including optically thick clumping. Statistical errors from the GA are provided alongside best-fit values, if the errors are underestimated by the GA we provide lower limits on the uncertainties instead. Stars separated at the bottom are those with no detectable wind features.}
	\label{tbl: Wind Parameters}
	\begin{tabular}{llllllllll}
		\hline
		\hline
		
		ID & SpT & log$(\frac{L}{L_{\odot}})$ & log$(\dot{M})$ & $\beta$  & $f_\mathrm{cl}$ & $f_\mathrm{vel}$ & $f_\mathrm{ic}$ & $v_\mathrm{cl}$ & $R_\mathrm{cl}$ \cr
		VFTS & & [dex] & [$M_{\odot}\rm{yr}^{-1}$] & & & & & [$v_{\infty}$] & [$R_{\rm{eff}}$] \\
		\hline
		180	&	O3If*	& $	5.98	\pm	0.07	$ & $	-5.72 \pm 0.05 $ & $	1.2 \pm 0.1 $ & $	22 \pm 0.1 $ & $	0.50	^{+	0.11	}_{-	0.05	}$ & $	0.12  \pm 0.05 $ & $	0.06 \pm 0.01 $ &	1.05	\\
		608	&	O4III(f)	& $	5.85	\pm	0.08	$ & $	-6.24 \pm 0.05 $ & $	2.0 \pm 0.1 $ & $	9 \pm 1 $ & $	0.11 \pm 0.1 $ & $	0.16 \pm 0.05 $ & $	0.08 \pm 0.01 $ &	1.04	\\
		096	&	O6V((n))((fc))z	& $	5.83	\pm	0.08	$ & $	-6.50	^{+	0.05	}_{-	0.45	}$ & $	2.1	^{+	0.1	}_{-	1.0	}$ & $	19	^{+	11	}_{-	14	}$ & $	0.90	^{+	0.02	}_{-	0.50	}$ & $	0.25	^{+	0.05	}_{-	0.20	}$ & $	0.20	^{+	0.01	}_{-	0.05	}$ &	1.11	\\
		216	&	O4V((fc))	& $	5.79	\pm	0.09	$ & $	-6.61	^{+	0.07	}_{-	0.16	}$ & $	1.8	^{+	0.1	}_{-	0.3	}$ & $	26	^{+	0	}_{-	17	}$ & $	0.43	^{+	0.56	}_{-	0.18	}$ & $	0.22	^{+	0.07	}_{-	0.16	}$ & $	0.13	^{+	0.01	}_{-	0.06	}$ &	1.08	\\
		422	&	O4III(f)	& $	5.69	\pm	0.08	$ & $	-6.48 \pm 0.05 $ & $	0.5	^{+	0.1	}_{-	0.5	}$ & $	23	^{+	0	}_{-	5	}$ & $	0.42 \pm 0.05 $ & $	0.01	^{+	0.09	}_{-	0.01	}$ & $	0.16	^{+	0.01	}_{-	0.07	}$ &	1.41	\\
		440	&	O6-6.5II(f)	& $	5.64	\pm	0.06	$ & $	-6.23 \pm 0.05 $ & $	1.6 \pm 0.1 $ & $	9 \pm 0.1 $ & $	0.14 \pm 0.05 $ & $	0.02	^{+	0.05	}_{-	0.01	}$ & $	0.04 \pm 0.01 $ &	1.02	\\
		244	&	O5III(n)(fc)	& $	5.58	\pm	0.08	$ & $	-6.40 \pm 0.05 $ & $	2.0	 \pm 0.1 $ & $	16	^{+	5	}_{-	0	}$ & $	0.19	^{+	0.81	}_{-	0.05	}$ & $	0.29	^{+	0.01	}_{-	0.07	}$ & $	0.09 \pm 0.01 $ &	1.04	\\
		664	&	O7II(f)	& $	5.55	\pm	0.07	$ & $	-6.39 \pm 0.05 $ & $	1.8	 \pm 0.2 $ & $	3	^{+	3	}_{-	0	}$ & $	0.11 \pm 0.05 $ & $	0.01	^{+	0.05	}_{-	0.00	}$ & $	0.03	^{+	0.02	}_{-	0.01	}$ &	1.02	\\
		143	&	O3.5V((fc))	& $	5.52	\pm	0.12	$ & $	-6.62 \pm 0.05 $ & $	2.0	 \pm 0.1 $ & $	3	\pm 1$ & $	0.74 \pm 0.05 $ & $	0.29	^{+	0.01	}_{-	0.05	}$ & $	0.06 \pm 0.01 $ &	1.03	\\
		586	&	O4V((n))((fc))z	& $	5.41	\pm	0.14	$ & $	-6.90	^{+	0.13	}_{-	0.03	}$ & $	1.6	^{+	0.1	}_{-	0.2	}$ & $	18	^{+	5	}_{-	8	}$ & $	0.46	^{+	0.05	}_{-	0.32	}$ & $	0.17	^{+	0.11	}_{-	0.10	}$ & $	0.11	^{+	0.03	}_{-	0.03	}$ &	1.08	\\
		385	&	O4-5V	& $	5.34	\pm	0.11	$ & $	-6.29 \pm 0.05 $ & $	0.6 \pm 0.1 $ & $	14 \pm 1 $ & $	0.13 \pm 0.05 $ & $	0.09 \pm 0.05 $ & $	0.03 \pm 0.01 $ &	1.05	\\
		087	&	O9.7Ib-II	& $	5.23	\pm	0.08	$ & $	-6.80	^{+	0.10	}_{-	0.00	}$ & $	3.7	^{+	0.1	}_{-	1.2	}$ & $	4 \pm 1 $ & $	0.13	^{+	0.08	}_{-	0.05	}$ & $	0.01	^{+	0.05	}_{-	0.00	}$ & $	0.17	^{+	0.01	}_{-	0.07	}$ &	1.05	\\
		184	&	O6.5Vnz	& $	5.05	\pm	0.16	$ & $	-8.29	^{+	0.00	}_{-	0.31	}$ & $	1.6	^{+	0.1	}_{-	0.5	}$ & $	16 \pm 5 $ & $	0.42	^{+	0.05	}_{-	0.14	}$ & $	0.21	^{+	0.05	}_{-	0.18	}$ & $	0.03	^{+	0.11	}_{-	0.01	}$ &	1.02	\\
		223	&	O9.5IV	& $	5.00	\pm	0.12	$ & $	-7.37	^{+	0.00	}_{-	0.47	}$ & $	1.0	^{+	0.8	}_{-	0.1	}$ & $	11 \pm 4 $ & $	0.13 \pm 0.05 $ & $	0.01	^{+	0.05	}_{-	0.00	}$ & $	0.05	^{+	0.04	}_{-	0.01	}$ &	1.05	\\ 
		\hline
		517	&	O9.5V-III((n))	& $	4.99	\pm	0.10	$ & $	-10.14	^{+	0.44	}_{-	0.00	}$ & $	1.2	^{+	0.5	}_{-	0.6	}$ & $	5.6	^{+	23	}_{-	3	}$ & $	0.10	^{+	0.33	}_{-	0.05	}$ & $	0.02	^{+	0.21	}_{-	0.00	}$ & $	0.07	^{+	0.06	}_{-	0.02	}$ &	1.06	\\
		280	&	O9V	& $	4.82	\pm	0.14	$ & $	-9.32	^{+	0.46	}_{-	0.34	}$ & $	0.7 \pm 0.1 $ & $	7	^{+	12	}_{-	3	}$ & $	0.14	^{+	0.17	}_{-	0.05	}$ & $	0.01	^{+	0.05	}_{-	0.00	}$ & $	0.18	^{+	0.01	}_{-	0.07	}$ &	1.35	\\
		235	&	O9.7III	& $	4.58	\pm	0.18	$ & $	-10.44	^{+	0.44	}_{-	0.06	}$ & $	1.2	^{+	0.8	}_{-	0.6	}$ & $	11	^{+	18	}_{-	7	}$ & $	0.92	^{+	0.05	}_{-	0.80	}$ & $	0.08	^{+	0.20	}_{-	0.06	}$ & $	0.17	^{+	0.02	}_{-	0.01	}$ &	1.17	\\
		627	&	O9.7V	& $	4.60	\pm	0.17	$ & $	-10.38	^{+	0.00	}_{-	0.22	}$ & $	1.7	^{+	0.2	}_{-	0.4	}$ & $	28	^{+	0	}_{-	25	}$ & $	0.12	^{+	0.79	}_{-	0.05	}$ & $	0.06	^{+	0.23	}_{-	0.05	}$ & $	0.03	^{+	0.15	}_{-	0.02	}$ &	1.02	\\
		\hline
		\\              
	\end{tabular}
\end{table}

\begin{landscape}
\begin{table}
	\tiny
	\caption{Best-fit photospheric \& wind parameters from GA fitting including optically thick clumping as in Table \ref{tbl: Wind Parameters}.}
	\vspace{2mm}
	\label{tbl:Parameters}
	\begin{tabular}{lllllllllllllll}
		\hline
		\hline
		ID & Sp Type & $T_{\rm{eff}}$ & log $g$ & $R_\mathrm{eff}$ & log$(\dot{M})$ & $v_{\infty}$ & $\beta$  & $v \sin i$ & $v_\mathrm{mac}$ & $N_{He}/N_{H}$ & $\epsilon$(C) & $\epsilon$(N) & $\epsilon$(O) \cr
		& & [K] & [cgs] & [$R_{\odot}$] & [$M_{\odot}\rm{yr}^{-1}$] & [$\rm{km}\, \rm{s^{-1}}$] &  & [$\rm{km}\, \rm{s^{-1}}$] & [$\rm{km}\, \rm{s^{-1}}$] & & &  \\
		\hline
180	&	O3If*	& $	44100	^{+	500	}_{-	500	}$ & $	3.62 \pm 0.05 $ & 	16.5	& $	-5.72 \pm 0.05 $ & $	2600	^{+	300	}_{-	100	}$ & $	1.2 \pm 0.1 $ & $	90 \pm 10$ & $	48	^{+	13	}_{-	10	}$ & $	0.30\pm 0.02$ & $	7.4	^{+	0.3	}_{-	0.9	}$ & $	8.9\pm 0.1$ & $	6.0	^{+	1.1	}_{-	0.1	}$ \\
608	&	O4III(f)	& $	43500	^{+	0	}_{-	1000	}$ & $	3.87	^{+	0.01	}_{-	0.09	}$ & 	14.8	& $	-6.24 \pm 0.05 $ & $	2600 \pm 100$ & $	2.0	 \pm 0.1 $ & $	127\pm 10$ & $	95	^{+	10	}_{-	89	}$ & $	0.13\pm 0.02$ & $	7.3\pm 0.1$ & $	8.3\pm 0.1$ & $	7.5	^{+	0.1	}_{-	0.4	}$ \\
096	&	O6V((n))((fc))z	& $	41400	^{+	1300	}_{-	700	}$ & $	3.95	^{+	0.03	}_{-	0.14	}$ & 	16.0	& $	-6.50	^{+	0.05	}_{-	0.45	}$ & $	2300	^{+	600	}_{-	300	}$ & $	2.1	^{+	0.1	}_{-	1.0	}$ & $	114\pm 20$ & $	85	^{+	14	}_{-	61	}$ & $	0.10	^{+	0.03	}_{-	0.02	}$ & $	7.1	^{+	0.1	}_{-	1.0	}$ & $	7.2	^{+	0.1	}_{-	1.1	}$ & $	7.1	^{+	1.8	}_{-	1.1	}$ \\
216	&	O4V((fc))	& $	44100	^{+	400	}_{-	1700	}$ & $	3.74 \pm 0.05 $ & 	13.4	& $	-6.61	^{+	0.07	}_{-	0.16	}$ & $	2900	^{+	100	}_{-	200	}$ & $	1.8	^{+	0.1	}_{-	0.3	}$ & $	116\pm 10$ & $	84	^{+	11	}_{-	14	}$ & $	0.11	^{+	0.02	}_{-	0.03	}$ & $	7.4	^{+	0.5	}_{-	0.1	}$ & $	7.3	^{+	0.2	}_{-	0.3	}$ & $	8.33	^{+	0.6	}_{-	0.3	}$ \\
422	&	O4III(f)	& $	42000	^{+	500	}_{-	600	}$ & $	3.51 \pm 0.05 $ & 	13.1	& $	-6.48 \pm 0.05 $ & $	2100 \pm 100$ & $	0.5	^{+	0.1	}_{-	0.5	}$ & $	200	^{+	31	}_{-	10	}$ & $	73\pm 10$ & $	0.09\pm 0.02$ & $	8.7\pm0.1$ & $	8.2	^{+	0.1	}_{-	0.3	}$ & $	7.0	^{+	0.5	}_{-	0.1	}$ \\
440	&	O6-6.5II(f)	& $	33400	^{+	500	}_{-	500	}$ & $	3.13 \pm 0.05 $ & 	19.7	& $	-6.23 \pm 0.05 $ & $	2400 \pm 100$ & $	1.6 \pm 0.1 $ & $	124\pm 10$ & $	91	^{+	7	}_{-	46	}$ & $	0.14\pm 0.02$ & $	7.9	^{+	0.1	}_{-	0.3	}$ & $	8.2	^{+	0.3	}_{-	0.1	}$ & $	7.6	^{+	0.2	}_{-	0.1	}$ \\
244	&	O5III(n)(fc)	& $	39700	^{+	500	}_{-	500	}$ & $	3.58 \pm 0.05 $ & 	13.0	& $	-6.40 \pm 0.05 $ & $	2600 \pm 100$ & $	2.0 \pm 0.1 $ & $	223	^{+	10	}_{-	17	}$ & $	98	^{+	10	}_{-	16	}$ & $	0.11\pm 0.02$ & $	7.3	^{+	0.1	}_{-	0.30	}$ & $	7.5\pm0.1$ & $	7.9	^{+	0.3	}_{-	0.1	}$ \\
664	&	O7II(f)	& $	35900	^{+	500	}_{-	500	}$ & $	3.46  \pm 0.05 $ & 	15.3	& $	-6.39 \pm 0.05 $ & $	2600	^{+	100	}_{-	400	}$ & $	1.8 \pm 0.2 $ & $	90	^{+	10	}_{-	17	}$ & $	98\pm 10$ & $	0.08\pm 0.02$ & $	8.0	^{+	0.1	}_{-	0.5	}$ & $	7.8	^{+	0.3	}_{-	0.1	}$ & $	7.9	^{+	0.8	}_{-	0.8	}$ \\
143	&	O3.5V((fc))	& $	44800	^{+	500	}_{-	500	}$ & $	3.90 \pm 0.05 $ & 	9.5	& $	-6.62 \pm 0.05 $ & $	2900	^{+	100	}_{-	100	}$ & $	2.0 \pm 0.1 $ & $	127	\pm 10$ & $	94\pm 10$ & $	0.11\pm 0.02$ & $	7.9\pm0.1$ & $	6.5\pm0.1$ & $	7.7\pm0.1$ \\
586	&	O4V((n))((fc))z	& $	46300	^{+	500	}_{-	500	}$ & $	3.92	^{+	0.06	}_{-	0.05	}$ & 	7.9	& $	-6.90	^{+	0.13	}_{-	0.03	}$ & $	2900 \pm 100$ & $	1.6	^{+	0.1	}_{-	0.2	}$ & $	78	^{+	23	}_{-	10	}$ & $	74	^{+	10	}_{-	63	}$ & $	0.11\pm 0.02$ & $	7.6	^{+	0.3	}_{-	0.1	}$ & $	6.4	^{+	0.1	}_{-	0.7	}$ & $	8.9\pm0.1$ \\
385	&	O4-5V	& $	37700	^{+	1400	}_{-	0	}$ & $	3.59	^{+	0.03	}_{-	0.07	}$ & 	10.9	& $	-6.29 \pm 0.05 $ & $	2700	^{+	100	}_{-	200	}$ & $	0.6 \pm 0.1 $ & $	119\pm 10$ & $	90\pm 10$ & $	0.06	^{+	0.03	}_{-	0.02	}$ & $	7.7	^{+	0.3	}_{-	0.1	}$ & $	7.2	^{+	0.2	}_{-	0.1	}$ & $	8.9\pm0.1$ \\
087	&	O9.7Ib-II	& $	31000	^{+	500	}_{-	500	}$ & $	3.35 \pm 0.05 $ & 	14.3	& $	-6.80	^{+	0.10	}_{-	0.00	}$ & $	2900 \pm 100$ & $	3.7	^{+	0.1	}_{-	1.2	}$ & $	66	^{+	12	}_{-	10	}$ & $	45	^{+	10	}_{-	15	}$ & $	0.10	^{+	0.03	}_{-	0.02	}$ & $	7.5	^{+	0.3	}_{-	0.1	}$ & $	7.8\pm0.1$ & $	7.5	^{+	0.9	}_{-	0.1	}$ \\
184	&	O6.5Vnz	& $	39600	^{+	2000	}_{-	0	}$ & $	4.04	^{+	0.01	}_{-	0.14	}$ & 	7.1	& $	-8.29	^{+	0.00	}_{-	0.31	}$ & $	2200 \pm 100$ & $	1.6	^{+	0.1	}_{-	0.5	}$ & $	323	^{+	32	}_{-	10	}$ & $	50	^{+	10	}_{-	40	}$ & $	0.10\pm 0.02$ & $	7.9\pm0.1$ & $	6.5\pm0.1$ & $	6.3	^{+	1.1	}_{-	0.1	}$ \\
223	&	O9.5IV	& $	33800	^{+	500	}_{-	500	}$ & $	3.86 \pm 0.05 $ & 	9.1	& $	-7.37	^{+	0.00	}_{-	0.47	}$ & $	2100	^{+	500	}_{-	100	}$ & $	1.0	^{+	0.8	}_{-	0.1	}$ & $	1	^{+	12	}_{-	1	}$ & $	59\pm 10$ & $	0.09\pm 0.02$ & $	7.8\pm0.1$ & $	7.5	^{+	0.2	}_{-	0.1	}$ & $	6.3	^{+	1.3	}_{-	0.2	}$ \\
\hline
517	&	O9.5V-III((n))	& $	31300	^{+	500	}_{-	500	}$ & $	3.62	^{+	0.13	}_{-	0.00	}$ & 	10.6	& $	-10.14	^{+	0.44	}_{-	0.00	}$ & $	2600	^{+	300	}_{-	500	}$ & $	1.2	^{+	0.5	}_{-	0.6	}$ & $	103	^{+	46	}_{-	48	}$ & $	76	^{+	9	}_{-	71	}$ & $	0.09	^{+	0.04	}_{-	0.02	}$ & $	8.1	^{+	0.6	}_{-	0.2	}$ & $	7.6	^{+	0.2	}_{-	0.6	}$ & $	7.5	^{+	1.0	}_{-	0.90	}$ \\
280	&	O9V	& $	33400	^{+	500	}_{-	500	}$ & $	3.74	^{+	0.07	}_{-	0.03	}$ & 	7.7	& $	-9.32	^{+	0.46	}_{-	0.34	}$ & $	700	^{+	400	}_{-	100	}$ & $	0.7 \pm 0.1 $ & $	155	^{+	15	}_{-	16	}$ & $	92\pm 10$ & $	0.10	^{+	0.05	}_{-	0.02	}$ & $	8.3	^{+	0.3	}_{-	0.3	}$ & $	6.7	^{+	0.6	}_{-	0.1	}$ & $	8.7	^{+	0.1	}_{-	2.4	}$ \\
235	&	O9.7III	& $	31500	^{+	2000	}_{-	0	}$ & $	4.11	^{+	0.21	}_{-	0.01	}$ & 	6.5	& $	-10.44	^{+	0.44	}_{-	0.06	}$ & $	2700	^{+	200	}_{-	600	}$ & $	1.2	^{+	0.8	}_{-	0.6	}$ & $	19	^{+	31	}_{-	17	}$ & $	50	^{+	19	}_{-	18	}$ & $	0.09	^{+	0.02	}_{-	0.03	}$ & $	8.1	^{+	0.3	}_{-	0.4	}$ & $	7.0	^{+	0.7	}_{-	1.0	}$ & $	7.3	^{+	1.6	}_{-	1.3	}$ \\
627	&	O9.7V	& $	32000	^{+	0	}_{-	1600	}$ & $	3.85	^{+	0.00	}_{-	0.15	}$ & 	6.5	& $	-10.38	^{+	0.00	}_{-	0.22	}$ & $	3000	^{+	100	}_{-	500	}$ & $	1.7	^{+	0.2	}_{-	0.4	}$ & $	29	^{+	49	}_{-	28	}$ & $	95	^{+	10	}_{-	22	}$ & $	0.12	^{+	0.02	}_{-	0.03	}$ & $	8.3	^{+	0.1	}_{-	0.4	}$ & $	7.9	^{+	0.1	}_{-	0.4	}$ & $	6.1	^{+	2.5	}_{-	0.1	}$ \\
		\hline
		\\         
	\end{tabular}
\end{table}
\end{landscape}

\vspace{2mm}

\twocolumn

We obtain synthetic spectral best-fit solutions to high resolution UV and optical spectra for 18 O-type stars. For each of these fits we have optimised 15 input parameters describing the stellar and wind properties using the GA based fitting technique described in Sect. \ref{sec: methods}. The final best-fit parameters are listed in Tables \ref{tbl: Wind Parameters} \& \ref{tbl:Parameters}. Individual fits and discussing thereof are included in Appendix \ref{appendix}.

Uncertainties are obtained by considering all models within a 95\% confidence interval of the best-fit model to be statistically equivalent solutions, as in previous GA studies (see e.g. \citealp{Abdul-Masih2019, Hawcroft2021}). As for the stellar radius ($R_{\rm{eff}}$, defined at Rosseland optical depth $\tau=2/3$) we follow the method of \cite{Repolust2004}, \cite{Mokiem2005} and find the uncertainty is dominated by the uncertainty in the K-band magnitude, which results in $\Delta R = 1.3 R_{\odot}$ generally corresponding to 10-20 \%, similar to the 15\% found by \cite{Repolust2004} who used V-band magnitudes.

A number of stars analysed here have been investigated in previous spectroscopic studies. We are therefore able to compare our best-fit parameters to literature values for some fundamental stellar parameters, keeping in mind that there may be systematic discrepancies as these previous studies have been made using only optical spectra, without consideration of optically thick clumps and often with differing optimisation techniques. The 4 supergiants (or bright giants) were previously analysed by \cite{RamirezAgudelo2017} using a GA approach on only optical spectra. 2 giants were also included in \cite{RamirezAgudelo2017}, and the 2 others were studied by \cite{Bestenlehner2014} using a grid-based approach with CMFGEN models. 8 of the dwarfs are included in \cite{SabinSanjulian2017} which uses a grid based approach to fit FASTWIND models to optical spectra. The 2 remaining dwarfs (VFTS143 and VFTS184) have not been included in any previous atmospheric analysis studies. A more thorough comparison with literature is presented on a star-by-star basis in Appendix \ref{appendix}, for this section we briefly compare to the fundamental spectral parameters, effective temperature and surface gravity. We focus on the effective temperature and surface gravity here as significant discrepancies in these parameters between studies, without sufficient explanation, would require further investigation.

For most of the dwarf stars we find effective temperatures that are in agreement with the spectral type calibrations of \cite{Walborn2014} within 1 kK. The supergiants and giants that have literature properties are in generally good agreement with our findings with differences in best-fit temperature and gravity on an individual basis that do not appear to be systematic, likely due to our inclusion of the UV diagnostics. This also seems to be the case for early dwarfs, however there may be a systematic difference between our method and the grid based approach of \cite{SabinSanjulian2017} for late dwarfs. We find generally lower temperatures and surface gravities. These differences are shown in Fig. \ref{fig: Kiel-diagram}. 

We include the surface abundances of carbon, nitrogen and oxygen as free parameters in our fit optimisation which provides important diagnostic information on CNO wind lines. We define e.g., $\epsilon$(C) as the ratio of carbon to hydrogen abundance by number through $\epsilon$(C) = log($n_{\rm{C}}/n_{\rm{H}}$)+12, where the baseline LMC CNO abundances are $\epsilon$(C)=7.75, $\epsilon$(N)=6.9 and $\epsilon$(O)=8.35 (\citealp{Kurt1998, Brott2011, Kohler2015}). These abundances are primarily used to estimate evolutionary status of massive stars as their ratios are expected to change as a result of the CNO cycle. We generally observe the expected enrichment with evolutionary stage (in that the ratios of N/C and N/O are larger for supergiants) but the only stars for which we can constrain these ratios relative to the values predicted for CNO processing, or even initial composition, are the supergiants and bright giants. For the rest of the sample the uncertainties are too large to comment on the CNO processing. For a comprehensive study on the evolutionary status and surface abundances a wider range of diagnostic line profiles for the relevant elements are required, and/or the contribution of these metal lines to the goodness-of-fit would need to be boosted relative to the strong wind line profiles. For our purposes including the abundances allows us to assess the degeneracies between abundances and wind parameters, mainly to determine whether the abundances are affecting wind parameters but this does not appear to be the case due to the aforementioned limited number of photospheric lines and their relative strengths.

For the mass-loss rates of stars with higher luminosity (5.2 < log$(L/L_{\odot})$ < 6.2) we find an average reduction of 6 times compared to theoretical predictions from \cite{Vink2001}. The mass-loss rates we determine here are within 20\% of the predictions from \cite{Bjorklund2021} on average. For low luminosity stars (log$(L/L_{\odot}$ < 5.2) we find very low mass-loss rates consistent with the 'weak-wind' problem (\citealp{Martins2005b, Marcolino2009, Najarro2011, deAlmeida2019}). However this does not appear to be a universal issue, as we find two dwarfs at the edge of the canonical weak-wind regime with fairly typical mass-loss rates (VFTS184 and VFTS223). 

\section{Discussion} \label{sec: discussion}

\subsection{Mass-loss rates}

Mass-loss rates determined with consistent fitting of UV and optical wind spectral features are thought to be closer to the true values, relative to mass-loss rates determined only using optical features. The mass-loss rate and clumping parameters are highly degenerate for diagnostics which are sensitive to the square of the density and so optical recombination lines (depending on $\rho^{2}$) can only be used to constrain a combined $\dot{M}\sqrt{f_{\rm{cl}}}$ factor. Simultaneous UV fitting breaks this degeneracy as UV resonance wind lines depend only linearly on density. Therefore, with coverage of both processes we can accurately constrain the mass-loss rate and clumping factor. However, there are still problems in determining the mass-loss rates when assuming that the wind clumps are optically thin. This is highlighted, for example, by difficulties in fitting the absorption components of P\,{\sc v} line profiles: generally the phosphorus abundance has to be reduced to unphysical values or the mass-loss rates reduced to such a degree that there are issues in reproducing other UV lines (\citealp{Pauldrach2001, Crowther2002, Hillier2003, Bouret2005, Bouret2012}). Recent progress has shown that an alternate solution is viable. It is possible to reproduce the P\,{\sc v} profiles if a more detailed wind structure is considered, specifically the limited velocity span of the clumps which allows for additional light leakage that can reduce the strength of absorption in high velocity blue-shifted components of P-Cygni UV lines (\citealp{Oskinova2007}, \citealp{Sundqvist2010}, \citealp{Sundqvist2011a}, \citealp{Surlan2013}). In addition, it has been shown that simultaneous fits to FUV, UV and optical wind features can be obtained when using a wind clumping prescription that allows clumps to become optically thick, considers clump velocity coverage, and the density of the interclump medium (see \citealp{Hawcroft2021} and \citealp{Brands2022}). The mass-loss rates presented here are determined with models including these effects, and the same method (GA fitting of UV and optical spectra) as outlined in these previous studies.

\begin{figure}[t]
    \centering
    \includegraphics[scale=0.28]{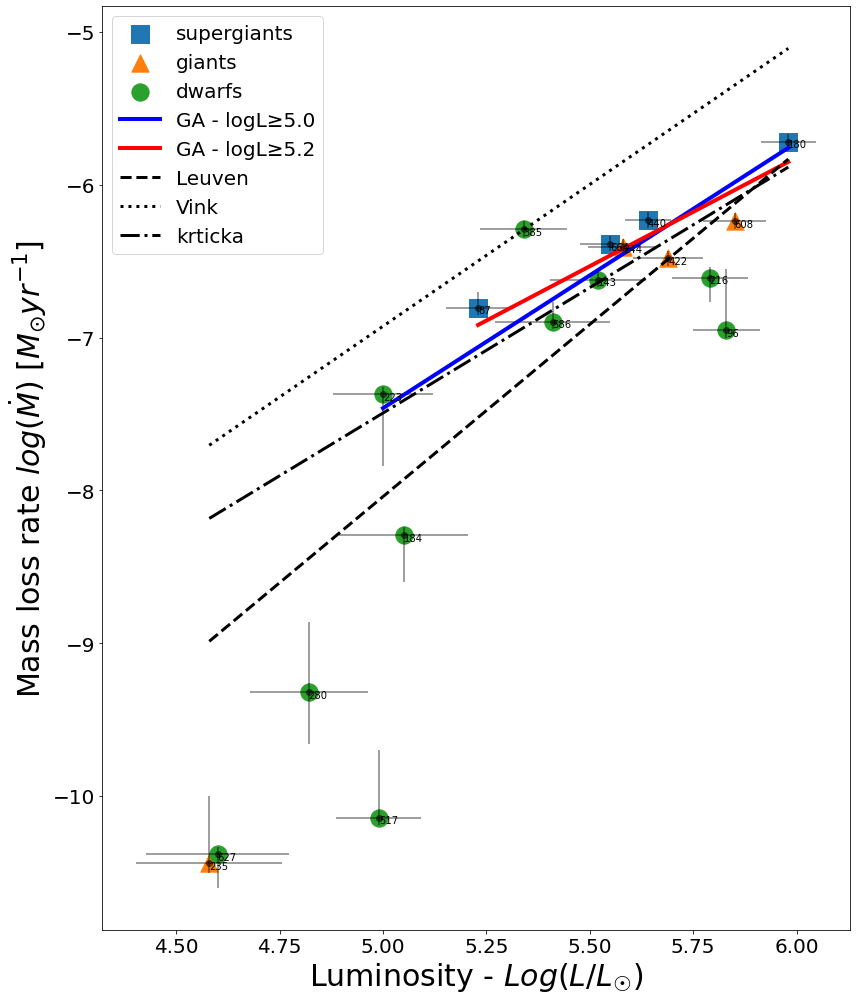}
    \caption{Mass-loss rates from GA best fits, supergiants are blue squares, giants are orange triangles and dwarfs are green circles. A linear fit to the GA best-fit mass-loss rates for stars with log$(L/L_{\odot}) \geq 5.0$ is shown by the solid line. Theoretical predictions tailored to our best-fit stellar parameters from Vink \citep{Vink2001}, Krti{\v{c}}ka \citep{Krticka2018} and Leuven \citep{Bjorklund2021} are shown in dotted, dot-dashed and dashed lines, respectively. } 
    \label{fig: Luminosity-Massloss}
\end{figure}

\begin{figure}[t!]
	\centering
	\includegraphics[scale=0.28]{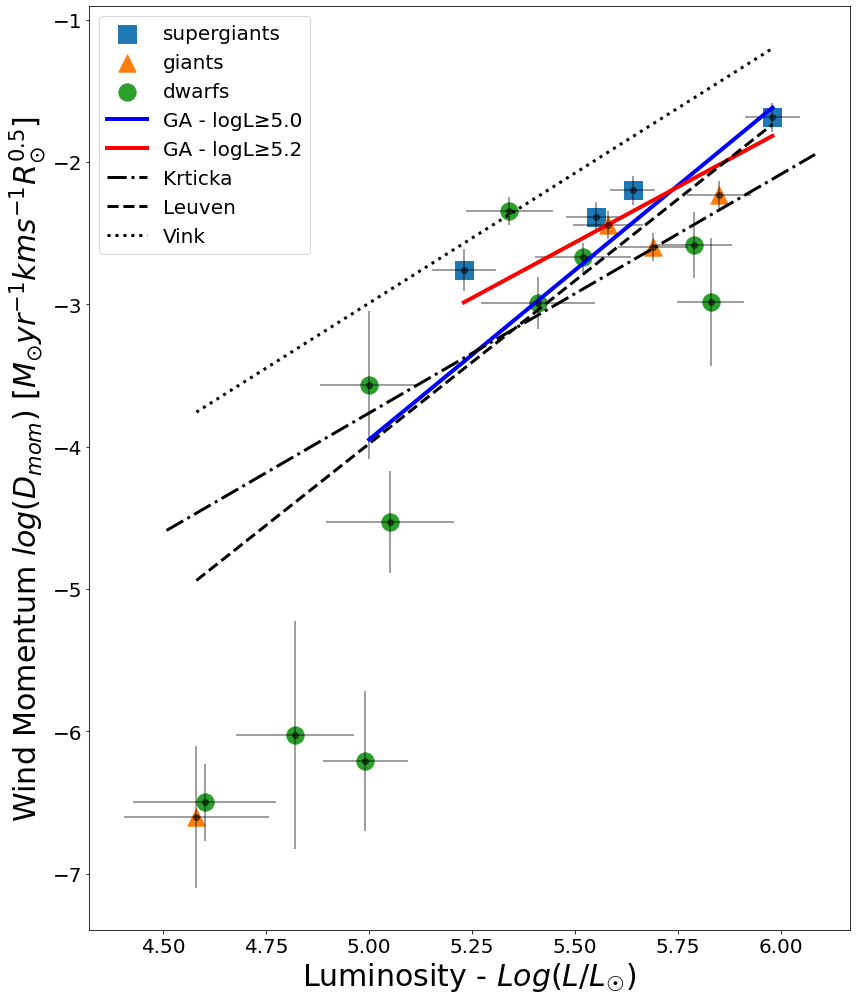}
	\caption{Modified wind-momentum rate computed using GA best fit values compared to the predictions made by Leuven (Bj{\"o}rklund et al. 2021), Krti{\v{c}}ka \citep{Krticka2018} and Vink (Vink et al. 2001) for each star in our sample. Supergiants are blue squares, giants are orange triangles and dwarfs are green circles. A linear fit to the wind-momentum calculated using the GA best-fit values for stars with log$(L/L_{\odot}) \geq 5.0$ is shown by the solid line. Theoretical predictions tailored to our best-fit stellar parameters from Vink \citep{Vink2001} and Leuven \citep{Bjorklund2021} are shown in dotted and dashed lines, respectively. Theoretical predictions for LMC stars from Krti{\v{c}}ka \citep{Krticka2018} are shown in dot-dashed line, although these are not tailored to the GA results.} 
	\label{fig: windmom_lmc}
\end{figure}

We compare mass-loss rates found with the GA to the mass-loss rates predicted from theoretical recipes using the relevant GA best-fit parameters as inputs, this is shown in Fig. \ref{fig: Luminosity-Massloss}. For the portion of the sample with typical O star winds (log$(L/L_{\odot}) > 5.2$) we find that the observationally derived rates are, on average, four times lower than those predicted using the \cite{Vink2001} rates. This is after accounting for the different metallicities used to compute the theoretical predictions, with \cite{Vink2001} using a solar metallicity of $Z_{\odot}=0.019$ from \cite{Anders1989} while \cite{Krticka2018} and \cite{Bjorklund2021} use $Z_{\odot}=0.013$ from \cite{Asplund2009}. However, this $\sim$40\% adjustment for input metallicity may be an over-correction, as \cite{Sundqvist2019} find only a $\sim$20\% difference in mass-loss rates computed with the two different input metallicities in their wind models. Therefore the empirical mass-loss rates we find could be up to 5 times lower than the predictions from \cite{Vink2001}. This factor of four is a slightly larger discrepancy than that found in previous studies. A reduction of factor three was found by \cite{Crowther2002} for LMC O-supergiants, a factor three by \cite{Bouret2005} who studied an O dwarf and supergiant in the Galaxy, a factor three by \cite{Bouret2003} for SMC O-dwarfs, a factor three found by \cite{Evans2004} for late O and early B-type supergiants in the LMC and SMC and a factor two to three by \cite{Bouret2012} and \cite{Hawcroft2021} for Galactic O-supergiants, a factor two by \cite{Brands2022} for O stars in the LMC, a factor three by \cite{Surlan2013}. \cite{Cohen2014} find a factor of three from X-ray diagnostics. Our results are also slightly lower than the factor of five reduction found by \cite{Massa2003} for LMC O-stars, which is a study using only UV spectra from the Far Ultraviolet Spectroscopic Explorer (FUSE). We find a mix of over- and under-predictions of mass-loss rates from \cite{Krticka2018} and \cite{Bjorklund2021}, although the agreement is generally good. We assess the ability of the predictions to reproduce the empirical results ($\chi^{2}_{\rm{red, Vink}} = 31.3$, $\chi^{2}_{\rm{red, Krticka}} = 5.4$ and $\chi^{2}_{\rm{red, Leuven}} = 12.9$), finding the best agreement with the predictions from \cite{Krticka2018}. This is similar to the results of \cite{Brands2022} who find a similar trend in goodness-of-fit between the predictions. We also compare the modified wind-momentum rate (Fig. \ref{fig: windmom_lmc}) and find similar results.

Below a luminosity of log$(L/L_{\odot}) =  5.2$ we notice a significant downward trend in derived mass-loss rates, entering the canonical weak-wind regime. Thanks to a combination of the flexibility of our fitting technique and sufficient computational resources, we are able to explore the lowest limit of mass-loss rate within the capabilities of FASTWIND. The mass-loss rates we find for the weak-wind stars are between $10^{-9}$ and $10^{-11}$ $M_{\odot}\rm{yr}^{-1}$. This is much lower than those found for similar stars in the LMC in \cite{Brands2022}, although this is likely only due to a lower range of mass-loss rates being considered, as these authors find uncertainties on the mass-loss rates for similar stars which extend to the lower limit of their parameter space. The rates we find are therefore also lower than equivalent Galactic stars, with dwarfs presented in \cite{Martins2005b} and \cite{Marcolino2009}, and giants analysed in \cite{Mahy2015} and \cite{deAlmeida2019}. Again lower than the mass-loss rates of $10^{-9}$ and $10^{-8}$ $M_{\odot}\rm{yr}^{-1}$ found in late-type SMC dwarfs by \cite{Bouret2003} and \cite{Martins2004}. The comparison between the mass-loss rates determined here and previous studies is generally more complex as these studies do not account for optically thick clumps or wind porosity in velocity space (except \citealp{Brands2022}), although in the case of weak-wind stars we are unable to constrain these additional clumping parameters and so perhaps a more straightforward comparison can be made. The mass-loss rates we find in this regime are on average 270, 140 and 30 times lower than theoretical predictions from \cite{Vink2001}, \cite{Krticka2018} and \cite{Bjorklund2021} respectively. 
Although for two dwarf stars with $5.0 <$ log$(L/L_{\odot}) < 5.2$ we find mass-loss rates which are not too far from the theoretical predictions perhaps suggesting a gradual onset of the weak-wind problem. If we move these stars out of the weak-wind sample for the purpose of measuring average scalings we have a reduction of 4.8 relative to \cite{Vink2001} for the typical O star, and a reduction of nearly 400 times in the weak-wind regime. There is little change to the average agreement with \cite{Bjorklund2021} and \cite{Krticka2018} for normal O stars ($\chi^{2}_{\rm{red, Vink}} = 35.2$, $\chi^{2}_{\rm{red, Krticka}} = 5.8$ and $\chi^{2}_{\rm{red, Leuven}} = 14.8$) but the discrepancy for weak-wind stars rises to a factor 50 compared to \cite{Bjorklund2021} and a factor 200 compared to \cite{Krticka2018}. We note that there is a more significant change in the empirical trend for the modified wind-momentum rate when these two stars are included. Due to the lack of a wind-momentum recipe from the \cite{Krticka2018} models we cannot make comparisons for all predictions and so cannot quantify the best agreement with our empirical results but qualitatively there is very good agreement with \cite{Bjorklund2021}.

\subsection{Clumping Properties}

\begin{figure}[t]
    \centering
    \includegraphics[scale=0.28]{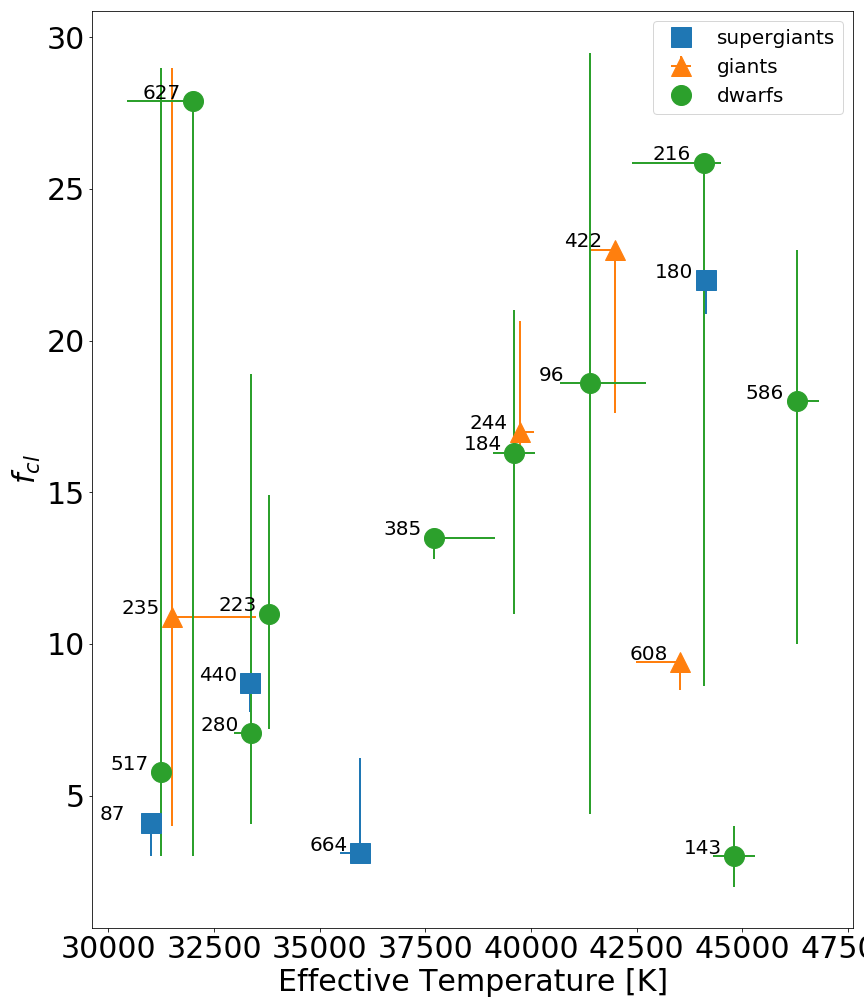}
    \caption{Best fit clumping factors from the GA, with $f_{\rm{cl}}$ for supergiants shown by blue squares, orange triangles for giants and green circles for dwarfs. We find weighted average values of $f_{\rm{cl}} = 16 \pm 1$ and $f_{\rm{cl}} = 7 \pm 1$ above and below 38 kK respectively, with an overall average $f_{\rm{cl}} = 11 \pm 1$. For weak-winds stars there are little to no wind signatures in the spectrum, and so $f_{\rm{cl}}$ cannot be reliably determined with the current method.} 
    \label{fig: clumping-factors}
\end{figure}

\begin{figure}[ht]
    \centering
    \includegraphics[scale=0.28]{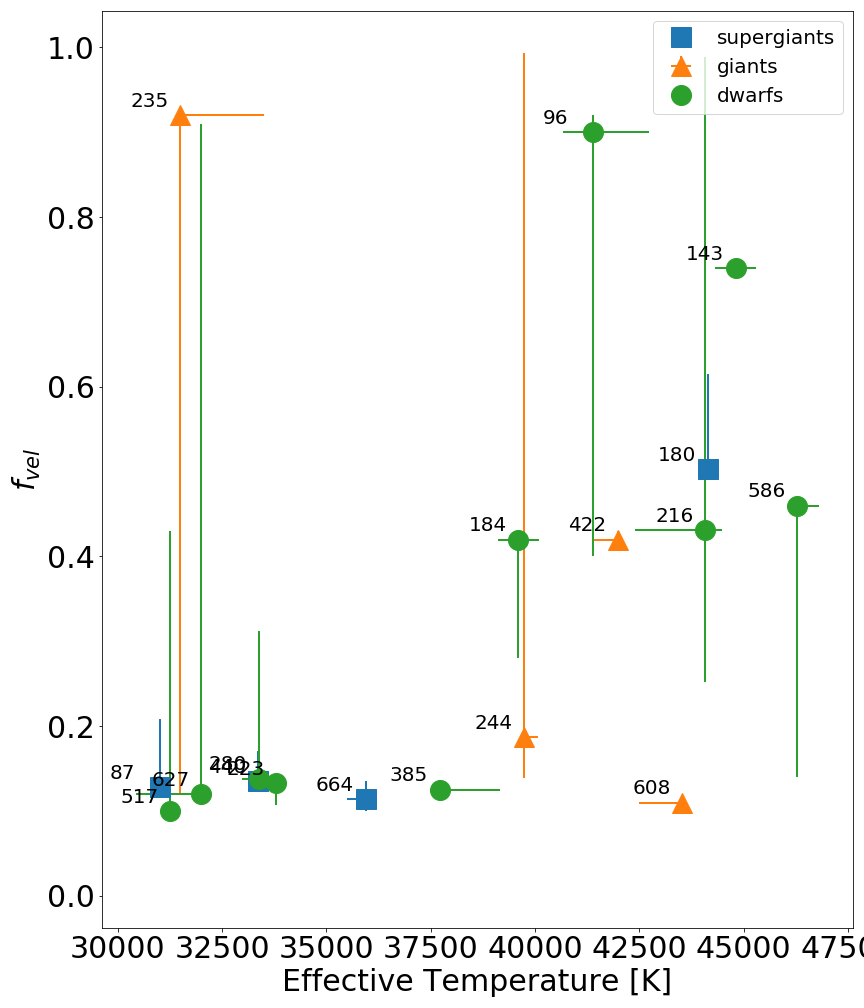}
    \caption{Best fit velocity filling factors from the GA, with $f_{\rm{vel}}$ for supergiants shown by blue squares, orange triangles for giants and green circles for dwarfs. We find average values of $f_{\rm{vel}} = 0.44 \pm 0.05$ and $f_{\rm{vel}} = 0.13 \pm 0.04$ above and below 38 kK respectively. } 
    \label{fig: Teff-fvel}
\end{figure}

\begin{figure}[ht]
    \centering
    \includegraphics[scale=0.28]{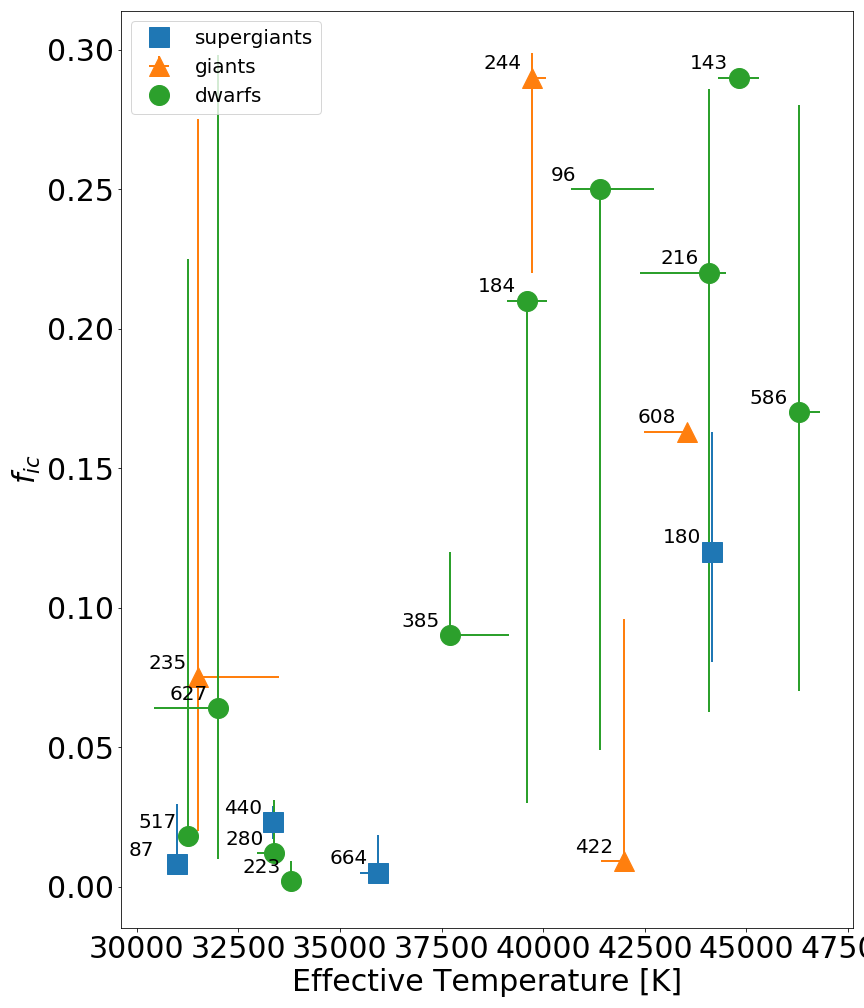}
    \caption{Best fit interclump density factors from the GA, with $f_{\rm{ic}}$ for supergiants shown by blue squares, orange triangles for giants and green circles for dwarfs. We find average values of $f_{\rm{ic}} = 0.18 \pm 0.04$ and $f_{\rm{ic}} = 0.03 \pm 0.04$ above and below 38 kK respectively.} 
    \label{fig: Teff-fic}
\end{figure}

\subsubsection{Stars with typical winds}

Best-fit clumping factors are shown in Fig. \ref{fig: clumping-factors} in relation to effective temperature. In our models we implement a linearly increasing clumping factor from an onset velocity $v_{\rm{cl}}$, which is also a free parameter, to the maximum clumping factor which is reached at 2$v_{\rm{cl}}$ and remains constant throughout the rest of the wind. We find clumping onset velocities generally close to the photosphere ($< 1.1 R_{\rm{eff}}$). Although there are a few exceptions with large uncertainties, which may be due to the loss of recombination line cores to nebular contamination. It is difficult to identify overall trends due to the differences in relations between luminosity classes. For late O-type supergiants we find distinctly low clumping factors, $f_{\rm{cl}} < 10$. The single early supergiant in the sample has $f_{\rm{cl}} \sim 20$, typical of Galactic supergiants \citep{Bouret2012, Hawcroft2021}. For the giants the clumping factors are intermediate, between 10-25, the low temperature weak-wind giant VFTS235 (O9.7III) may be closer to a dwarf with a high log $g$. For the dwarf stars there appears to be a preference towards intermediate values centred around $f_{\rm{cl}} \sim 15-20$ with high uncertainties. Two hot dwarfs have very well constrained clumping factors, one has a lower limit on the clumping factor of 25 and the other an upper limit of 6. Two of the outliers in Fig. \ref{fig: clumping-factors} (VFTS143 and VFTS608) are reported as single-lined spectroscopic binaries with large amplitude radial velocity variability in \cite{Sana2013} and so we have difficulties in fitting these objects, discussed further in Appendix \ref{appendix}. To quantify our search for trends, we compute a number of correlation metrics including the Pearson ($r_{p}$), Spearman ($r_{s}$) and Kendall ($\tau_{k}$) coefficients with corresponding p-values ($p_{p}$, $p_{s}$ and $p_{\tau}$, the probability that an uncorrelated data-set would produce the $r$ value found, using the relevant functions from SciPy, \citealp{SciPy}), between the clumping parameters and a range of stellar parameters. We do not find any statistically significant correlations between $f_{\rm{cl}}$ and any other parameters but the stellar parameter with the largest correlation coefficient with clumping factor is the effective temperature ($r_{p} = 0.37: p_{p} = 0.13$, $r_{s} = 0.35: p_{s} = 0.15$, $\tau_{k} = 0.33: p_{\tau} = 0.06$). We find strong and statistically significant correlations with temperature for $f_{\rm{vel}}$ and $f_{\rm{ic}}$ ($f_{\rm{vel}}$, $\tau_{k} = 0.38, p_{\tau} = 0.03$; $f_{\rm{ic}}$, $\tau_{k} = 0.42, p_{\tau} = 0.02$). \cite{Brands2022} also look for correlations between $f_{\rm{cl}}$ and other stellar parameters and do not find any statistically significant trends. The strongest relation found by these authors is a decrease in clumping factor in stars with log$(\dot{M}) > -6$, these stars correspond with log$(L/L_{\odot}) > 6.0$ which is outside of our parameter range and so we cannot comment on such a trend. However, it is suggested that at very high luminosity the characteristics of the wind launching mechanism changes which can lead to lower clumping factors \citep{Moens2022}. \cite{Brands2022} do find significant correlations between $f_{\rm{vel}}$ and $f_{\rm{ic}}$ with other stellar parameters, including luminosity, effective temperature and mass-loss rate. This is in good agreement with our findings, and may suggest that our inclusion of low luminosity stars (log$(L/L_{\odot}) < 5.2$) with higher effective temperatures (relative to the sample of \citealp{Brands2022}) allows us to highlight the fact that these wind parameters have a stronger correlation with temperature than other stellar parameters.

It is not immediately clear why the effective temperature would have such an effect on these parameters but there is evidence for a trend with temperature as defined in \cite{Driessen2019}, their equation 6. These authors find that the growth rate of the line-deshadowing instability ($\Omega$) is related to effective temperature as $\Omega \propto T_{\rm{eff}}^{4}$. This is in qualitative agreement with the increasing trend of clumping parameters with temperature which we find empirically, assuming that the growth rate generally reflects the wind structure as parameterised in our models.

The 2D LDI simulations of O stars at low metallicity from \cite{Driessen2022} show that clumping factors derived from 2D LDI simulations are lower than those found from 1D simulations as clumps of dense gas can expand into additional physical space provided by the second dimension, effectively allowing some wind smoothing. These authors also find a metallicity dependence of the clumping factor $f_{\rm{cl}} \propto Z^{0.15 \pm 0.01}$, and further predict average clumping factors around 14 in their simulations at LMC metallicity. This is fairly close to the average clumping factor we find for this sample, from stars with temperatures above 38 kK, of $16\pm1$. 

Using a weighted average (where values with lower uncertainties are given higher weight) we find intermediate velocity filling for stars with temperatures above 38 kK (of $f_{\rm{vel}} = 0.44\pm0.05$), which matches the average found for Galactic O-supergiants ($f_{\rm{vel}} = 0.44$) in \cite{Hawcroft2021}. For the relatively low temperature part of the sample ($T_{\rm{eff}} < 38$ kK) we find an average of $f_{\rm{vel}} = 0.13 \pm 0.04$ Although, as we discuss in Sect. \ref{sec: weak-wind}, it might be that we are not actually constraining $f_{\rm{vel}}$ in these stars as, for the spectral lines available, there may be little to no remaining high velocity material in the outflow. In this case, the statistical constraint on $f_{\rm{vel}}$ is a result of changes in the model profiles that may not be representative of the true physical processes shaping the observed line profile, but nevertheless cause similar morphological changes.

There is a similar trend in inter-clump density as we observe in velocity filling, suggesting again that there is no diagnostic for inter-clump density in the UV for weak-wind stars and the values obtained are not likely to be quantitatively reliable. For the stars with strong winds we find a large range of inter-clump densities from 10-30\% of the mean wind density. This is in agreement with \cite{Hawcroft2021}, although it is generally difficult to accurately constrain $f_{\rm{ic}}$ for O-supergiants even with high quality UV and optical spectra, as it has a similar but weaker effect on unsaturated UV spectral lines as $f_{\rm{vel}}$, and so it is only more difficult in the generally lower S/N observations of this LMC sample. 

\subsubsection{Weak-wind stars} \label{sec: weak-wind}

For the four definite weak-wind stars (log$(L/L_{\odot}) < 5.0$, VFTS517, VFTS280, VFTS235 and VFTS627) we find no constraint of $f_{\rm{cl}}$. It is well established that the optical features we rely on, e.g. H$\alpha$, are not sensitive to wind conditions at low $\dot{M}$ and that near-infrared and infrared diagnostics are required (\citealp{Puls2006}; \citealp{Najarro2011}). Additionally, UV lines are equally uncertain wind diagnostics at low $\dot{M}$ if the X-ray properties of the star are not well constrained. Although for two of the cooler stars (VFTS280 and VFTS223) we do find statistically significant constraints on the clumping factors, and in these cases the factors are low. 

Simulations of the LDI in OB stars show that, for a relatively cool O-dwarf ($\sim$ 30 kK), the wind is less dense and comprised of large shock-heated regions which are unable to cool, with increased ionisation resulting in lower UV line opacities and thus the spectral appearance of a weak wind \citep{Lagae2021}. It may be that this process is also driving the lower clump filling in velocity space we empirically find here. 

Fundamentally, a low velocity filling of high-density material means more blue-shifted light in strong UV P-Cygni line profiles can escape, reducing absorption at in the blue-shifted component of the P-Cygni profile and in extreme cases returning the feature to continuum levels. However, it might be that our low values of $f_{\rm{vel}}$ are here actually mimicking the effects of a highly ionised bulk wind. Namely, if the bulk wind is hot and highly ionised at large velocities, there will be no remaining feature of the strong UV metal lines we rely on, the only signature available is photospheric. This is essentially the same spectral signature in synthetic spectra as no clump coverage, explaining the convergence to low values. Studies of the solar corona such as \cite{Mazzotta1998} predict carbon to become rapidly ionised at temperatures greater than $10^{6}$ K, while the LDI simulations of \cite{Lagae2021} find the average temperature of the wind to be at least $10^{6}$ K with large pockets of gas (spanning $\approx 0.1 R_{*}$) reaching even higher temperatures. These profile morphologies and fitness distributions are consistent in all low-temperature dwarfs. There are no low-temperature giants to compare with. 
As for the 3 supergiants with low temperatures we find also low velocity filling but there are clear P-Cygni profiles in these stars. The presence of wind features is to be expected in supergiants as the wind cooling is predicted to be much more efficient, meaning the typical diagnostic lines remain and the wind signatures are still visible. 

This change in wind feature morphology (from visible P-Cygni profiles to only absorption) corresponds well with similar studies of SMC dwarf stars. \cite{Bouret2013} notice this change in morphology around spectral type O6-7V at 38-40 kK. The only exception being AvZ 429 as an O7V star with wind features, although for this star there is no optical spectrum. In other studies of SMC dwarfs (\citealp{Bouret2013}, \citealp{Bouret2021}, \citealp{Marcolino2022}, and in our study) the models do not reproduce the strong absorption in C\,{\sc iv}\,$\lambda \lambda$1548-1551 in the near photospheric profile. It is therefore unclear what is missing in the models to create this strong absorption at low velocity. 

The lack of wind signatures in this context raises the question, whether the mass loss in weak-wind stars is truly as low as we measure here or whether we cannot constrain mass loss with the usual diagnostics. While the mass-loss rate cannot be high, otherwise wind signatures would be observed, it is possible that the extremely low mass-loss rates allowed in the GA solutions are under-estimates. Note that previously it has not been computationally worthwhile to explore these extremely low mass-loss rate solutions, due to lack of diagnostics in optical studies, and this is likely why we find statistically acceptable solutions down to the lowest end of the capabilities of the FASTWIND code and previous studies do not. It might be that high degree of ionisation means we lose diagnostics in the UV and would need to search at even higher ionisation states in the FUV or X-rays. Indeed, some studies suggest weak-wind stars have been shown to host larger X-ray luminosities than giants and supergiants (\citealp{Oskinova2006}, \citealp{NebotGomez-Moran2018}). Additionally, there is observational evidence that, when using X-ray diagnostics, the mass-loss rates determined for 'weak-wind' stars are in good agreement with those found for typical main sequence OB stars (\citealp{Huenemoerder2012}). This problem could also be alleviated with near-infrared and infrared observations, the hydrogen Brackett series at these wavelengths is far more sensitive to wind physics, for stars with weak winds, than those in the optical \citep{Najarro2011}. The Brackett lines are also formed in close photospheric layers, and so might be relatively unaffected by the shock-heated outer wind. Or it may be that the low density of these winds means that coupling between the accelerated metal ions and the abundant hydrogen and helium ions is no longer valid (\citealp{Castor1976, Springmann1992, Babel1996, Krticka2001, Owocki2002}), and so the mass-loss rates are actually capable of becoming as low as we observe in this sample.

\subsection{Terminal Wind Speeds} \label{sec:lmc-vinf}

For 12 of the stars in this sample the P-Cygni profile of C\,{\sc iv}\,$\lambda \lambda$1548-1551 is strong enough to show extended blue-shifted absorption from which we can easily diagnose the terminal wind speed. 6 of the 8 late-type subgiants and dwarfs (those with spectral types of O7 or later) do not have sufficient wind signatures to be able to measure a terminal wind speed using the diagnostics covered by observations used in this work. These parameters are left free in the GA fitting but our ability to constrain them are discussed on a case-by-case basis for the relevant stars in Appendix \ref{appendix}. We compare the terminal wind from the GA to the value computed using the recipe provided in \cite{Hawcroft2023}. These authors have derived a new empirical calibration for the terminal wind speed as a function of temperature and metallicity, based on an analysis of 150 OB stars in the LMC and SMC from the ULLYSES programme and the Galactic sample of \cite{Prinja1990}. For 9 of the 12 stars we find good agreement between the calibration and GA results, within errors provided by the GA, four of these agree with the best-fit value within 100 $\rm{km}\, \rm{s^{-1}}$ and the others generally have large errors on the terminal wind speed from the global fitness metric but the fitness distributions of C\,{\sc iv}\,$\lambda \lambda$1548-1551 peak closer to prediction. The three stars in disagreement (VFTS440, VFTS385 and VFTS087), each seem to be discrepant for different reasons. VFTS087 has a much larger terminal wind speed than predicted from \cite{Hawcroft2023} and actually reaches the top of the parameter space in the GA, although it is possible that the main issue here is the normalisation in the UV as the GA fit to the bluest edge of the profile is poor. We can test this by directly measuring the velocity of maximum absorption in the P-Cygni trough. VFTS440 has a much better GA fit to C\,{\sc iv}\,$\lambda \lambda$1548-1551 but there is a secondary peak in the exploration between 2000 and 2100 $\rm{km}\, \rm{s^{-1}}$, close to the lower boundary. The prediction underestimates $v_{\infty}$ relative to the GA fit. Again here it is useful to check with an alternate method. In VFTS385 we again have a good fit to a saturated profile although there is some disagreement between the best-fit minimum $\chi^{2}$ and maximum fitness. This disagreement between fitness metrics is clearest in the temperature, as the maximum fitness is close to 40 kK, around 2 kK higher than the minimum $\chi^{2}$. Using 40 kK as input to the prediction of \cite{Hawcroft2023} would increase $v_{\infty}$ by around 200 $\rm{km}\, \rm{s^{-1}}$, bringing the discrepancy down to near 300 $\rm{km}\, \rm{s^{-1}}$. 

\section{Conclusions} \label{sec: conclusions}

We have determined stellar and wind properties for a sample of 18 O-type stars in the LMC including dwarfs, giants and supergiants with early to late spectral types. This is achieved through simultaneous spectroscopic fitting of UV and optical stellar and wind diagnostic line profiles using the code FASTWIND, optimised with a genetic algorithm. Our main goal was to investigate mass-loss rates and clumping properties of these stars. With this goal in mind we include the effects of optically thick clumping and clump coverage in velocity space. We find:

- Well defined empirical mass-loss rates that agree well with the theoretical predictions from \cite{Krticka2018} and \cite{Bjorklund2021} for stars within a luminosity range 5.2 < log$(L/L_{\odot})$ < 6.2, and the best agreement is with the predictions from \cite{Krticka2018}. These empirical rates are roughly four to five times lower than those predicted by \cite{Vink2001}. This is a slightly larger discrepancy than the average scaling factors (around a factor 3) found in previous studies which do not include the effects of optically thick clumping.

- For dwarf-type stars with low luminosity, log$(L/L_{\odot})$ < 5.2, mass-loss rates are orders of magnitude lower than predicted. This is attributed to the 'weak-wind' problem. It is suggested that this discrepancy is due to shock-heating ionising metals used to diagnose wind properties, and so no discernible wind features remain in the UV spectra. Physical motivations for this are outlined in theoretical studies of the relatively low density winds of O dwarfs (\citealp{Lucy2012}, \citealp{Lagae2021}). In this case it is likely that mass-loss rates determined using typical UV diagnostics are lower limits, generally we find mass-loss rates lower than previous empirical studies as we explore lower mass-loss rate solutions, this is supported by our weak-wind mass-loss rates upper error limits approaching the lower estimates in the literature.

- The strongest and most significant trend we find for the clumping factor with other stellar parameters is a positive correlation with the effective temperature. The average throughout the sample is found to be  $f_{\rm{cl}} = 11 \pm 1$, this depends somewhat on luminosity class, due to the uneven distribution of stars throughout the effective temperature range available. The average is in agreement with theoretical predictions of the clumping factor for hot ($\sim 40$ kK) O-supergiants from 2D LDI simulations \citep{Driessen2022}. However if we consider splitting the sample into hot stars ($T_{\rm{eff}} > 38$ kK) with typical winds (log$(L/L_{\odot})$ > 5.2) and cooler weak-wind stars, we find $f_{\rm{cl}} = 16 \pm 1$ and $f_{\rm{cl}} = 7 \pm 1$ respectively.

- For hot stars ($T_{\rm{eff}} > 38$ kK) with typical winds (log$(L/L_{\odot})$ > 5.2), we find intermediate velocity filling factors, $f_{\rm{vel}} = 0.44 \pm 0.05$ on average. This suggests that just under half of the wind velocity profile is covered by clumps. This is in good agreement with previous results for Galactic O-supergiants \citep{Hawcroft2021} and for O stars in the LMC \citep{Brands2022}. For cooler weak-wind stars the clump velocity filling converges to very low values ($f_{\rm{vel}} = 0.13 \pm 0.04$), in an effort to fit the photospheric UV profiles that remain due to a lack of wind signatures. This indicates that either the wind is sufficiently porous to remove all high-velocity blue-shifted absorption, or that there is no diagnostic available due to the aforementioned wind ionisation.

- A similar trend in interclump density as for velocity filling; hotter stars with typical winds have relatively larger interclump densities, around 20\% of the mean wind density, similar to results for Galactic O-supergiants \cite{Hawcroft2021}. For the cooler weak-wind stars we find interclump densities approaching the lower limit of the parameter space, just a few percent or even lower, of the mean wind ($f_{\rm{ic}} = 0.03 \pm 0.04$). Although we note this related to the aforementioned issues in diagnosing the wind structure properties associated with optically thick clumping for weak-wind stars.

This analysis will be extended to lower metallicity as we measure stellar and wind properties for a similar sample in the SMC. We also await the full release of the optical follow-up to the HST-ULLYSES survey \citep{RomanDuval2020}, X-Shooting ULLYSES. These surveys combined will provide high quality UV and optical spectra for more than 200 stars across the LMC and SMC \citep{Vink2023, Sana2024}. This will be an unprecedented opportunity to investigate the wind properties of hot, massive stars at low metallicity.

\begin{acknowledgements}
We would like to thank the anonymous referee, for their helpful comments which improved the content and clarity of this paper.This project has received funding from the  KU Leuven Research Council (grant C16/17/007: MAESTRO), the FWO through a FWO junior postdoctoral fellowship (No. 12ZY520N) as well as the European Space Agency (ESA) through the Belgian Federal Science Policy Office (BELSPO). LM thanks ESA and BELSPO for their support in the framework of the PRODEX programme (MAESTRO). The computational resources and services used in this work were provided by the VSC (Flemish Supercomputer Center), funded by the Research Foundation - Flanders (FWO) and the Flemish Government. This research has made use of the SIMBAD database, operated at CDS, Strasbourg, France.
\end{acknowledgements}

\bibliography-style{aa} 
\bibliography{main.bib}

\break

\begin{appendix}

\section{Appendix - Star-by-Star} \label{appendix}

\subsection{VFTS180}

This star sits on the border of O-type supergiant and Wolf-Rayet star but has been classed as O3 If* by \cite{Evans2011} and confirmed as apparently singe \citep{Shenar2020}. We find a well-defined temperature $T_{\rm{eff}} = 44.1 \pm 0.5$ kK, 4 kK higher than \cite{RamirezAgudelo2017}. The temperature in our fit is constrained primarily from the optical hydrogen and helium lines, although the He\,{\sc i} lines are slightly too strong. The wings of the Balmer lines are reproduced very well to log $g = 3.62 \pm 0.05$, and the higher temperature relative to \cite{RamirezAgudelo2017} is consistent with the log $g$ we find which is 0.2 dex higher. However, it is difficult to compare the fits as the fit quality is deemed poor and left out of extended analysis of \cite{RamirezAgudelo2017}. This is the most luminous star in our sample with log$(L/L_{\odot}) = 5.98 \pm 0.07$ and the highest mass-loss rate log($\dot{M}$ [$M_{\odot}\rm{yr}^{-1}$]) $= -5.72 \pm 0.05$, thanks to the strong wind features this star also has some of the most well-defined fitness distributions. $\beta$ is well constrained by the optical emission features to $1.2 \pm 0.1$ and $v_{\infty}$ is drawn towards an erroneously high value of 2650 $\rm{km}\, \rm{s^{-1}}$, it should be closer to 2000 $\rm{km}\, \rm{s^{-1}}$ from the saturated C\,{\sc iv}\,$\lambda \lambda$1548-1551 line. We find moderately low rotation of $v\sin\,i$ = 90 $\rm{km}\, \rm{s^{-1}}$ and $v_\mathrm{mac}$ = 50 $\rm{km}\, \rm{s^{-1}}$, fairly consistent with \cite{RamirezAgudelo2017}. We find a significant helium enrichment $N_{He}/N_{H} = 0.3$, as do \cite{RamirezAgudelo2017} although the enrichment found in their study is higher, this is mainly constrained by the optical hydrogen and helium lines but the UV helium lines help as well. We find significant nitrogen enrichment from emission in the optical nitrogen lines, which are well fitted, at $\epsilon$(N) = 8.93 $\pm 0.1$, essentially at the upper limit of our parameter space. Although there is emission in  N\,{\sc v}\,$\lambda$4603 which is not in the model. The wind parameters found for this star are consistent with those from \cite{Hawcroft2021} for Galactic O-type supergiants, suggesting the metallicity is not influencing these parameters much, at least for this star. The clumping factor is around 20, the velocity filling between 0.5 - 0.6, one-tenth average density in the interclump medium ($f_{\rm{ic}} = 0.12 \pm 0.04$) and the clumping onset is around 150 $\rm{km}\, \rm{s^{-1}}$. 

\subsection{VFTS143}

VFTS143 was determined to be O3.5V((fc)) by \cite{Walborn2014}, indicating C\,{\sc iii} emission in the optical and absorption in He\,{\sc ii}\,$\lambda$4686. This star has not been included in previous atmosphere modelling studies but our derived $T_{\rm{eff}} = 44.8$ kK agrees with \cite{Walborn2014} within 0.1 kK and, with a log $g$ = 3.9 $\pm$ 0.05, it sits within the parameter range of a typical O-dwarf in 30 Dor \citep{SabinSanjulian2017}. It is also categorised as a large amplitude SB1 \citep{Sana2013} and significant radial velocity variation can be seen in the GA best fit, mainly in the optical He\,{\sc ii} lines. The mass loss for this object lies comfortably in the trend for the full sample, and is well constrained by UV resonance lines e.g. C\,{\sc iv}\,$\lambda \lambda$1548-1551, although we find a very low clumping factor ($f_{\rm{cl}} = 3 \pm 3$), with this low clumping constrained by optical lines such as H$\alpha$. The fitness distributions for $f_{\rm{vel}}$ and $f_{\rm{ic}}$ are quite flat, reducing somewhat our confidence in the high values found of $\sim$ 0.7 and $\sim$ 0.3, respectively. This star is among those with the highest terminal wind speeds found in this sample, at 2900 $\mathrm{km s^{-1}}$. The $\beta$ value pushes to the upper edge of the parameter space, due to the Balmer series especially H$\gamma$ and H$\delta$. Rotation and macroturbulent velocities are both moderate, near 100 $\rm{km}\, \rm{s^{-1}}$. While we do not model the optical carbon features key to the spectral class of this object we do find a fairly standard carbon abundance for the LMC (\citealp{Dopita2019}, \citealp{Brott2011}). From the GA fitness distributions it is difficult to say much about the oxygen abundance and nitrogen, they can be typical or depleted but certainly not enriched. 

\subsection{VFTS608}

VFTS608 is an early-mid giant (O4III) only extensively studied in \cite{Bestenlehner2014}. We find a temperature  $T_{\rm{eff}} = 43.0 \pm 0.5$ kK, which is 1kK higher than \cite{Bestenlehner2014} and fairly well constrained with a good optical fit, there are some issues due to significant line profile variability as identified in \cite{Sana2013}, although the shift in e.g. the optical helium lines is not as severe as in VFTS143. The statistical significance of the line profile variability detection is similar between VFTS143 and VFTS608 but the magnitude of the variation is $\sim$40 $\rm{km}\, \rm{s^{-1}}$ lower for VFTS608. The UV fit is not perfect but the main issue is that the Si\,{\sc iv}\,$\lambda \lambda$1394, 1403 lines are much too weak in our model, it appears the only way to reproduce these from the GA fitness distributions is to massively lower the temperature and or mass-loss rate. The surface gravity is well defined with a good fit (log $g = 3.84 \pm 0.05$). 
The mass-loss rate is well constrained (log($\dot{M}$)$ = -6.24 \pm 0.05$) with a strong wind signature in C\,{\sc iv}\,$\lambda \lambda$1548-1551 but is lower than predicted, \cite{Bjorklund2021} over-predict the mass-loss rate by a factor 1.3. $\beta$ is high, hitting the top of our parameter space at 2. Terminal wind speed is well constrained from C\,{\sc iv}\,$\lambda \lambda$1548-1551 to 2640 $\rm{km}\, \rm{s^{-1}}$ which agrees with our prediction from \cite{Hawcroft2023}. The clumping factor is well constrained ($f_{\rm{cl}} = 9 \pm 1$) from the optical and, along with VFTS143 is one of the two stars with a low clumping factor and relatively high temperature, although much of the H$\alpha$ core had to be removed. The velocity filling is at the lower bound $f_{\rm{vel}} \sim 0.1$ and the interclump density is intermediate $f_{\rm{ic}} \sim 0.15$ , both appearing to be mainly constrained by C\,{\sc iv}\,$\lambda \lambda$1548-1551 and He\,{\sc ii}\,$\lambda$4686. Rotation is normal with unconstrained macro. Helium abundance is normal, only slightly higher than most of the sample. Nitrogen is well constrained and enriched, carbon is maybe depleted and oxygen is lower than baseline \footnote{LMC baseline abundances are assumed to be $\epsilon$(C) $\sim$ 7.75, $\epsilon$(N) $\sim$ 6.9 and $\epsilon$(O) $\sim$ 8.35 (\citealp{Kurt1998, Brott2011, Kohler2015})}.

\subsection{VFTS216}

VFTS216 is an early-type dwarf (O4V) with fairly strong wind signatures in the UV. It was previously studied by \cite{SabinSanjulian2017}. We obtain a very good fit, aside from issues with normalisation around the N\,{\sc iii}\,$\lambda \lambda$-4634-4640-4642 lines, resulting in a temperature ($T_{\rm{eff}} = 44^{+0.5}_{-1.7} kK$ and log $g =$ 3.74 $\pm$ 0.05, which agree well with \cite{SabinSanjulian2017}. $\dot{M}$, $\beta$ and $v_{\infty}$ are fairly well constrained, $\dot{M}$ is half that predicted by \cite{Bjorklund2021}, $\beta$ is quite high at $1.6 \pm 0.2$ and $v_{\infty}$ agrees well with our predictions. The clumping factors are not very well constrained, $f_{\rm{cl}} \sim 9 - 26$, $f_{\rm{vel}} \sim 0.25 - 1.0$ and $f_{\rm{ic}} \sim 0.06 - 0.28$, so it is difficult to comment on these parameters for this star. 
Rotation and macroturbulence velocities are moderate, around 100 $\rm{km}\, \rm{s^{-1}}$ and 80 $\rm{km}\, \rm{s^{-1}}$, respectively. The helium abundance is normal. The nitrogen abundance is baseline and well constrained. The carbon abundance is fairly typical, the error bars including both baseline values, and oxygen also appears to be baseline.

\subsection{VFTS184}

VFTS184 is mid-late type dwarf (O6.5V) which is just on the edge of the weak-wind regime and has not been previously analysed. We obtain a good fit in the optical but the fit is not as good in the UV, for example the C\,{\sc iv}\,$\lambda \lambda$1548-1551 line which has an unusual morphology. The mass-loss rate hit the lower boundary of our fitting range so we launched another run, finding that the mass-loss rate is fairly well defined at log($\dot{M}$)$ = -8.45 \pm 0.15$, a factor of 2 lower than predicted by \cite{Bjorklund2021}. This star is also a relatively fast rotator with $v \sin i \sim 350$ $\rm{km}\, \rm{s^{-1}}$. $\beta$ is typical ($\sim 1.1$) but is not well peaked, the same can be said for $v_{\infty}$, which has a flat fitness distribution along with a poor fit to C\,{\sc iv}\,$\lambda \lambda$1548-1551 but the value found agrees with a direct measurement and our predictions. The fitness distributions for the clumping parameters are quite flat but with good constraints $f_{\rm{cl}} \sim 16 \pm 5$, $f_{\rm{vel}} \sim 0.35 \pm 0.05$ and $f_{\rm{ic}} \sim 0.1 \pm 0.1$. Normal helium abundance. Low to standard nitrogen and carbon abundances with oxygen again not present in the UV and unconstrained.

\subsection{VFTS244}

VFTS244 is a mid O-type giant (O5III) also modelled by \cite{RamirezAgudelo2017}. We find a temperature 1 kK lower and surface gravity 0.1 dex lower than \cite{RamirezAgudelo2017} which is fair agreement ($T_{\rm{eff}} = 39.7 \pm 0.5$ kK and log $g$ = 3.58 $\pm$ 0.05). The fit quality is high, there are some issues in fitting the Si\,{\sc iv}\,$\lambda \lambda$1394, 1403 lines and matching the strength of the absorption in He\,{\sc ii}\,$\lambda$4686. We also have to remove the core of H$\alpha$. 
Our mass-loss rate (log($\dot{M}$)$ = -6.40 \pm 0.05$) is lower than \cite{RamirezAgudelo2017}, with the same luminosity, so this discrepancy seems to be due to the inclusion of clumping, after correction the mass-loss rates are in agreement. Our mass-loss rate is double that of the prediction from \cite{Bjorklund2021} and 4.5 times lower than the prediction of \cite{Vink2001}. We have a good constraint on the terminal wind speed from a saturated C\,{\sc iv}\,$\lambda \lambda$1548-1551 profile, we find $v_{\infty} = 2550$ $\rm{km}\, \rm{s^{-1}}$ which is around 300 $\rm{km}\, \rm{s^{-1}}$ higher than our prediction. $\beta$ converges to the upper limit of the parameter space around a value of 2.0, driven by the optical hydrogen lines. The clumping factor ($f_{\rm{cl}} = 16 \sim 20$) is close to that found for Galactic O-supergiants \citep{Hawcroft2021}, and also well constrained by optical hydrogen. It is generally difficult to disentangle the effects of velocity filling and interclump density, though here we find that $f_{\rm{vel}} (= 0.15 \sim 1.0$) is unconstrained and $f_{\rm{ic}} (= 0.22 \sim 0.3$) must be high. This star has is one of the highest rotation rates in this sample ($v \sin i =$ 200 - 250 $\rm{km}\, \rm{s^{-1}}$) which agrees with the findings of \cite{RamirezAgudelo2017}. Helium abundance is normal. It appears nitrogen is slightly enhanced and carbon depleted. Oxygen looks normal, but maybe slightly lower than baseline.

\subsection{VFTS664}

VFTS664 is a mid-late bright giant (O7II) and is also included in \cite{RamirezAgudelo2017}. The fit is not bad but not as good as most of the stars in this sample, mainly the fits to C\,{\sc iv}\,$\lambda \lambda$1548-1551 and Si\,{\sc iv}\,$\lambda \lambda$1394, 1403 are poor, and the cores of all the optical hydrogen lines are lost to nebular contamination. We find good agreement with \cite{RamirezAgudelo2017} in temperature and surface gravity. C\,{\sc iv}\,$\lambda \lambda$1548-1551 is the only line with a clear wind signature, but the mass-loss rate is well constrained and found to be a factor 3.4 lower than the prediction of \cite{Vink2001} and a factor 2.6 larger than the prediction of \cite{Bjorklund2021}. $\beta$ is fairly high, between 1.6-1.9, and terminal wind speed is hard to determine due to the very low gradient of the blue edge. Interestingly, the clumping factor is decidedly low $f_{\rm{cl}} = 3 \sim 6$, and the velocity filling and interclump density converge to the lower bound of the parameter space, $f_{\rm{vel}} \sim 0.1$ and $f_{\rm{ic}} \sim 0.01$. Rotation is again moderate, close to 100 $\rm{km}\, \rm{s^{-1}}$ with a matching macroturbulence. Nitrogen is enriched with typical carbon. Oxygen is not well constrained and the helium abundance is normal.

\subsection{VFTS087}

\begin{figure}[t!]
	\centering
	\includegraphics[scale=0.28]{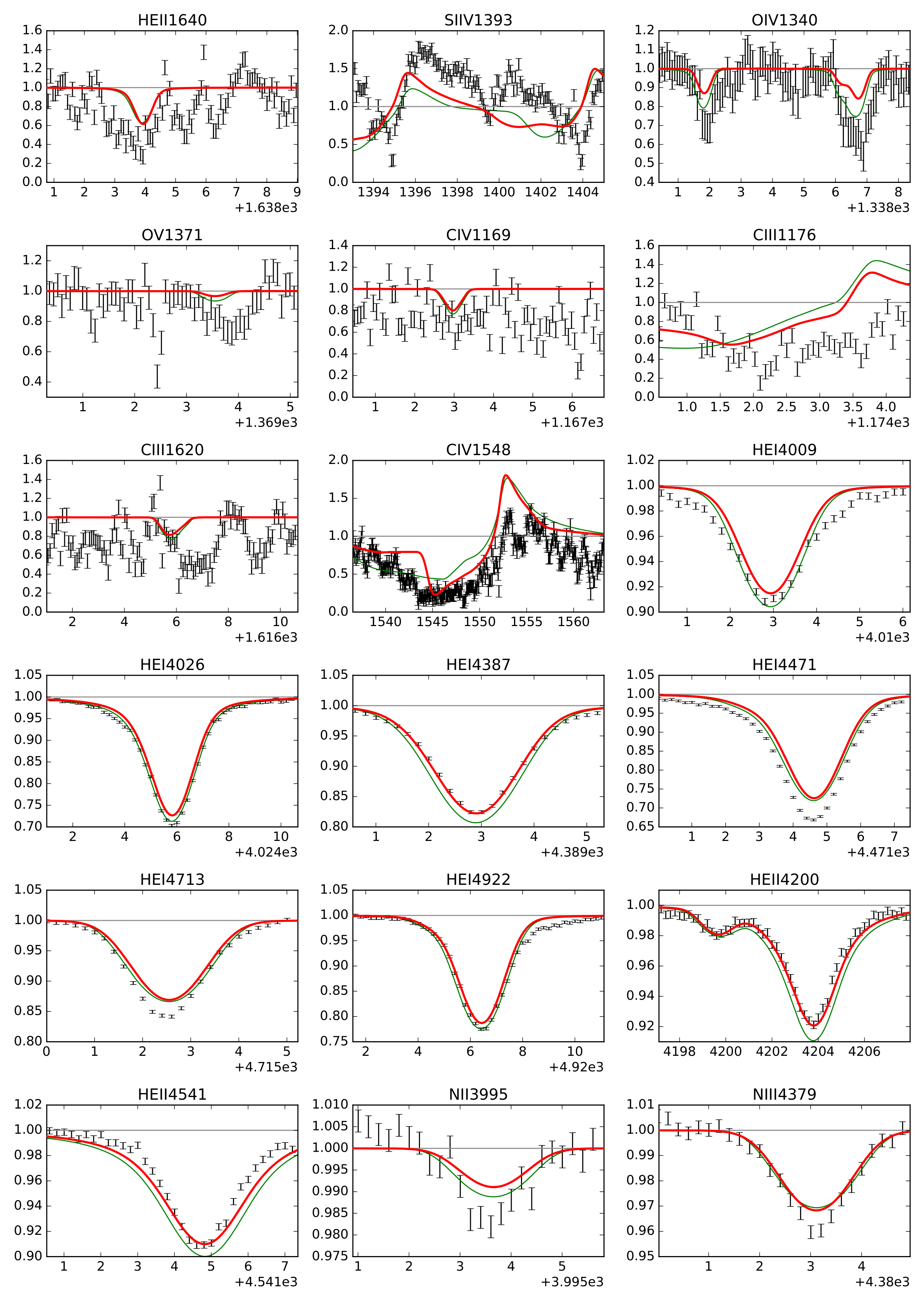}
	\includegraphics[scale=0.14]{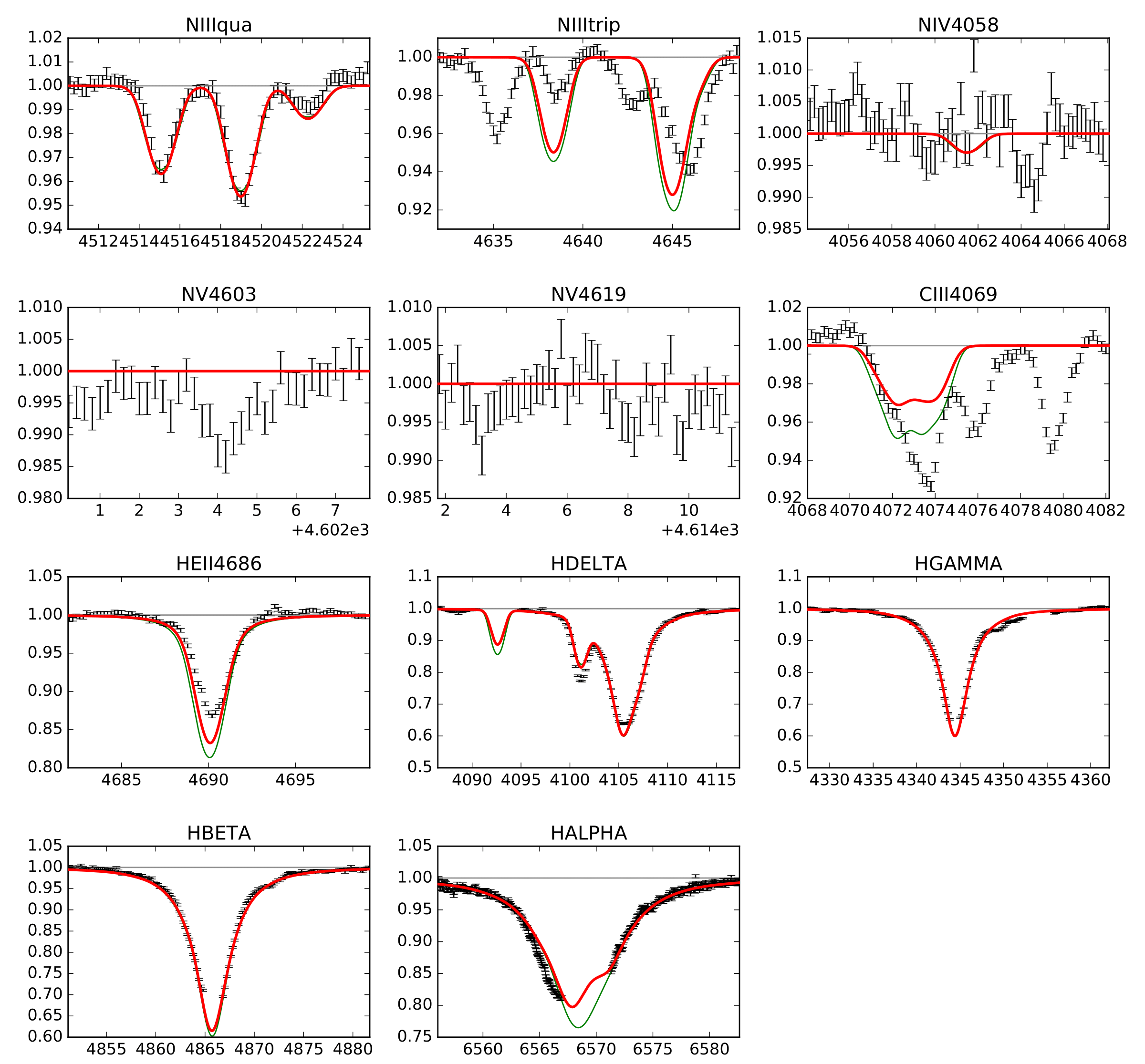}
	\caption{Best fit for VFTS087 O9.7Ib-II from GA with optically thick clumping.}
	\label{fig: Best-fit-VFTS087}
\end{figure}

VFTS087 is a (super/bright-)giant (O9.7I-II), and is also included in \cite{RamirezAgudelo2017}. The fit to the optical spectrum is very good, aside from a few metal lines, in the UV we may have some issues with normalistion that could be affecting the fit. We find this star to have the lowest temperature in our sample $T_{\rm{eff}} = 31.0 \pm 0.5$ kK and log $g$ = 3.35 $\pm$ 0.05, in very good agreement with \cite{RamirezAgudelo2017}. The terminal wind speed is found to be very high as discussed in Sect. \ref{sec:lmc-vinf}, perhaps due to normalisation or related to the high $\beta$ (= 3.0 $\pm$ 0.7), the fitness distribution of which peaks below 1.0 in the fit to C\,{\sc iv}\,$\lambda \lambda$1548-1551. The mass-loss rate is constrained mainly by the optical hydrogen lines in this fit, and with log($\dot{M}$)$ = -6.75 \pm 0.05$ it lies between predictions, a factor 2 lower than \cite{Vink2001} and a factor 5 higher than \cite{Bjorklund2021}. All clumping properties are towards the lower limits ($f_{\rm{cl}} = 4 \pm 1$, $f_{\rm{vel}} = 0.17 \pm 0.05$ and $f_{\rm{ic}} = 0.01 \pm 0.05$). The clumping factor is strongly constrained by the optical hydrogen lines, and the nebular contamination is not too bad in this star so most of the cores remain. The velocity filling and interclump density factors are strongly constrained by the UV, in particular the Si\,{\sc iv}\,$\lambda \lambda$1394, 1403 lines. Part of the fitting region of these lines is signficantly above the continuum but the morphology is unusual (see \ref{fig: Best-fit-VFTS087}), if there is no emission component and this morphology is only due to issues with normalisation, it is hard to say what effect this would have on the parameters but they may change. Rotation is well constrained as moderate to low, and the macroturbulence is similar, 70 $\pm$ 10 $\rm{km}\, \rm{s^{-1}}$ and 30 $\pm$ 10 $\rm{km}\, \rm{s^{-1}}$. The helium abundance is normal, nitrogen is clearly enriched, nitrogen and carbon are well constrained. Carbon and oxygen abundances seem normal.

\subsection{VFTS440}

VFTS440 is a mid-type bright giant (O6-6.5II), and is also included in \cite{RamirezAgudelo2017}. The fit is interesting as there is disagreement between best-fit values between the optical hydrogen lines while much of the core and red-wings must be removed, we also have issues fitting Si\,{\sc iv}\,$\lambda \lambda$1394, 1403, O\,{\sc iv}\,$\lambda \lambda$1340-1344 and He\,{\sc ii}\,$\lambda$4686. There could also be issues due to binarity as this star is flagged as an SB1 in \cite{Almeida2017}. The temperature and surface gravity ($T_{\rm{eff}} = 33.4 \pm 0.5$ kK and log $g$ = 3.13 $\pm$ 0.05) are consistent with \cite{RamirezAgudelo2017} but both are heavily constrained by H$\alpha$ and the He\,{\sc ii} lines. Looking at the fitness distributions for the other Balmer lines and some He\,{\sc i} lines, they mostly peak towards another solution with log $g$ $\sim$ 3.7 and $T_{\rm{eff}} \sim 40$ kK. The terminal wind speed could be lower, as discussed in Sect \ref{sec:lmc-vinf}. $\beta$ ($\sim$  1.6) is fairly high, again constrained by optical hydrogen. The mass-loss rate (log($\dot{M}$)$ = -6.23 \pm 0.05$) is well constrained by the optical, and the fit to UV is not bad, it is a factor 4 lower than \cite{Vink2001} and a factor 2 higher than \cite{Bjorklund2021}. The clumping factors again are low ($f_{\rm{cl}} = 9 \pm 1$, $f_{\rm{vel}} = 0.14 \pm 0.05$ and $f_{\rm{ic}} = 0.02 \pm 0.05$), with some possible secondary peak at $f_{\rm{cl}} \sim 15$ in H$\alpha$. The velocity filling and interclump density factors are low although it is not clear that any lines in particular are driving this, and here the normalisation is good unlike VFTS087. Again we find fairly standard broadening for this sample. Helium, nitrogen and carbon abundances are well constrained. Helium is normal to slightly high. We find significant nitrogen enrichment, while the carbon abundance is normal. Oxygen abundance is normal but the O\,{\sc iv}\,$\lambda \lambda$1340-1344 lines are clearly underestimated in our fit.

\subsection{VFTS385}

VFTS385 is a mid O-type dwarf (O4-5V) also modelled by \cite{SabinSanjulian2017}. We find a temperature and surface gravity lower (2 kK and 0.25 dex) than \cite{SabinSanjulian2017} ($T_{\rm{eff}} = 37.7 - 39.1$ kK and log $g$ = 3.55 $\pm$ 0.05). The fit quality is high, although there are some issues in fitting the Si\,{\sc iv}\,$\lambda \lambda$1394, 1403 lines and we remove the cores of the Balmer lines. 
Our mass-loss rate (log($\dot{M}$)$ = -6.29 \pm 0.05$) is a factor 1.6 lower than the \cite{Vink2001} prediction and a factor 9.8 higher than the prediction from \cite{Bjorklund2021}. There is some discrepancy in the terminal wind speed, discussed in Sect. \ref{sec:lmc-vinf}. $\beta$ is found to be between 0.6 - 0.7, which is a bit low. The clumping factor distribution is not very well defined but we find $f_{\rm{cl}} = 13 \pm 1$. We find a fairly low velocity filling factor $f_{\rm{vel}} (= 0.13 \pm 0.05$) and fairly typical interclump density ($f_{\rm{ic}} = 0.10 \pm 0.05$). 
Rotation and macroturbulence velocities are moderate, around 120 $\rm{km}\, \rm{s^{-1}}$ and 90 $\rm{km}\, \rm{s^{-1}}$, respectively. The helium abundance is normal. It appears that the nitrogen and carbon abundances are baseline to slightly reduced and oxygen is enhanced, converging to the upper limit of our parameter space.

\subsection{VFTS096}

VFTS096 is a mid O-type dwarf (O6V) and is included in \cite{SabinSanjulian2017}. The fit is good, although there are issues fitting the carbon lines, and the fitness distributions are very flat with a number of unconstrained parameters. Our best fit temperature and surface gravity agree well with \cite{SabinSanjulian2017} and at least the surface gravity is well defined by the Balmer lines, ($T_{\rm{eff}} = 41.4^{+1.3}_{-0.7}$ kK and log $g$ = 3.90 $\pm$ 0.1). Terminal wind speed is low here, with $v_{\infty} \sim 2300$ $\mathrm{km s^{-1}}$, which agrees with the predictions of \cite{Hawcroft2023} $v_{\infty} \sim 2400$ $\mathrm{km s^{-1}}$ but could actually be close to 2800 $\mathrm{km s^{-1}}$ from the C\,{\sc iv}\,$\lambda \lambda$1548-1551 line. $\beta$ is high (1.5 $\pm$ 0.5), which is consistent with other O dwarfs due to optical hydrogen, but with a large range the typical value is included. Mass-loss rate has a strong upper limit of log($\dot{M}$)$ \sim $ -6.5, and less strong lower limit at log($\dot{M}$)$ \sim $ -7.0, very low compared to theoretical predictions, a factor 12 lower than \cite{Vink2001} and a factor 6 lower than \cite{Bjorklund2021} in the worst case and a factor 5 and 2 lower in at the upper limit of $\dot{M}$. There is only a wind signature in C\,{\sc iv}\,$\lambda \lambda$1548-1551 which has a fairly atypical morphology, it is difficult to constrain any wind parameters within the large range of mass-loss rates available, and the error margins encompass almost the full parameter space for clumping factor, interclump density and velocity filling. 
Rotation is typical and moderate ($v \sin i \sim $ 90 - 140 $\mathrm{km s^{-1}}$), with macroturbulence unconstrained.
The helium, nitrogen and carbon abundances have some of the more well constrained fitness distributions. Helium abundance is typical. Carbon is typical to low and converges to the lower limit of the parameter space. Nitrogen features are very weak, with a low nitrogen abundance, again towards the lower edge of the parameter space. Oxygen is unconstrained but UV diagnostics peak in fitness around LMC baseline or slightly higher.

\subsection{VFTS586}

VFTS586 is an earlier O-type dwarf (O4V) and is included in \cite{SabinSanjulian2017}. For this star we achieve a high quality fit to all lines apart from  Si\,{\sc iv}\,$\lambda \lambda$1394, 1403, and \cite{SabinSanjulian2017} also have a good optical fit. The hydrogen line cores are removed but most of the wings remain. We find a well constrained temperature and surface gravity 1 kK higher and 0.1 dex lower than \cite{SabinSanjulian2017} with $T_{\rm{eff}} = 46.2 \pm 0.5$ kK and log $g$ = 3.90 $\pm$ 0.05. Terminal wind speed is well constrained by C\,{\sc iv}\,$\lambda \lambda$1548-1551 and agrees very well with our prediction. A high $\beta (= 1.5 \pm 0.1)$ results from the optical hydrogen lines while the UV diagnostics tend to prefer a lower beta. A strong upper limit is placed on the mass-loss rate (log($\dot{M}$)$ = -6.85 \pm 0.07$) by the optical hydrogen and helium lines and the peak agrees well with UV diagnostics. This $\dot{M}$ is a factor 9.5 lower than predicted by \cite{Vink2001} and a factor 1.7 higher than \cite{Bjorklund2021}. The clumping factor can not be too high or too low, $f_{\rm{cl}} \sim 15 \pm 5$. The velocity filling factor is in the lower half of the parameter space and interclump density is not well constrained but must be fairly high ($f_{\rm{vel}} (= 0.30 \pm 0.15$), $f_{\rm{ic}} \sim 0.07 - 0.28$).
Rotation is typical and moderate ($v \sin i \sim $ 80 - 100 $\mathrm{km s^{-1}}$), with macroturbulence unconstrained. All the abundances are well constrained. Helium abundance is typical, carbon is very consistent with LMC baseline, nitrogen is maybe low and the features are very weak. Oxygen abundance is high due to strong UV optical features. 

\subsection{VFTS422}

With $T_{\rm{eff}}$ = 42kK and log $g$ = 3.5, VFTS422 star is an early-mid giant (O4III). Although the fit is not very good due to relatively low S/N and we have to remove the hydrogen line cores and fairly significant parts of the wings due to nebular contamination and exclude H$\alpha$. As a result the fitness distributions are fairly flat and the peaks in surface gravity and mass-loss rate are not in great agreement with the best fit. It could be that our surface gravity is up to 0.3 dex too low which would also impose a significant upper bound on the effective temperature. This star is included in the sample of \cite{Bestenlehner2014}, who find a comparatively low $T_{\rm{eff}}$ = 40kK, although these authors use the optical nitrogen lines to determine $T_{\rm{eff}}$ and we would likely see a reduction in our best fit value if our focus was shifted to these lines. There are again a number of issues with radial velocity typical of a large amplitude SB1, as confirmed by \cite{Sana2013}. This star has fairly high rotation at 200 $\mathrm{km s^{-1}}$, although this is significantly lower than the 356 $\mathrm{km s^{-1}}$ reported in \cite{Bestenlehner2014}. Again the nebular contamination and radial velocity variation can affect this.
The largest discrepancy comes when comparing the mass-loss from the GA and \cite{Bestenlehner2014} which find -6.48 and -5.6 respectively. \cite{Bestenlehner2014} include a modest clumping correction but still reproduce a mass-loss rate close to the \cite{Vink2001} prediction. We find the clumping factor to be similar to that found for Galactic O-supergiants, $f_{\rm{cl}} \sim 21 \pm 3$. Note that we find a luminosity matching exactly that of \cite{Bestenlehner2014} and a mass-loss rate in very good agreement with the prediction of \cite{Bjorklund2021}. We find a terminal wind speed of 2100 $\mathrm{km s^{-1}}$ which is 300 $\mathrm{km s^{-1}}$ lower than our prediction and nearly 700 $\mathrm{km s^{-1}}$ lower than \cite{Bestenlehner2014}, although these authors calibrate this from an overall wind strength due to a lack of UV diagnostics. It is also worth noting the decidedly low $\beta$ value in our best fit, perhaps due to the difficulty in obtaining a high quality fit to the optical, a low beta, as is commonly preferred in the UV, is converged upon. 
Despite issues in the optical, the fit quality is higher in the UV, the main issue is fitting  Si\,{\sc iv}\,$\lambda \lambda$1394, 1403. We find fairly low velocity filling and interclump density factors ($f_{\rm{vel}} \sim 0.4$, although the fitness distribution peaks towards the lower end, mostly due to N\,{\sc iv}\,$\lambda$1718 so this could be an upper limit, and $f_{\rm{ic}} \sim 0.01 - 0.10$)
We find normal to low helium abundance, significant carbon and nitrogen enhancements compared to the LMC baseline and oxygen depletion.

\subsection{VFTS627}

VFTS627 is the latest type O dwarf in our sample (O9.7V) and a weak-wind star, also included in \cite{SabinSanjulian2017}. We are able to obtain a very high quality fit for all lines, except for C\,{\sc iv}\,$\lambda \lambda$1548-1551, the strength of the absorption is underestimated in this line, as is common for weak-wind stars. The fit to the optical in \cite{SabinSanjulian2017} is also very good. Although we find a temperature 2 kK lower which is highly constrained by UV lines, which may also help to explain our 0.35 dex lower surface gravity ($T_{\rm{eff}} = 30.5 - 32.0$ kK and log $g$ = 3.70 - 3.85). Although the difference in wind parameters may also contribute here, we can see that we have much stronger absorption in H$\alpha$, which is hard to confirm as the Balmer line cores are removed but more of the core remains in the higher level Balmer lines, the strength of which are matched well. There is no diagnostic for the terminal wind speed and essentially no constraint for $\beta$, the fitness distributions are very flat.
We find a mass-loss rate of log($\dot{M}$)$ = -10.5 \pm 0.1$, the lowest in the sample and very related to the weak-wind problem (see discussion in Sect. \ref{sec: weak-wind}). There is no constraint on any of the clumping factors.
Rotation is low ($v \sin i < 80 $ $\mathrm{km s^{-1}}$) and macroturbulence fairly high ($v_{\rm{mac}} = 85 \pm 10 $ $\mathrm{km s^{-1}}$). 
Helium, nitrogen and carbon abundances are well constrained while oxygen is not, with little to no visible diagnostics. Helium abundance is normal. Carbon abundance is typical for the LMC, nitrogen may be a bit high as found by N\,{\sc iii}\,$\lambda \lambda$-4634-4640-4642.

\subsection{VFTS280}

VFTS280 is a late O dwarf (O9V) and is another weak-wind star, also included in \cite{SabinSanjulian2017}. The fit is quite good, especially in the optical, except of course to C\,{\sc iv}\,$\lambda \lambda$1548-1551. We find very well constrained temperature and surface gravity, consistent with \cite{SabinSanjulian2017} at ($T_{\rm{eff}} = 33.4 \pm 0.5$ kK and log $g$ = 3.76 $\pm$ 0.05). 
$\beta$ and $v_{\infty}$ are low but fairly unconstrained due to lack of diagnostics. Mass-loss rate is low log($\dot{M}$)$ = -9.32^{+	0.46}_{-0.34}$. Clumping factor is essentially unconstrained ($f_{\rm{cl}} \sim 4 - 19$), it just cannot be high. Here the velocity filling and interclump density factor converge to the lower bound of the parameter space due to C\,{\sc iv}\,$\lambda \lambda$1548-1551 (see discussion in Sect. \ref{sec: weak-wind}).
Rotation is a bit higher ($v \sin i = 155 \pm 20 $ $\mathrm{km s^{-1}}$) and the macroturbulence is fairly high ($v_{\rm{mac}} = 90 \pm 5 $ $\mathrm{km s^{-1}}$).
All abundances apart from oxygen are fairly well constrained and normal.

\subsection{VFTS235}

VFTS235 is a very late O giant (O9.7II) and is included in \cite{RamirezAgudelo2017}. All of our findings are very close to VFTS627. The fit here is very good, even the C\,{\sc iv}\,$\lambda \lambda$1548-1551 absorption strength is fairly close. We have well defined temperature and surface gravity peaks in full agreement with \cite{RamirezAgudelo2017} at ($T_{\rm{eff}} = 32.5 \pm 1.0$ kK and log $g$ = 4.2 $\pm$ 0.1) but the surface gravity is the highest in the sample, higher than all the stars classed as dwarfs. Again there is no diagnostic for the terminal wind speed or $\beta$ so these parameters are completely unconstrained.
The mass-loss rate converges to the lowest capabilities of FASTWIND, with only an upper bound constrained log($\dot{M}$)$ < -10.0$. Again there is no constraint on any of the clumping factors. Rotation is low ($v \sin i < 50 $ $\mathrm{km s^{-1}}$) and macroturbulence is moderate ($v_{\rm{mac}} = 50 \pm 20 $ $\mathrm{km s^{-1}}$). 
Helium abundance is normal. Nitrogen abundance is low in our best fit but with large error bars, from the spectrum it appears the nitrogen abundance should be closer to our upper error limit, around 7.5 which is in good agreement with \cite{Grin2017}. A bit higher than LMC baseline around 6.9-7.2. Carbon abundance is low but may actually be close to baseline if we were to focus on CIII lines in the UV. Oxygen is not constrained and it is difficult to distinguish features in the UV.

\subsection{VFTS223}

VFTS223 is a late sub-giant star (O9.5IV), also included in \cite{SabinSanjulian2017} and has very similar spectral morphology as the weak-wind stars. Again the fit is very good, temperature and surface gravity are well constrained ($T_{\rm{eff}} = 33.8 \pm 0.5$ kK and log $g$ = 3.86 $\pm$ 0.05) and both are significantly lower than \cite{SabinSanjulian2017}. Terminal wind speed and $\beta$ are similar to other weak-wind stars, there is little to no diagnostic  and these parameters are essentially unconstrained. Perhaps some wind is seen in C\,{\sc iv}\,$\lambda \lambda$1548-1551, and the mass-loss rate is actually constrained to log($\dot{M}$)$ = -7.6 \pm 0.25$.  Clumping factor cannot be too high ($f_{\rm{cl}} \sim 11 \pm 4$). Much like VFTS280, the velocity filling and interclump density factor converge to the lower bound of the parameter space due to C\,{\sc iv}\,$\lambda \lambda$1548-1551 (again see discussion in Sect. \ref{sec: weak-wind}).
Rotation is very low ($v \sin i < 13 $ $\mathrm{km s^{-1}}$) and macroturbulence is moderate ($v_{\rm{mac}} \sim 60 \mathrm{km s^{-1}}$). 
All abundances have some constraint, the metal lines are generally clear and well fitted giving well peaked fitness distributions for carbon and nitrogen but oxygen is not clear in the UV so that is low and fairly unconstrained. We find a normal helium abundance and nitrogen is typical to slightly enhanced, carbon is typical to slightly depleted.

\subsection{VFTS517}

VFTS517 is a late O dwarf/giant (O9.5V-III), appears to be another weak-wind star and is included in \cite{SabinSanjulian2017}. We find significantly lower temperature and surface gravity than \cite{SabinSanjulian2017}, much like for VFTS223 ($T_{\rm{eff}} = 31.2 \pm 0.5$ kK and log $g$ = 3.67 $\pm$ 0.07) but we have high quality fits and well peaked fitness distributions for these parameters. Further repetition for the terminal wind speed and $\beta$, there is little to no diagnostic and these parameters are essentially unconstrained. The mass-loss rate converges to the lowest capabilities of FASTWIND, with only an upper bound constrained log($\dot{M}$)$ < -9.7$, although H$\alpha$ could be fitted with a higher mass-loss rate, which disagrees with all other diagnostics. There is no constraint on the clumping factor, and the velocity filling factors are slightly better constrained but essentially only offer upper limits ($f_{\rm{vel}} < 0.4$, $f_{\rm{ic}} < 0.23$).
Rotation is moderate ($v \sin i = 100 \pm 50 $ $\mathrm{km s^{-1}}$) and the macroturbulence is unconstrained.
Again all abundances have some constraint. The helium abundance is normal. Nitrogen may be slightly increased from baseline, but carbon is typical, oxygen is fairly absent from the spectrum so either depleted or unconstrained. 

\section{Best fits} \label{sec:app-lmc-fits}

In this appendix we present models from our fits with optically thick clumping, showing only the line profiles used in our GA analysis. The observed spectrum is shown by the black points, the solid red line is our best fitting model, and green lines represent any models generated during the GA iterations that lie within the error regions.

\begin{figure}[t!]
	\centering
	\includegraphics[scale=0.3]{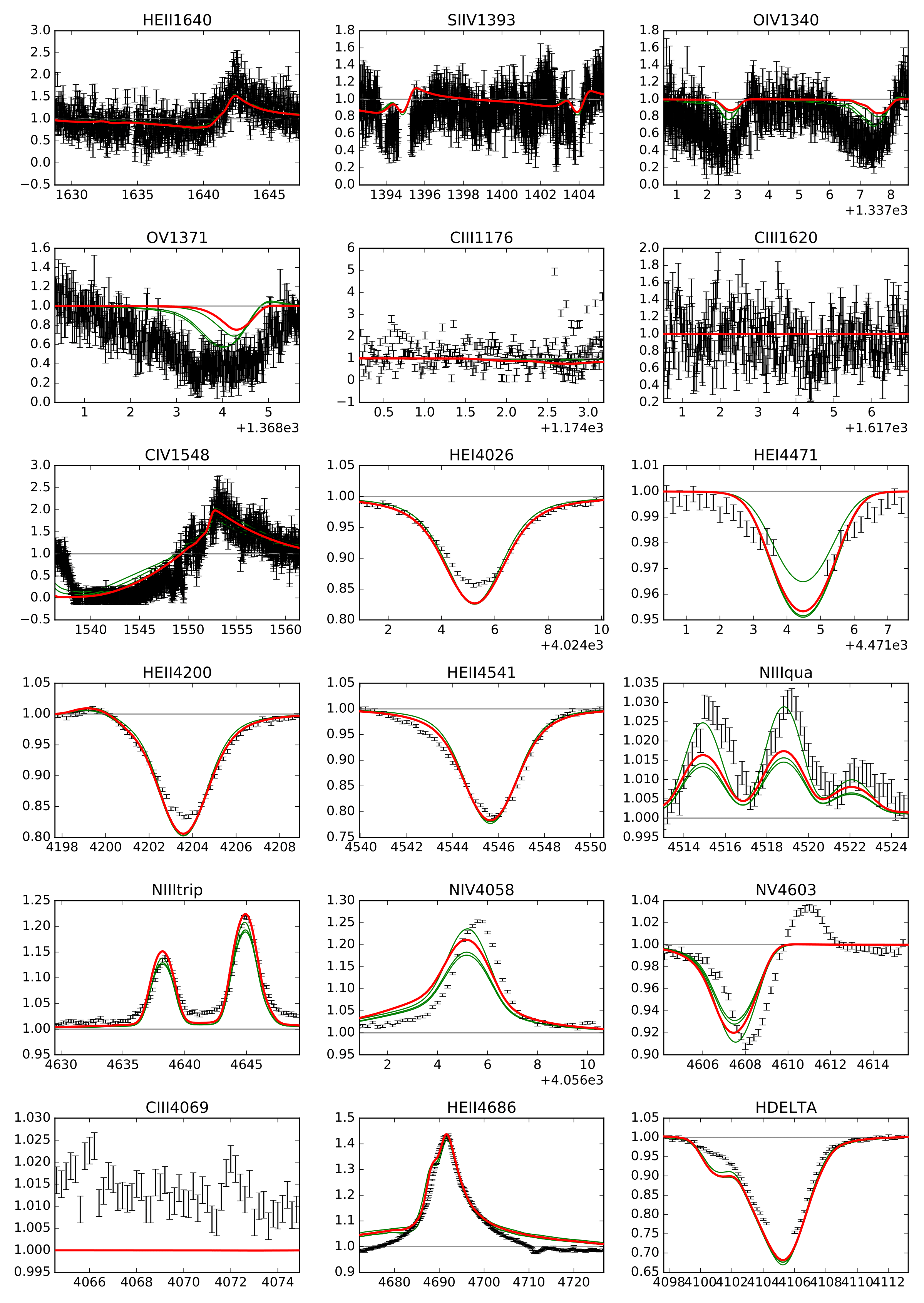}
	\includegraphics[scale=0.145]{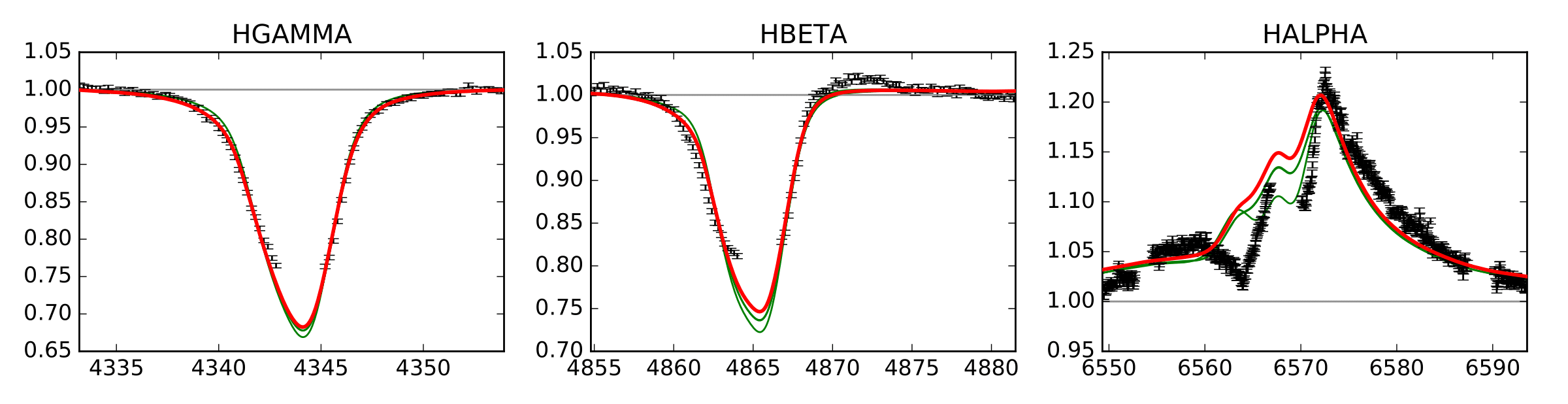}
	\caption{Best fit for VFTS180 O3If* from GA with optically thick clumping.}
	\label{fig: Best-fit-VFTS180}
\end{figure}
 
 \begin{figure}[t!]
	\centering
	\includegraphics[scale=0.3]{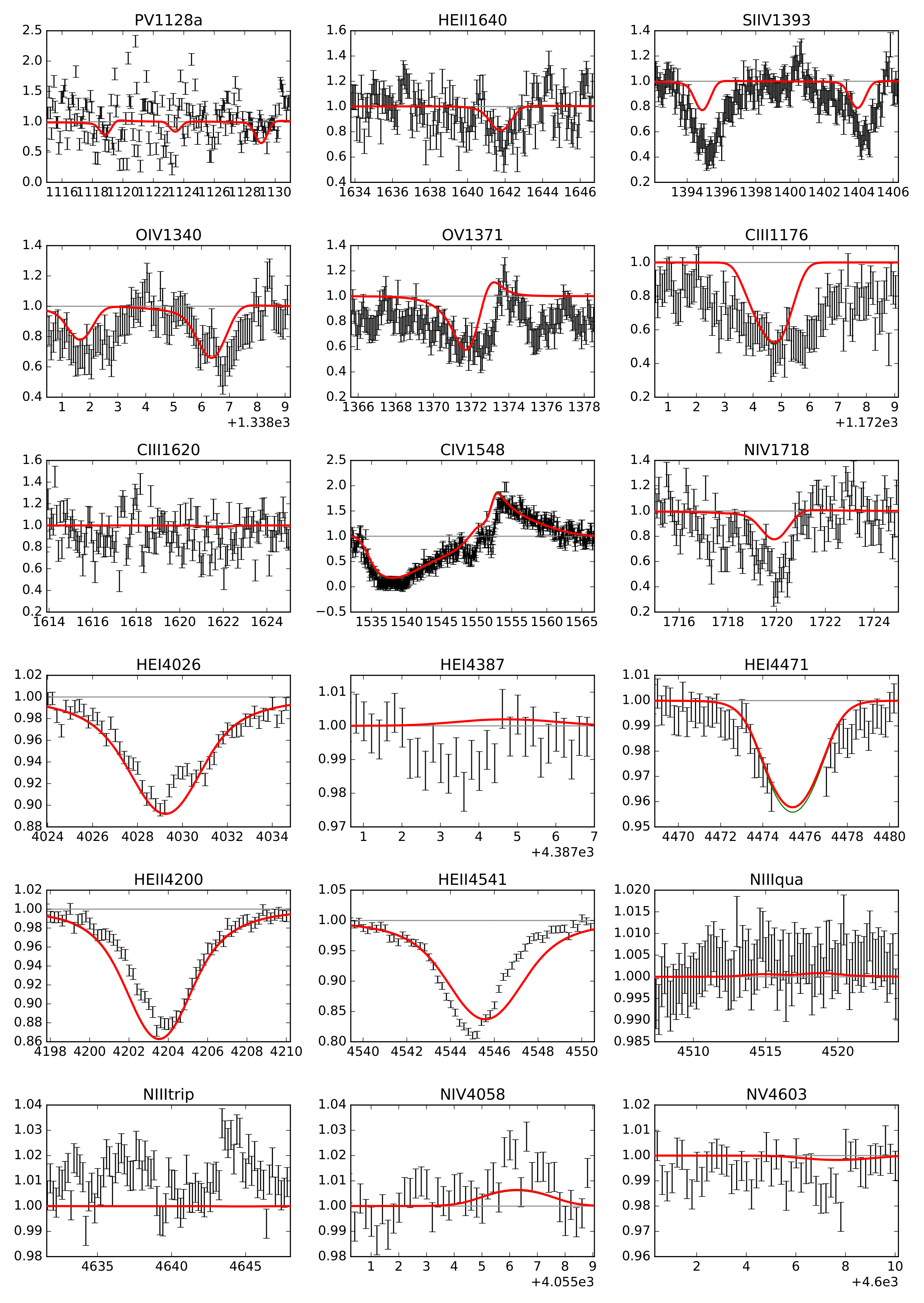}
	\includegraphics[scale=0.145]{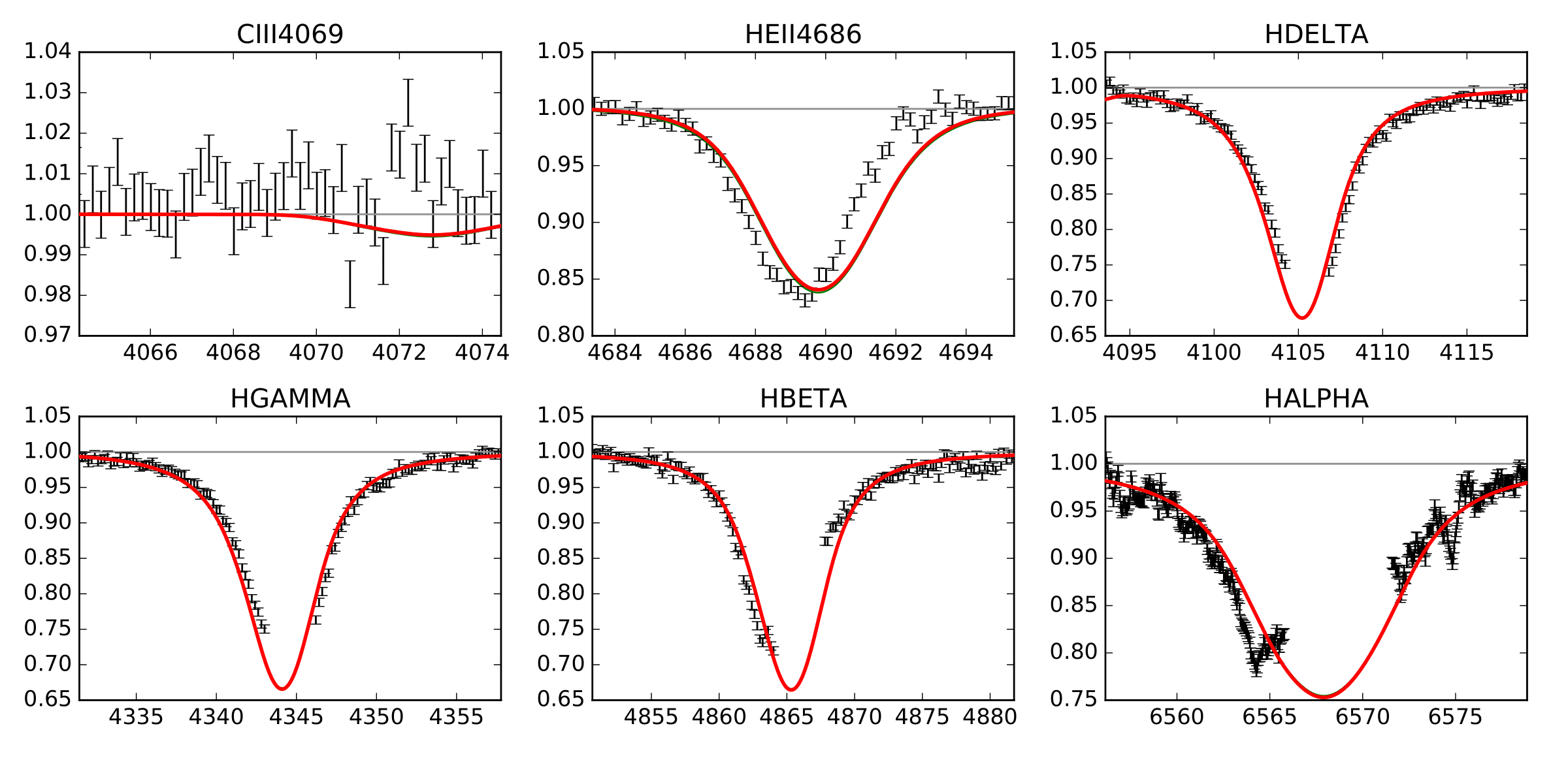}
	\caption{Best fit for VFTS143 O3.5V((fc)) from GA with optically thick clumping.}
	\label{fig: Best-fit-VFTS143}
\end{figure}

\begin{figure}[t!]
	\centering
	\includegraphics[scale=0.3]{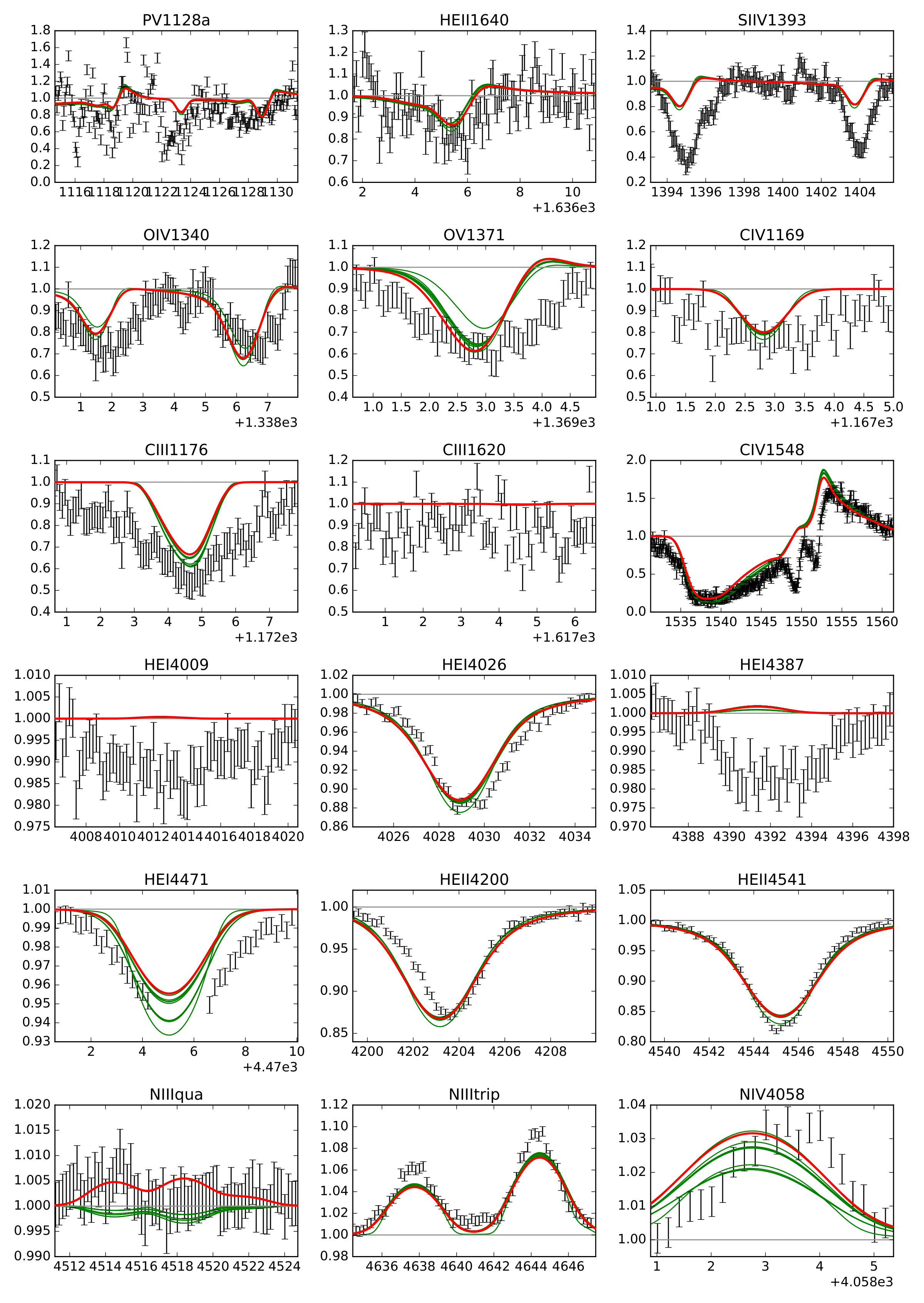}
	\includegraphics[scale=0.145]{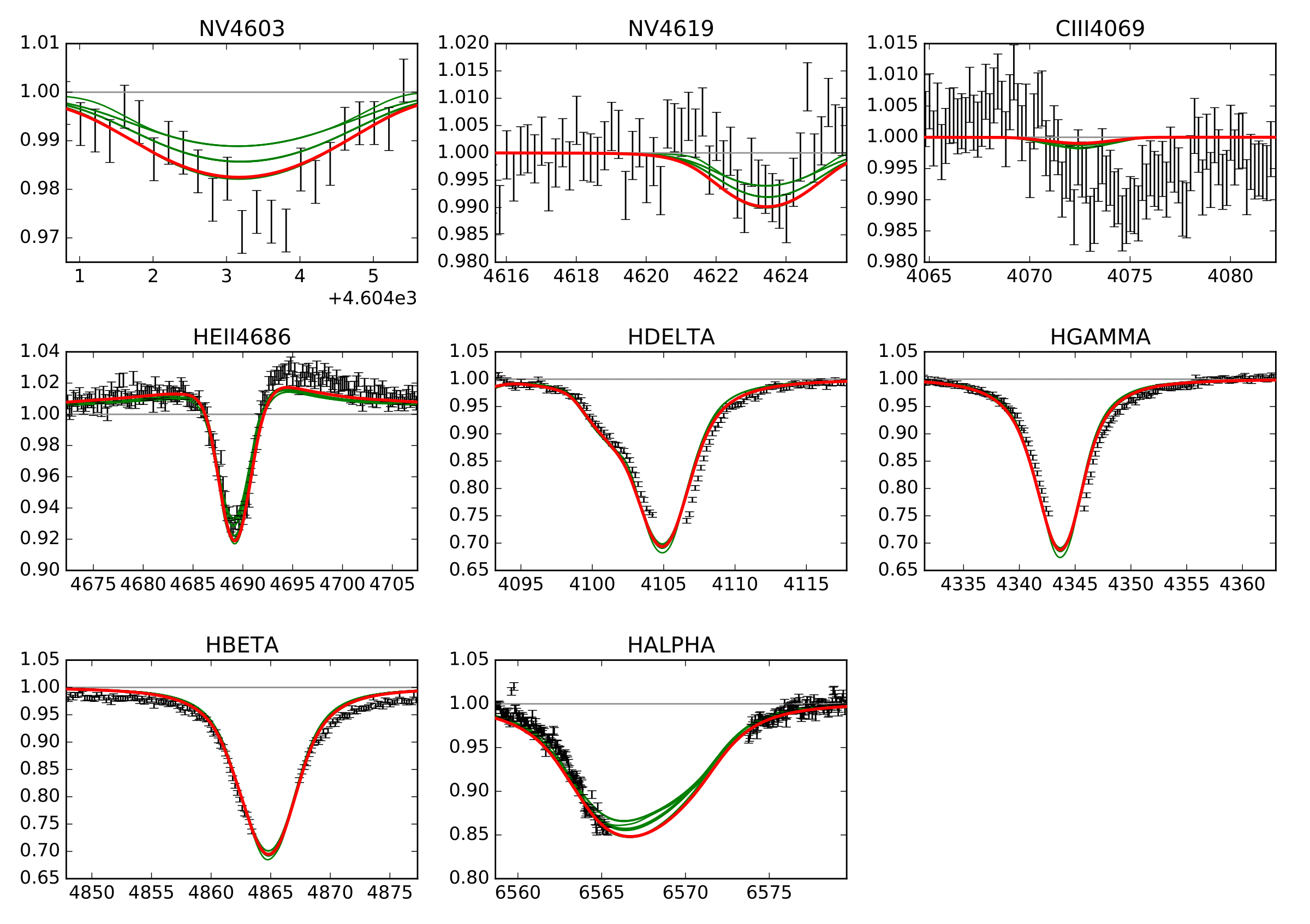}
	\caption{Best fit for VFTS608 O4III(f) from GA with optically thick clumping.}
	\label{fig: Best-fit-VFTS608}
\end{figure}

\begin{figure}[t!]
	\centering
	\includegraphics[scale=0.3]{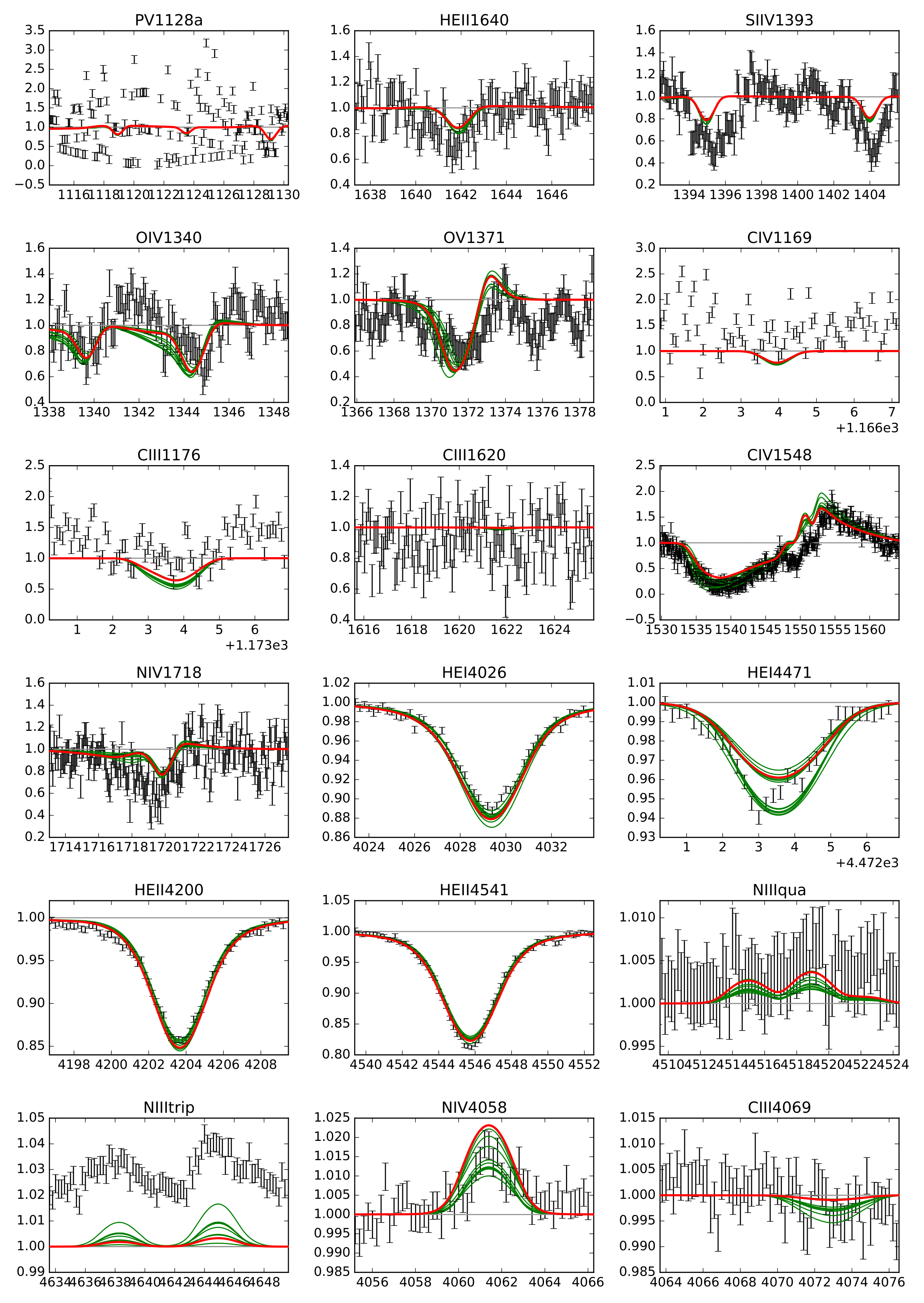}
	\includegraphics[scale=0.145]{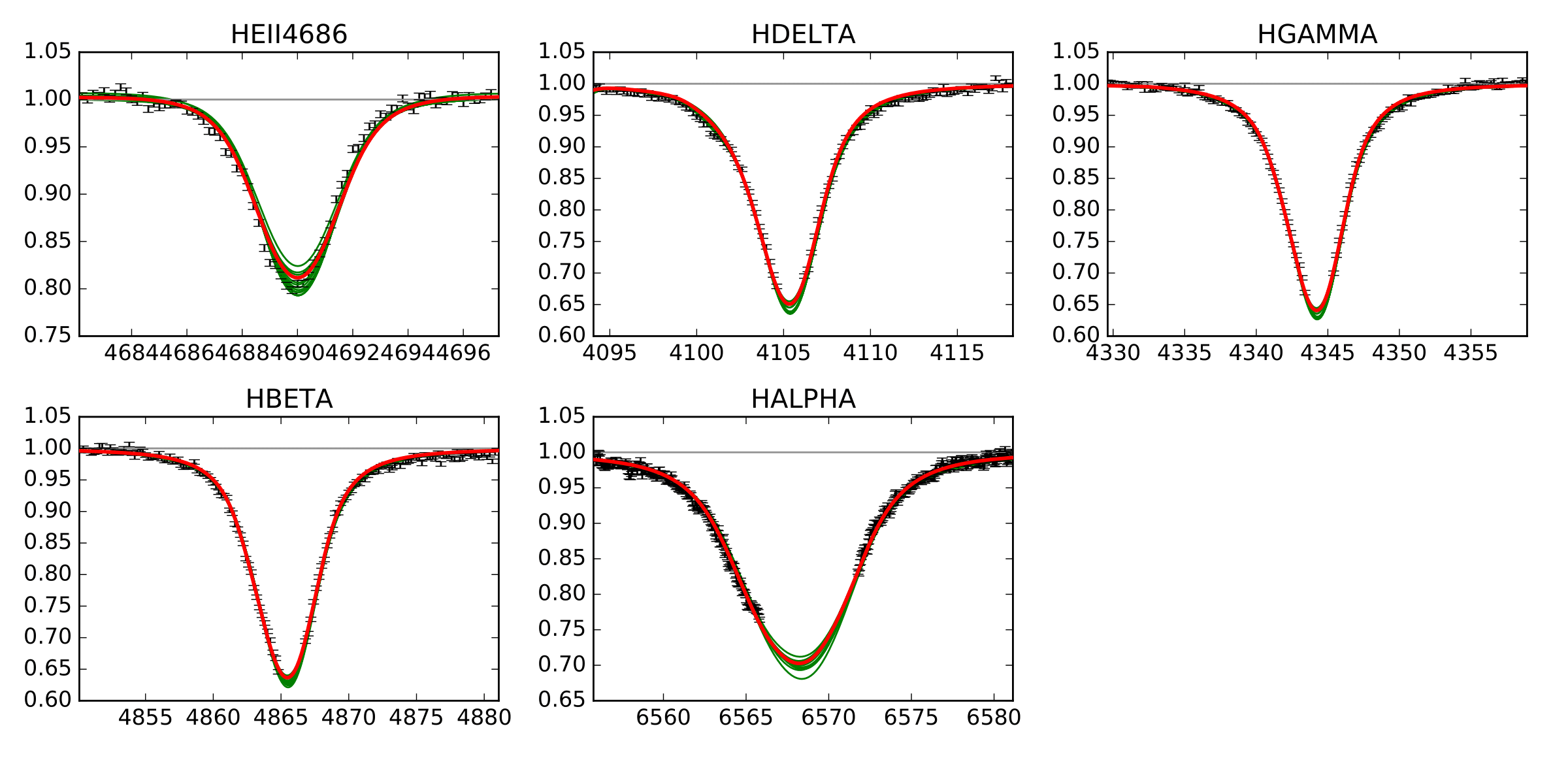}
	\caption{Best fit for VFTS216 O4V((fc)) from GA with optically thick clumping.}
	\label{fig: Best-fit-VFTS216}
\end{figure}

\begin{figure}[t!]
	\centering
	\includegraphics[scale=0.3]{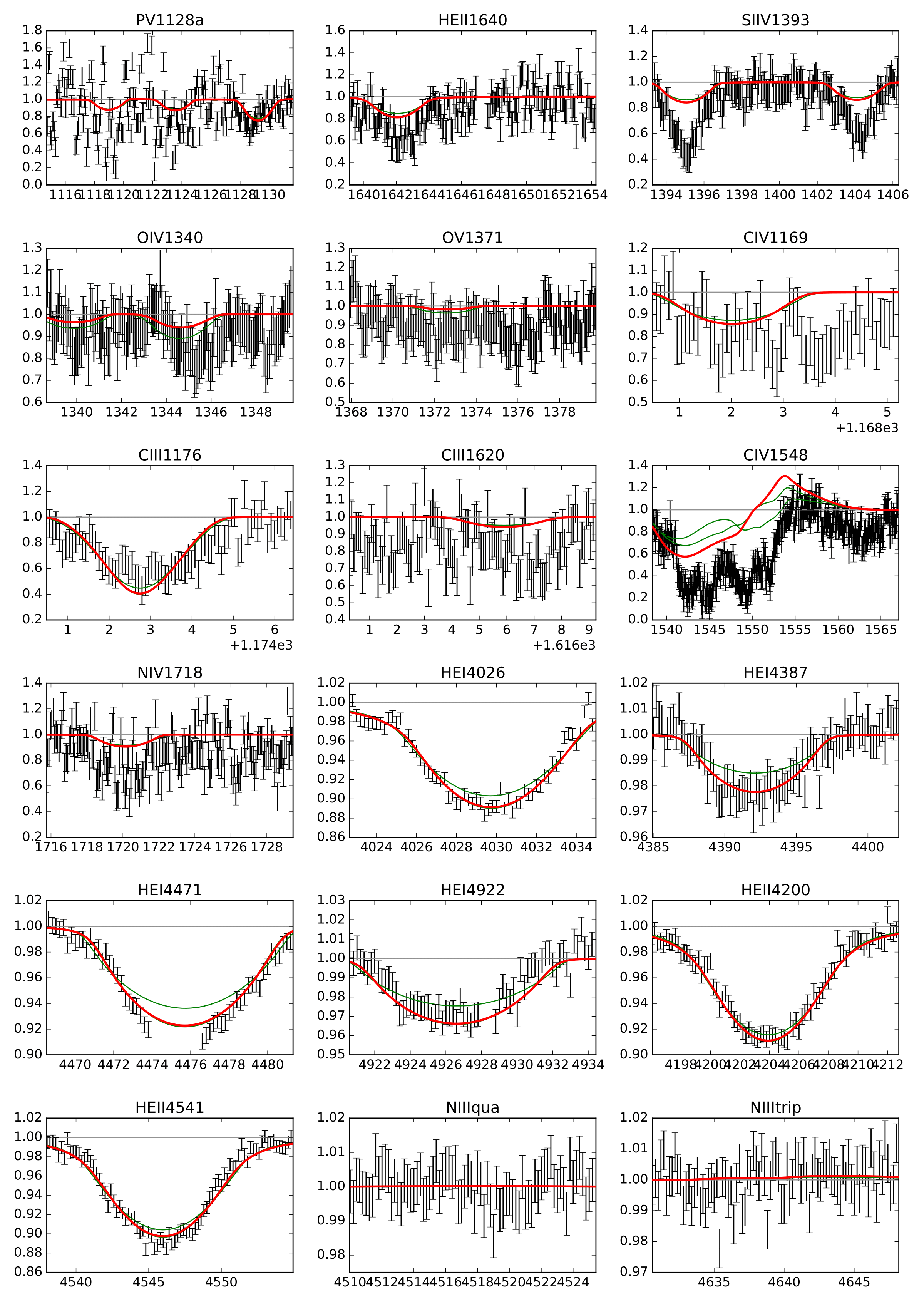}
	\includegraphics[scale=0.145]{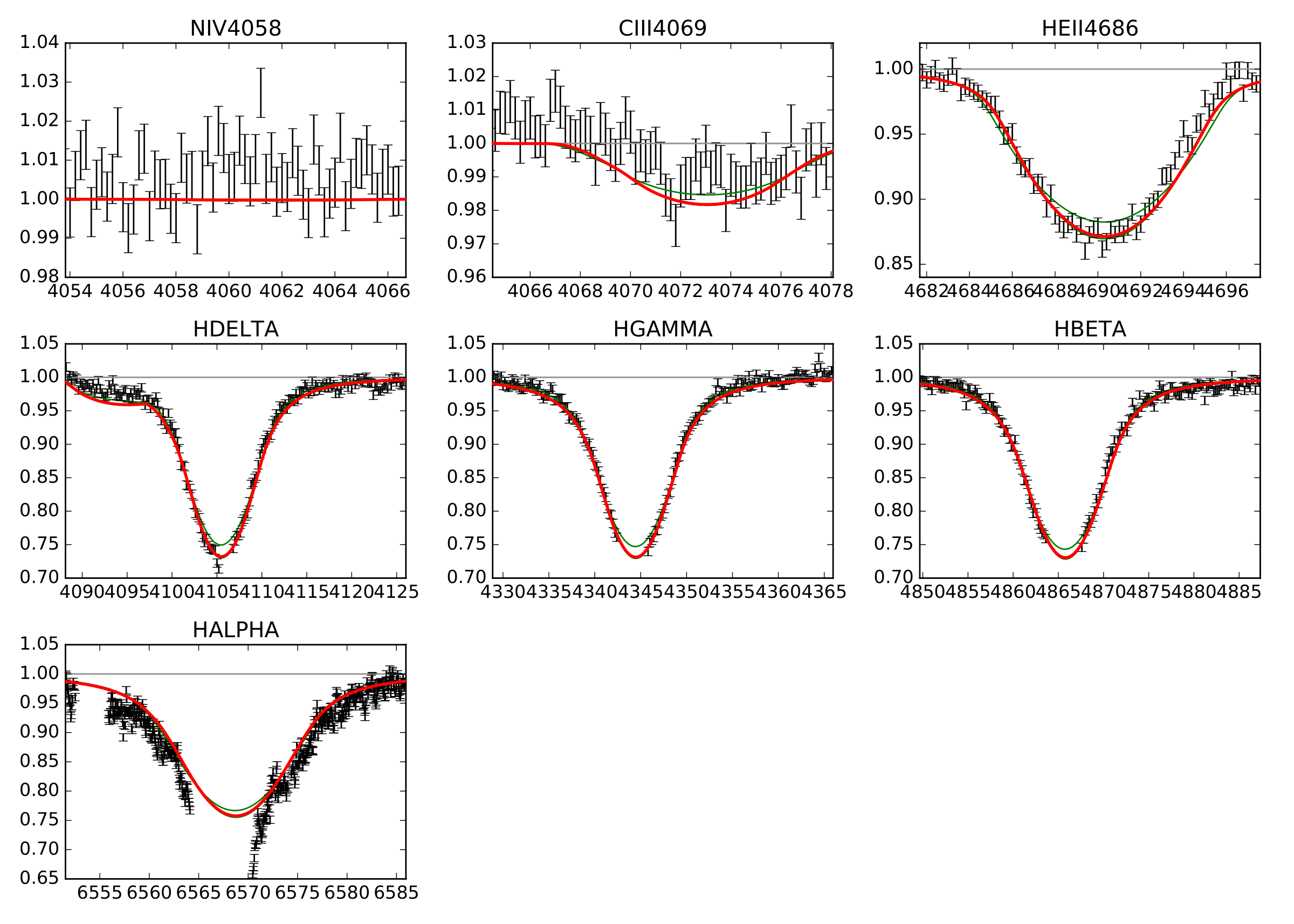}
	\caption{Best fit for VFTS184 O6.5Vnz from GA with optically thick clumping.}
	\label{fig: Best-fit-VFTS184}
\end{figure}

\begin{figure}[t!]
	\centering
	\includegraphics[scale=0.3]{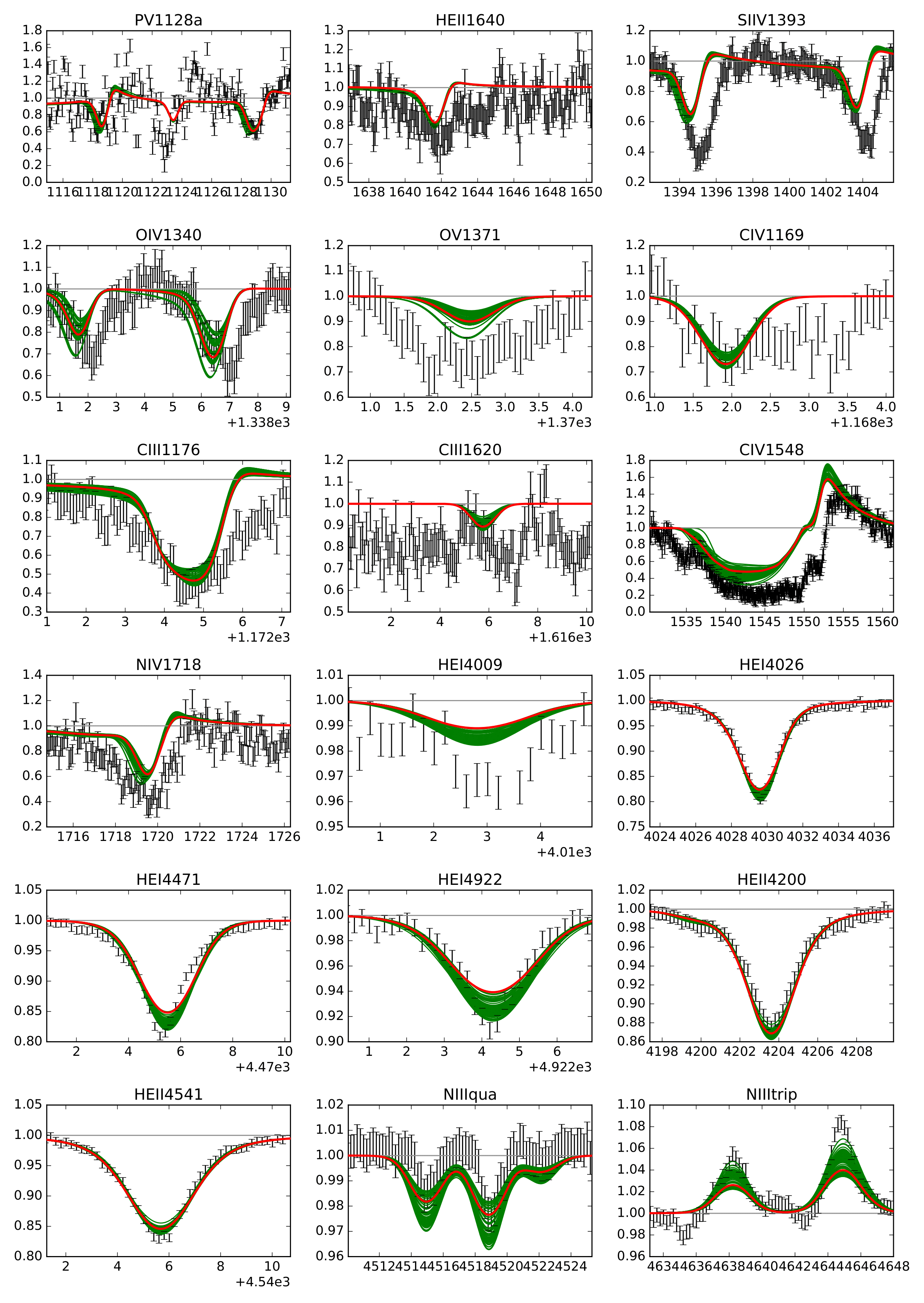}
	\includegraphics[scale=0.145]{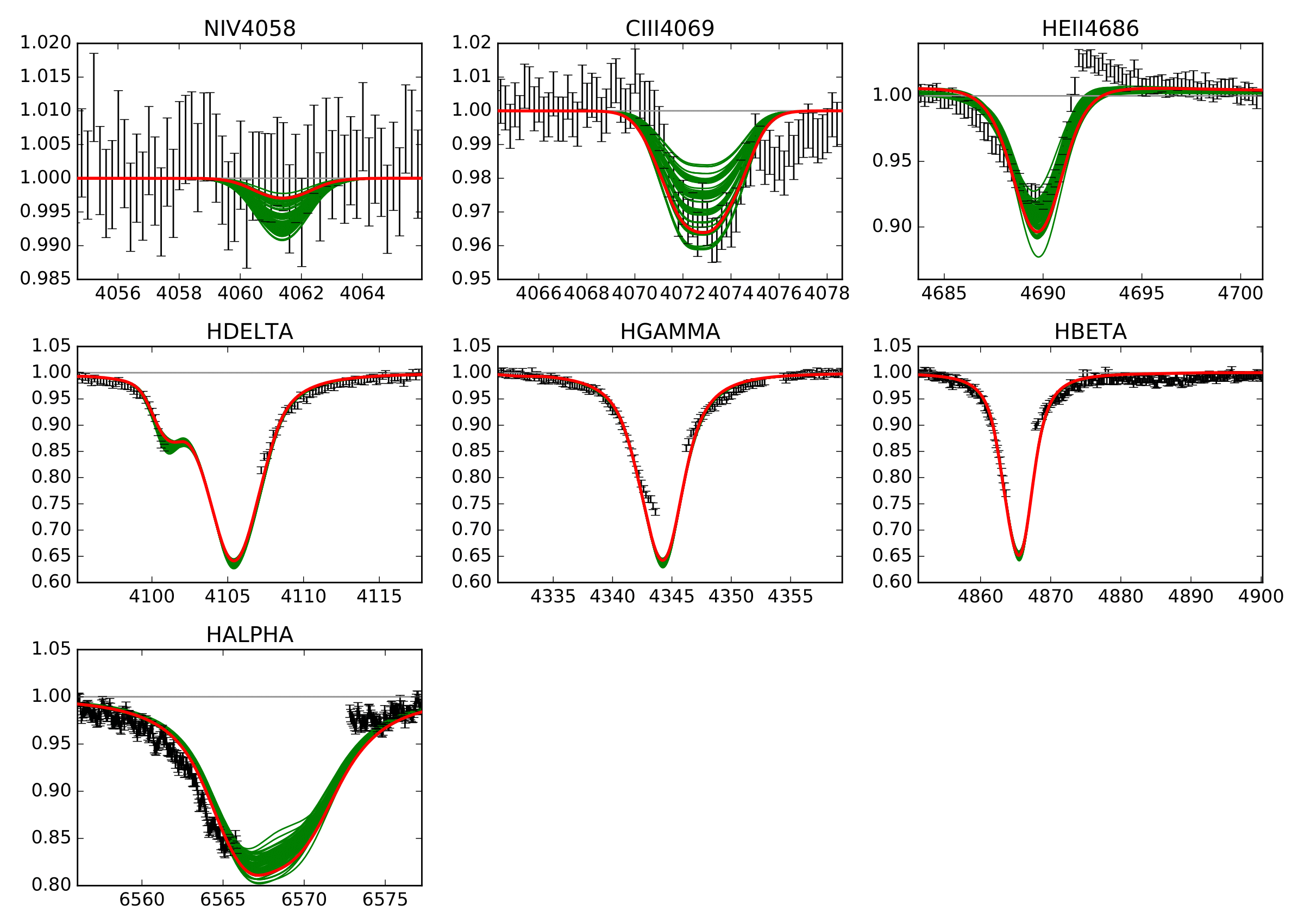}
	\caption{Best fit for VFTS664 O7II(f) from GA with optically thick clumping.}
	\label{fig: Best-fit-VFTS664}
\end{figure}

\begin{figure}[t!]
	\centering
	\includegraphics[scale=0.3]{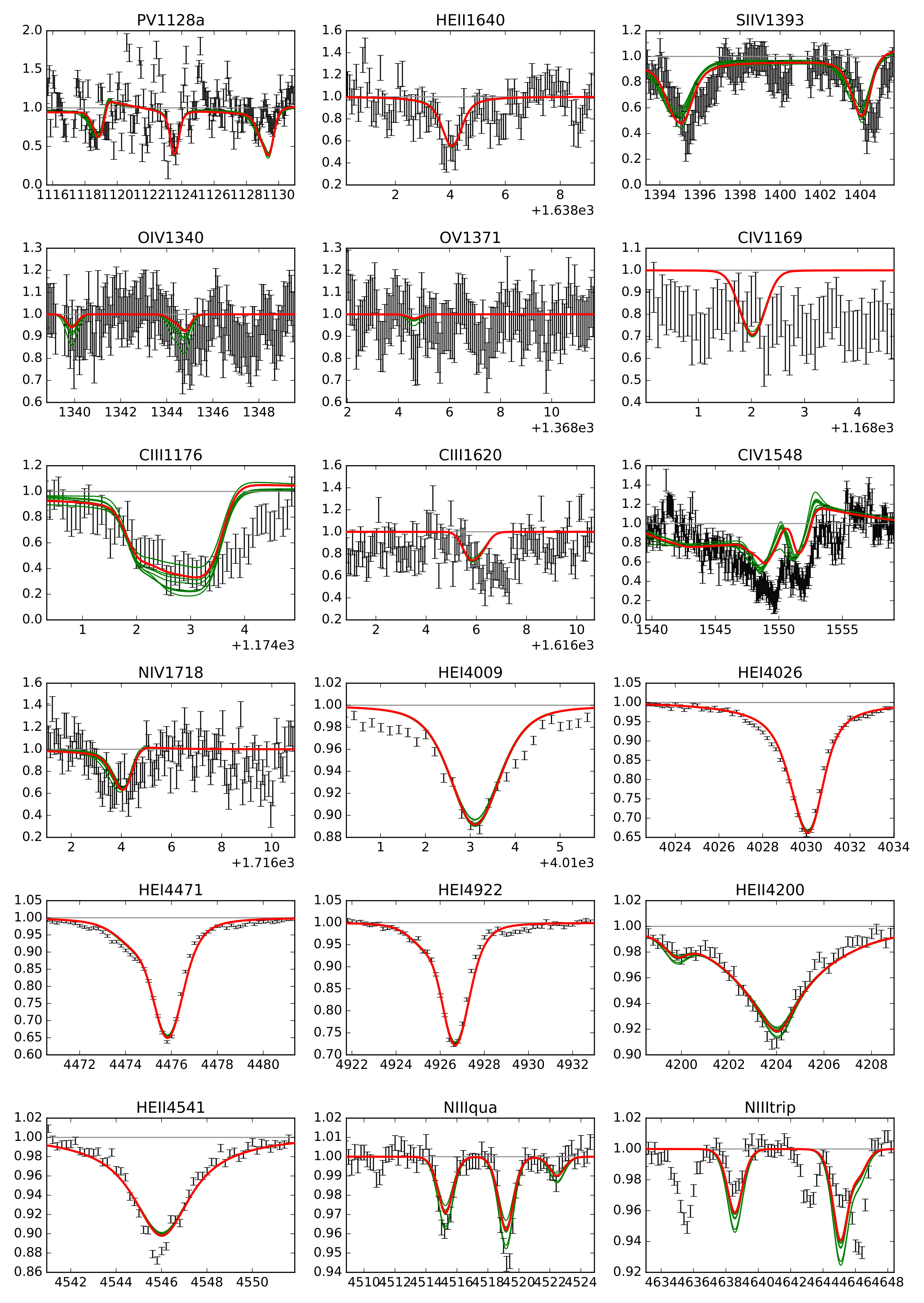}
	\includegraphics[scale=0.145]{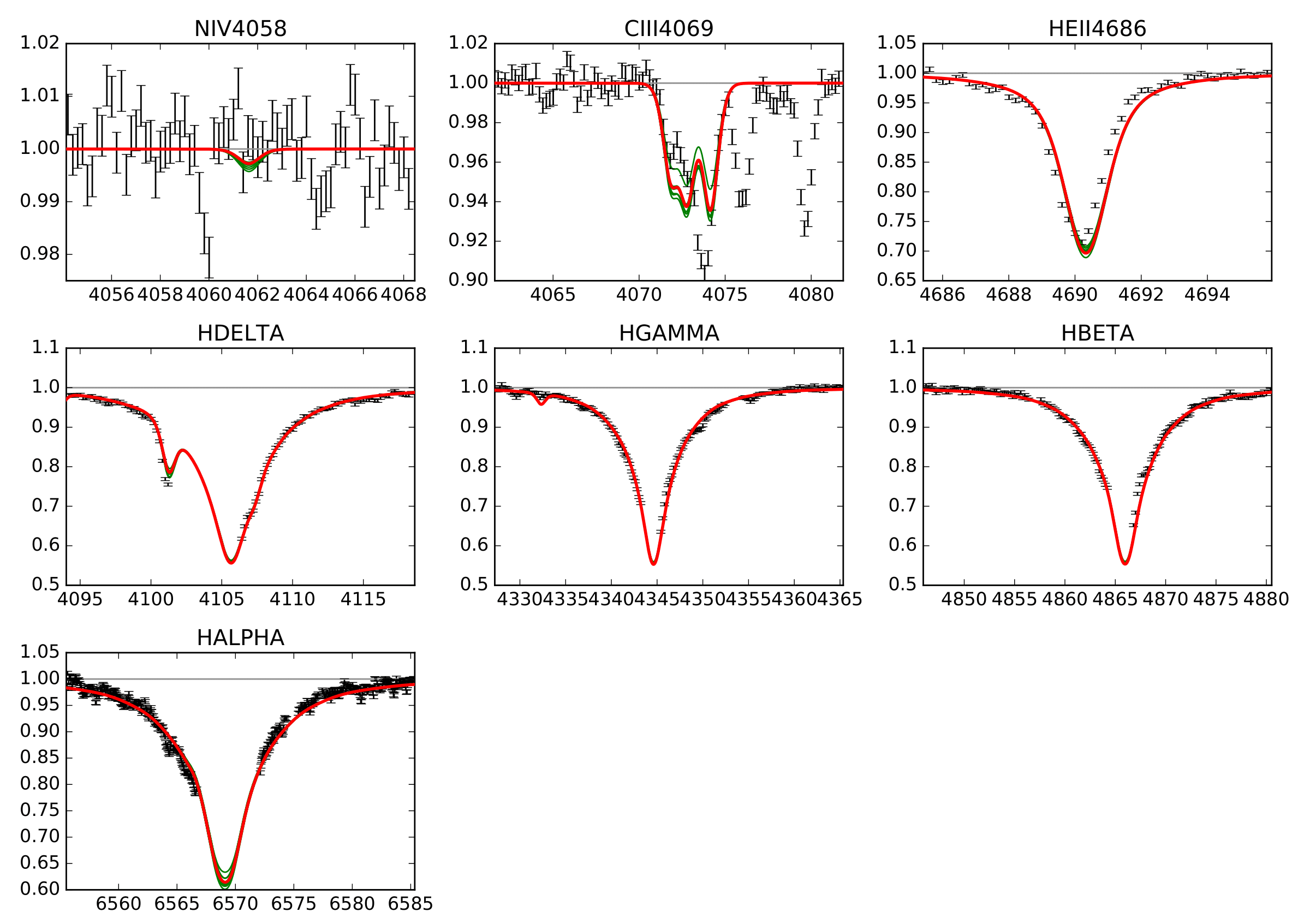}
	\caption{Best fit for VFTS223 O9.5IV from GA with optically thick clumping.}
	\label{fig: Best-fit-VFTS223}
\end{figure}

\begin{figure}[t!]
	\centering
	\includegraphics[scale=0.3]{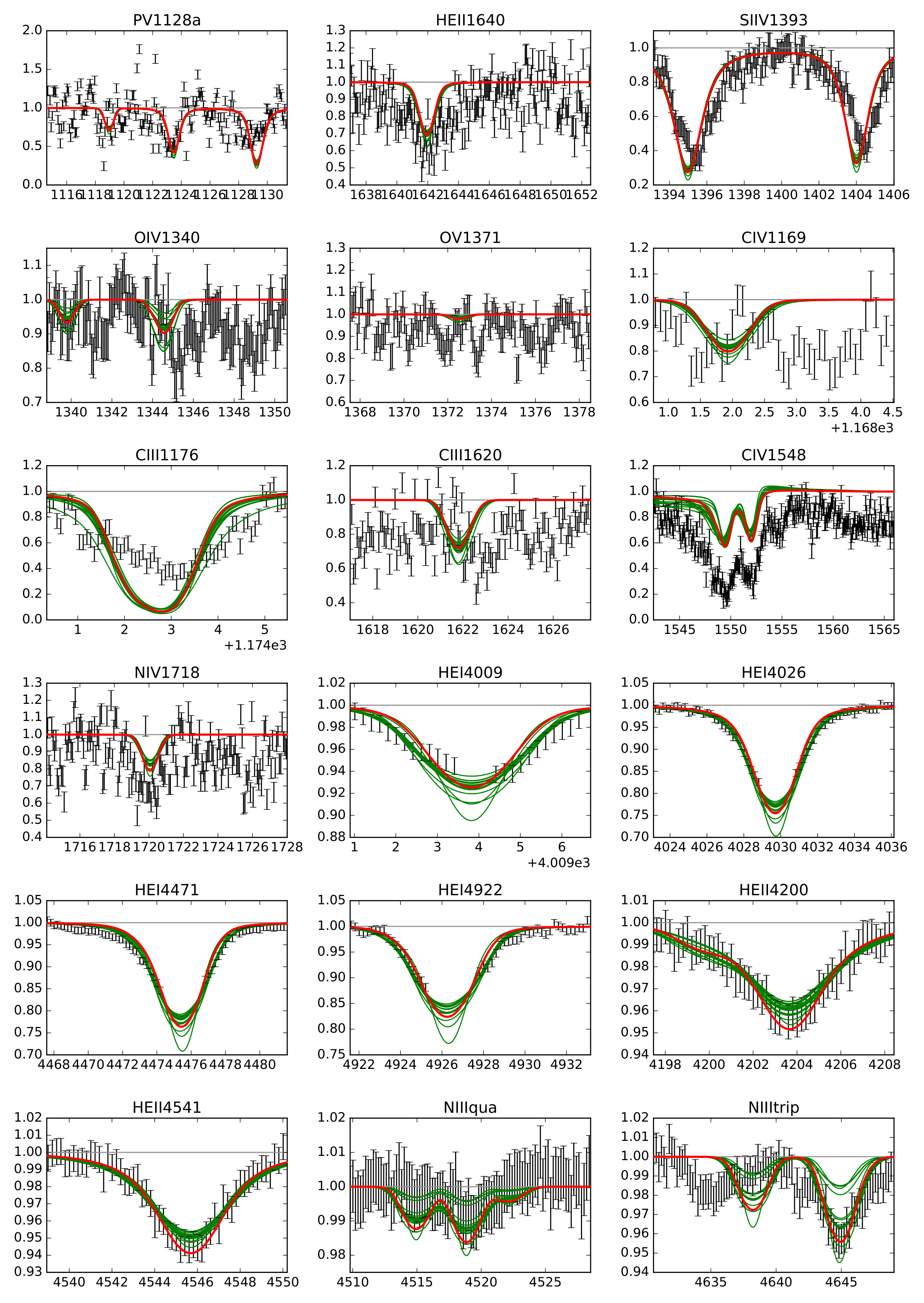}
	\includegraphics[scale=0.145]{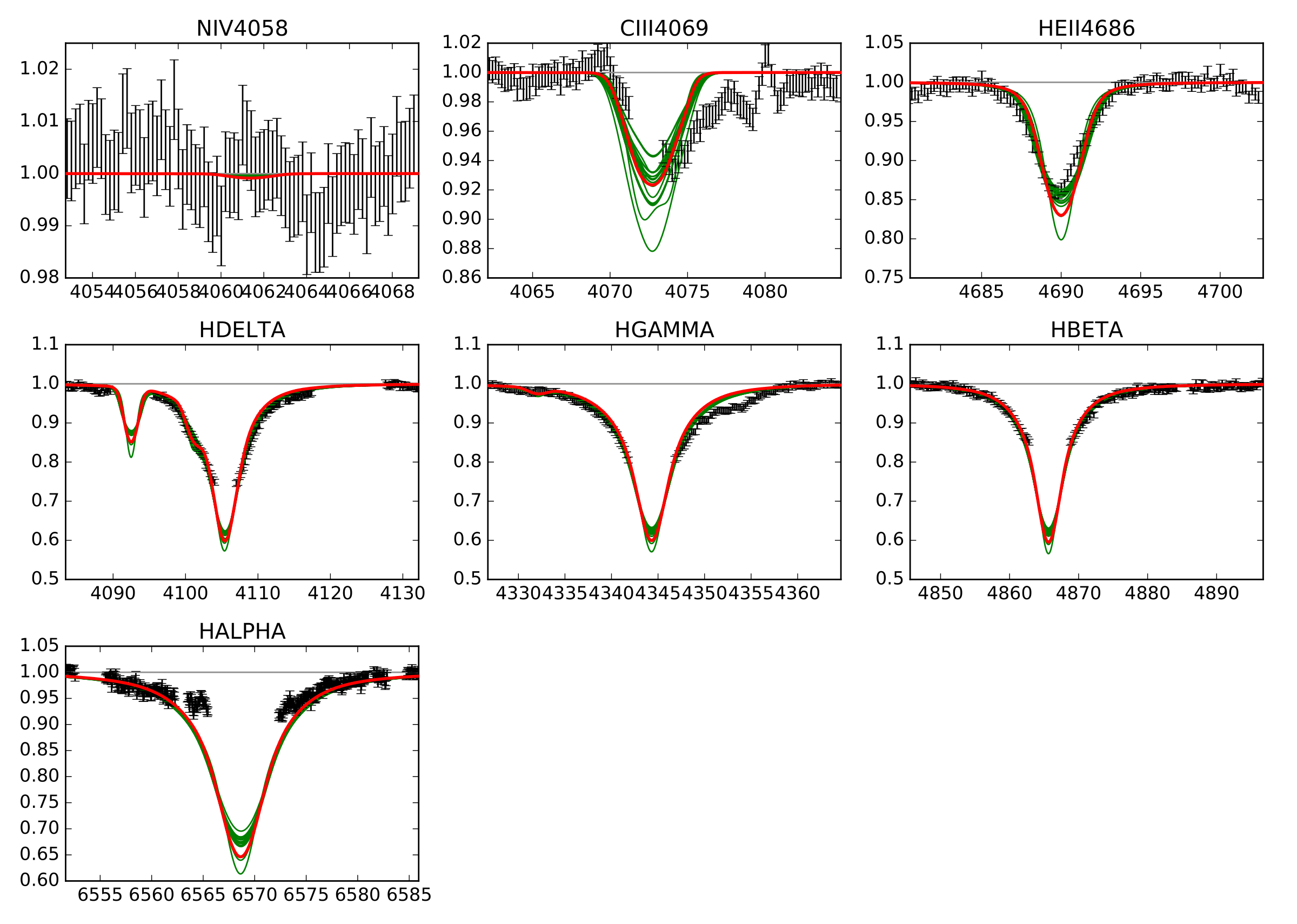}
	\caption{Best fit for VFTS517 O9.5V-III((n)) from GA with optically thick clumping.}
	\label{fig: Best-fit-VFTS517}
\end{figure}

\begin{figure}[t!]
	\centering
	\includegraphics[scale=0.3]{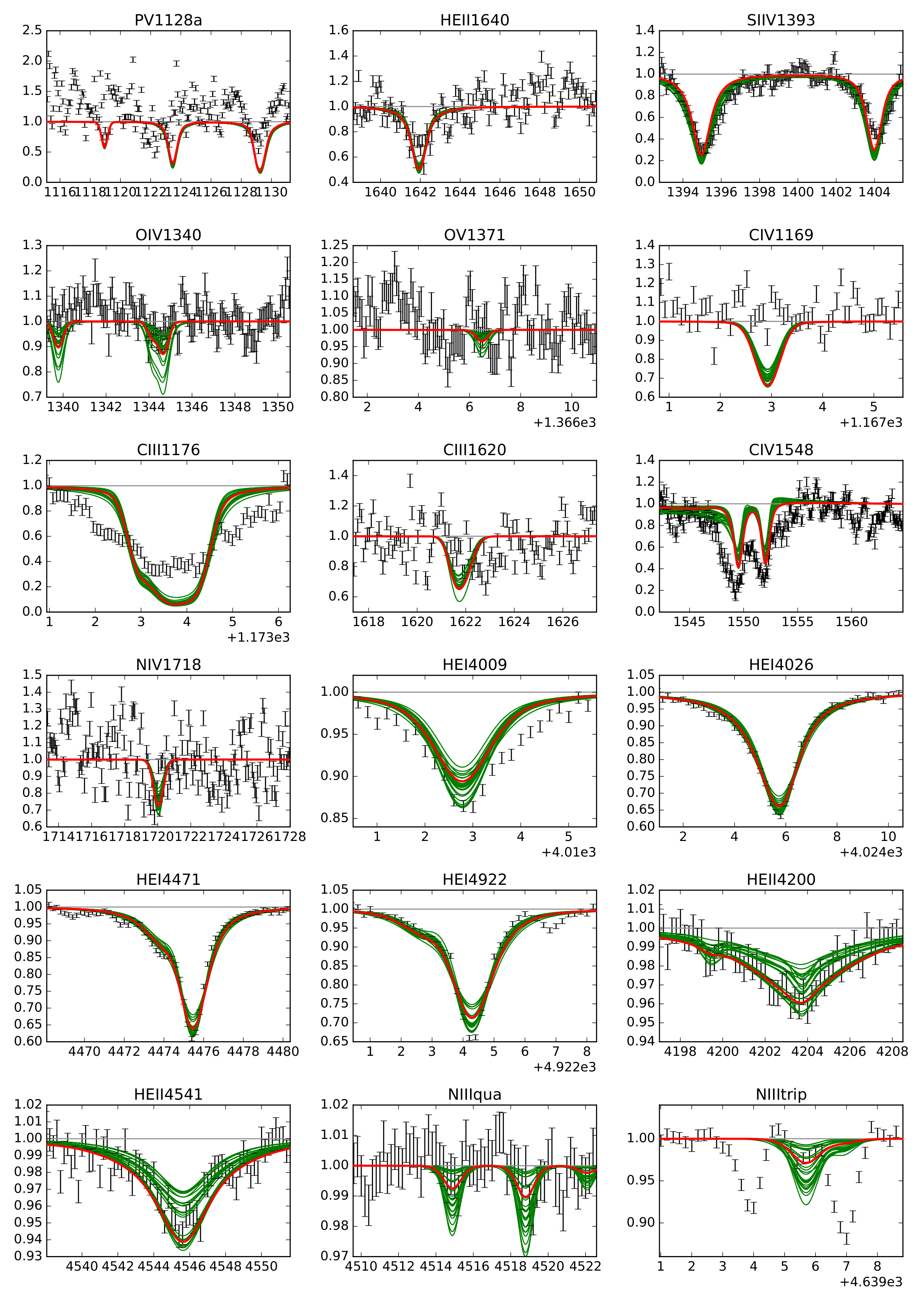}
	\includegraphics[scale=0.145]{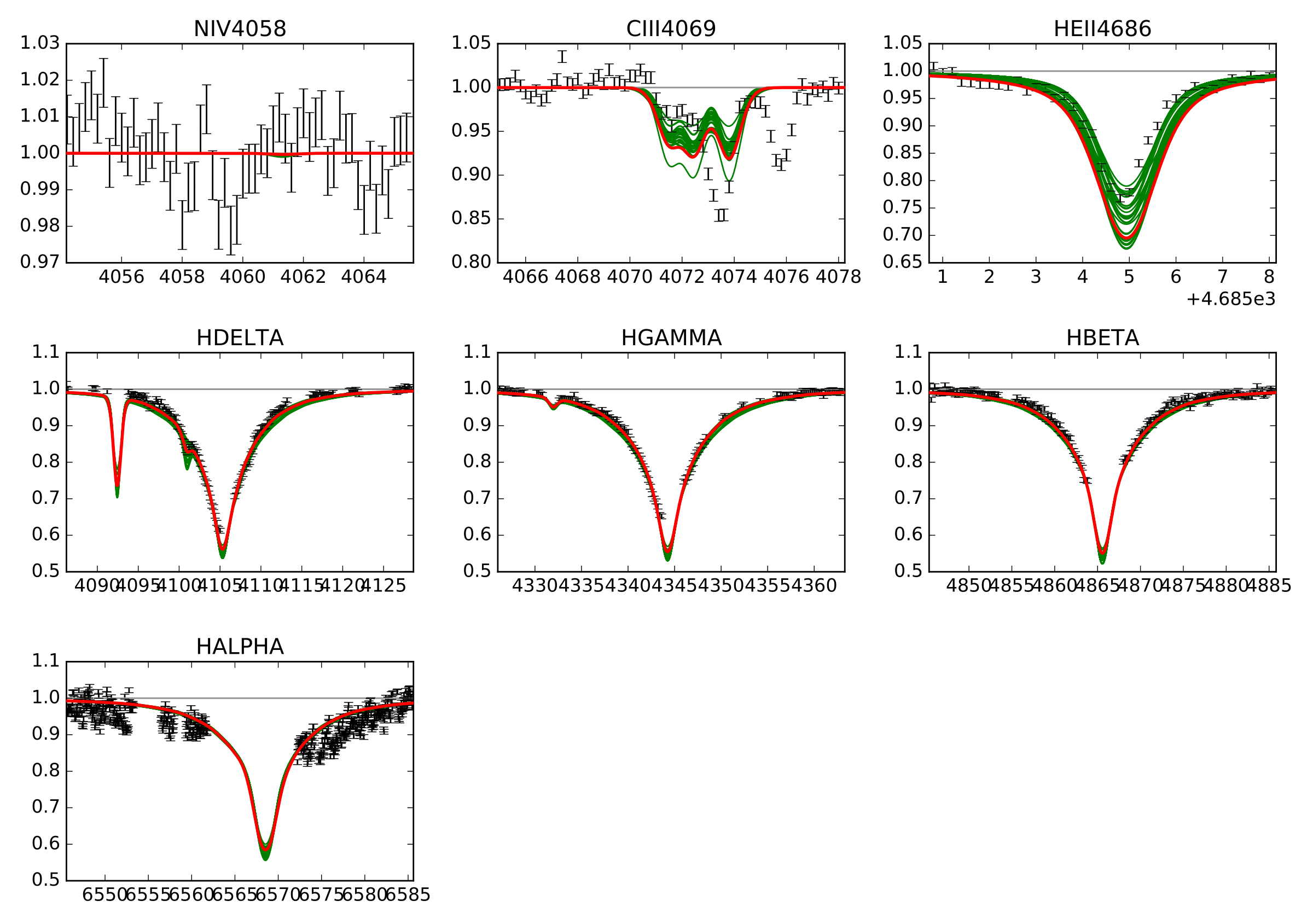}
	\caption{Best fit for VFTS235 O9.7III from GA with optically thick clumping.}
	\label{fig: Best-fit-VFTS235}
\end{figure}

\begin{figure}[t!]
	\centering
	\includegraphics[scale=0.3]{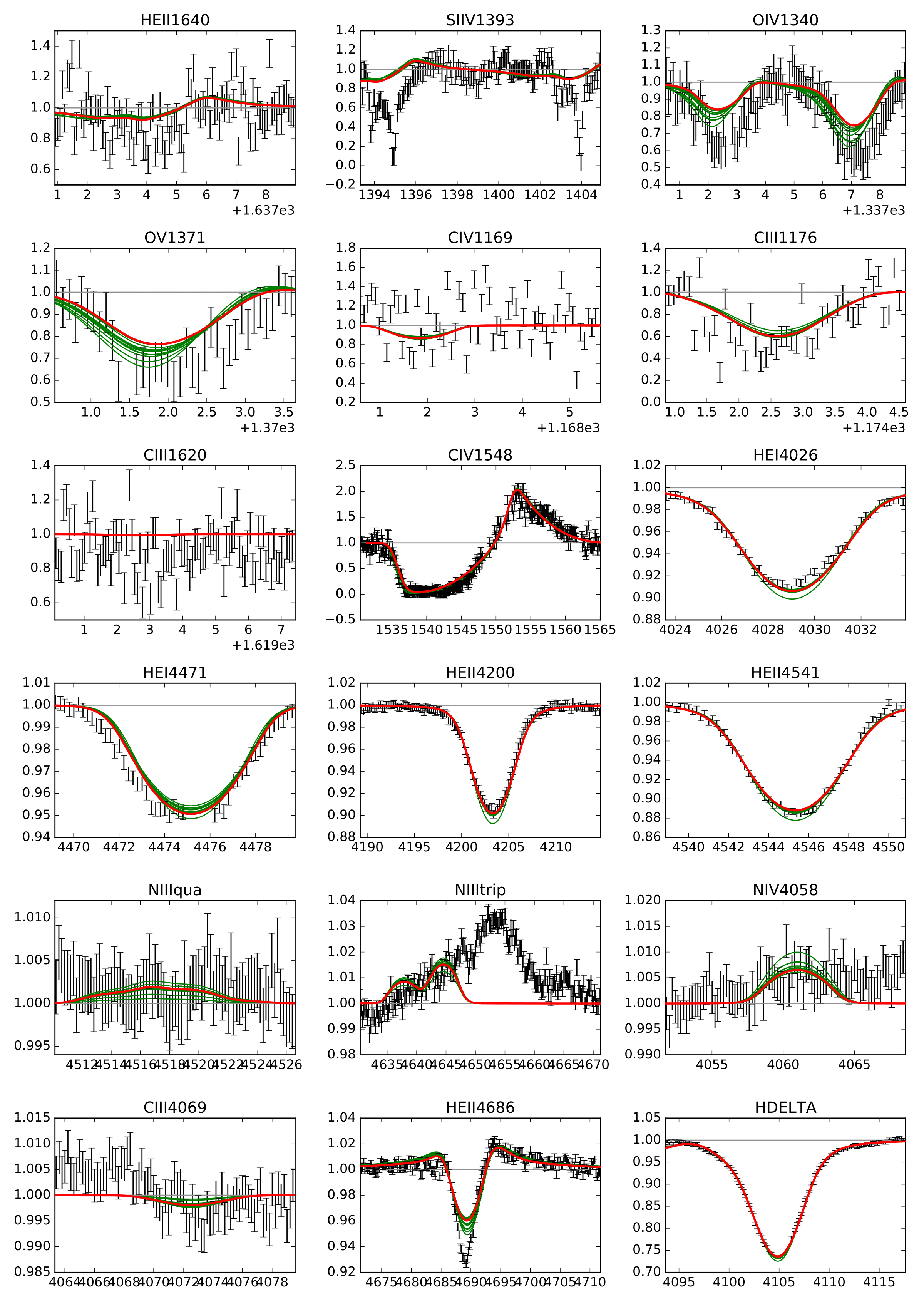}
	\includegraphics[scale=0.145]{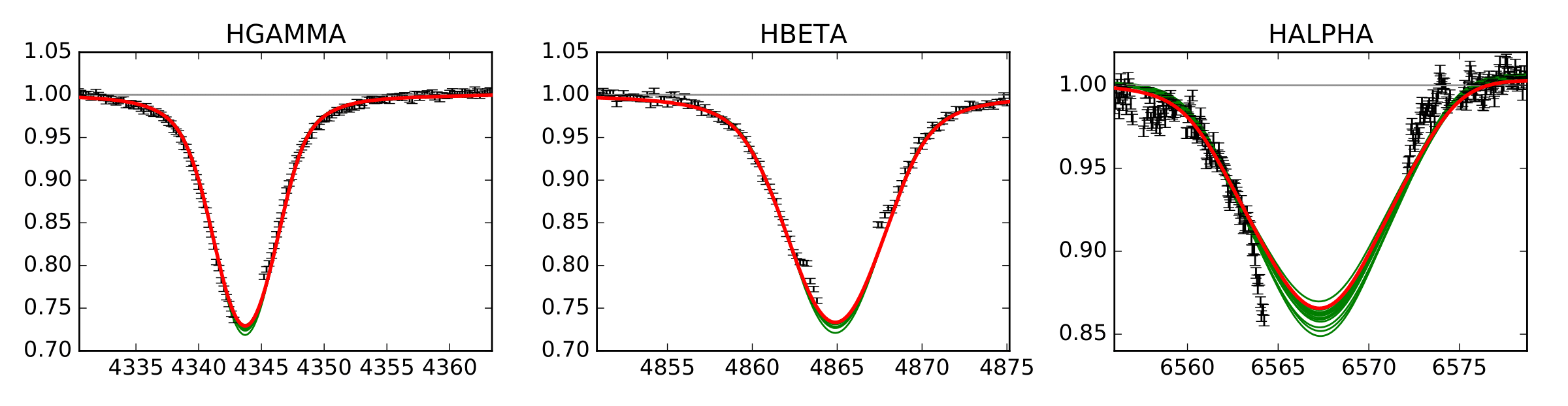}
	\caption{Best fit for VFTS244 O5III(n)(fc) from GA with optically thick clumping.}
	\label{fig: Best-fit-VFTS244}
\end{figure}

\begin{figure}[t!]
	\centering
	\includegraphics[scale=0.3]{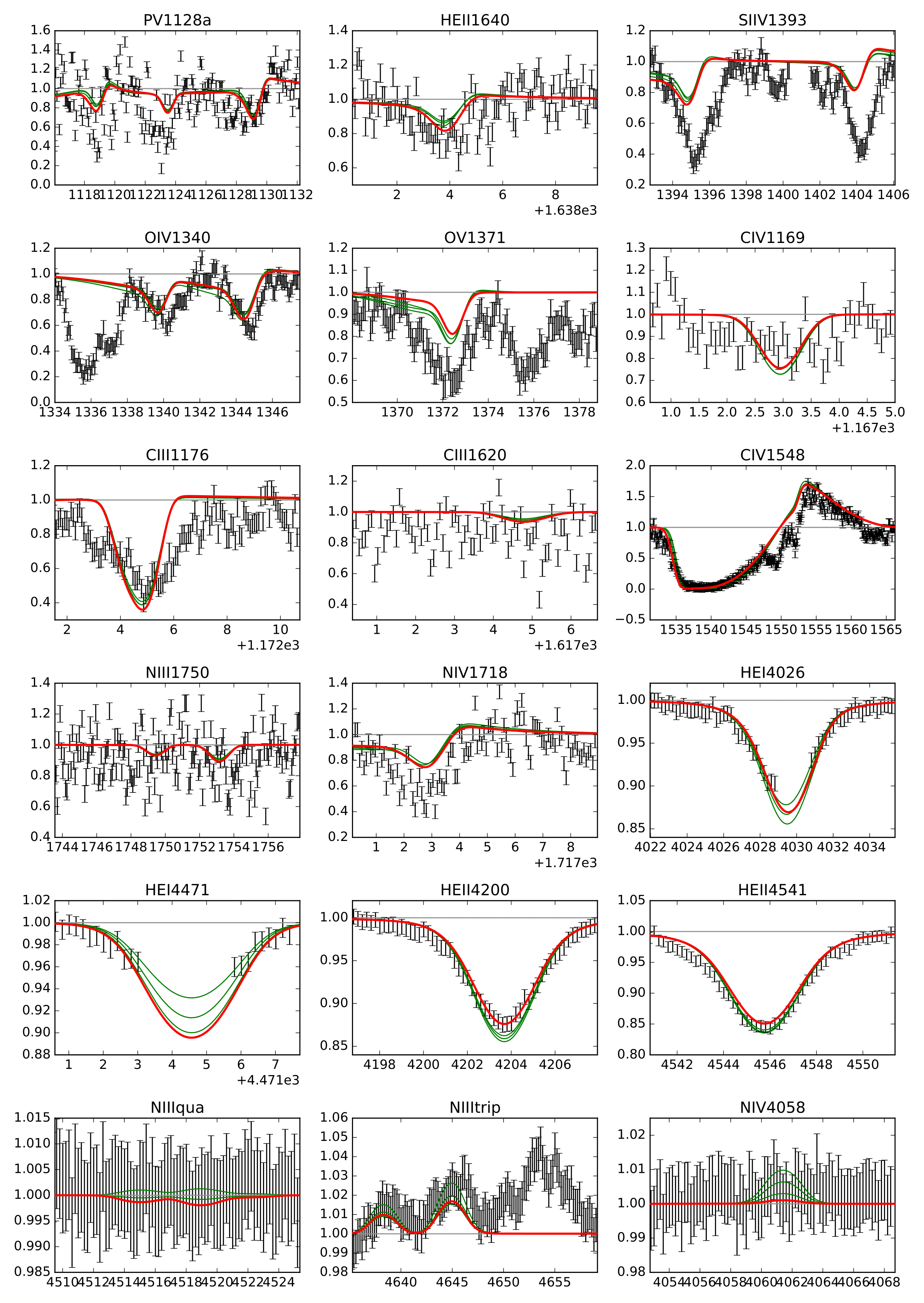}
	\includegraphics[scale=0.145]{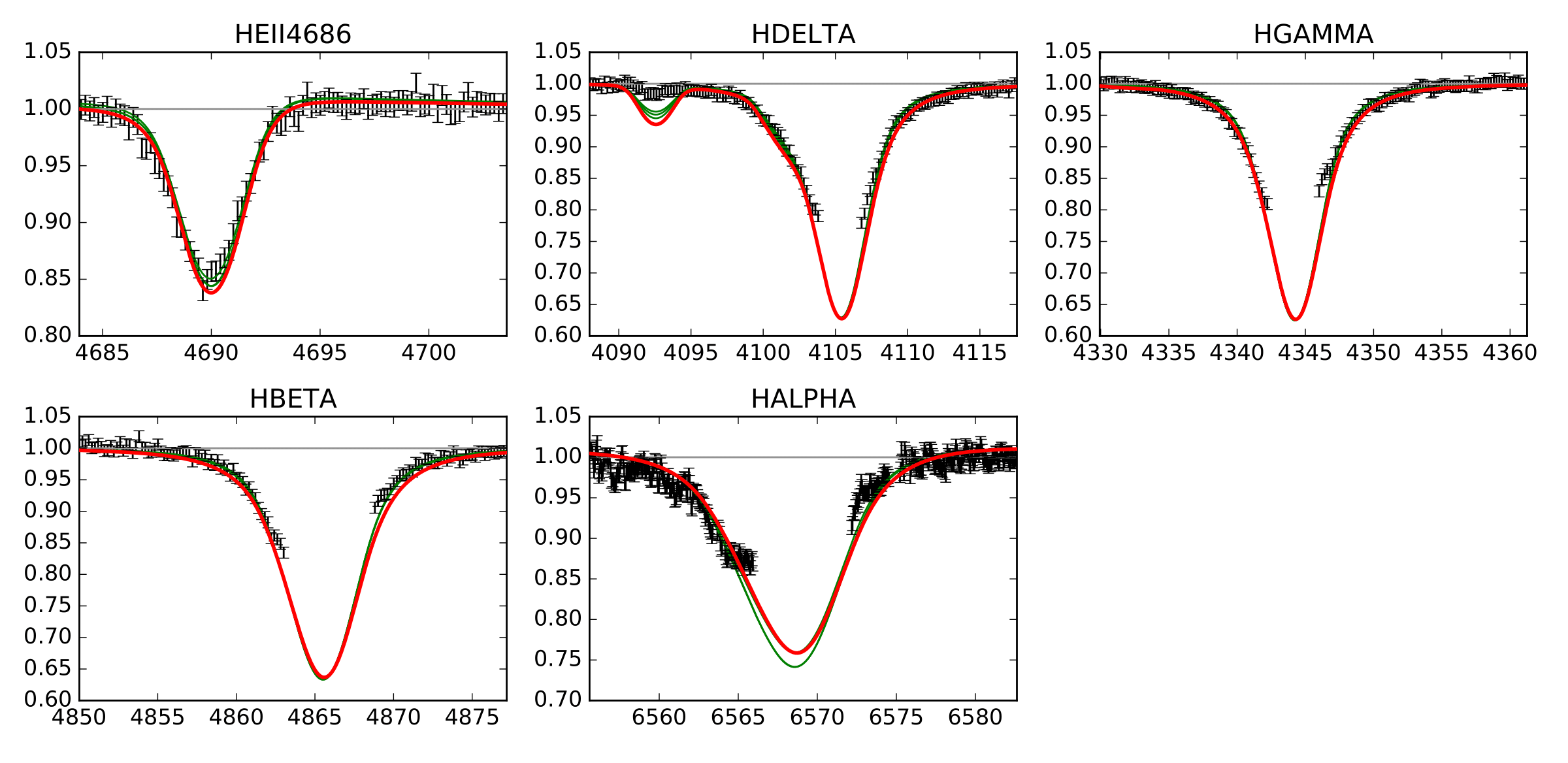}
	\caption{Best fit for VFTS385 O4-5V from GA with optically thick clumping.}
	\label{fig: Best-fit-VFTS385}
\end{figure}

\begin{figure}[t!]
	\centering
	\includegraphics[scale=0.3]{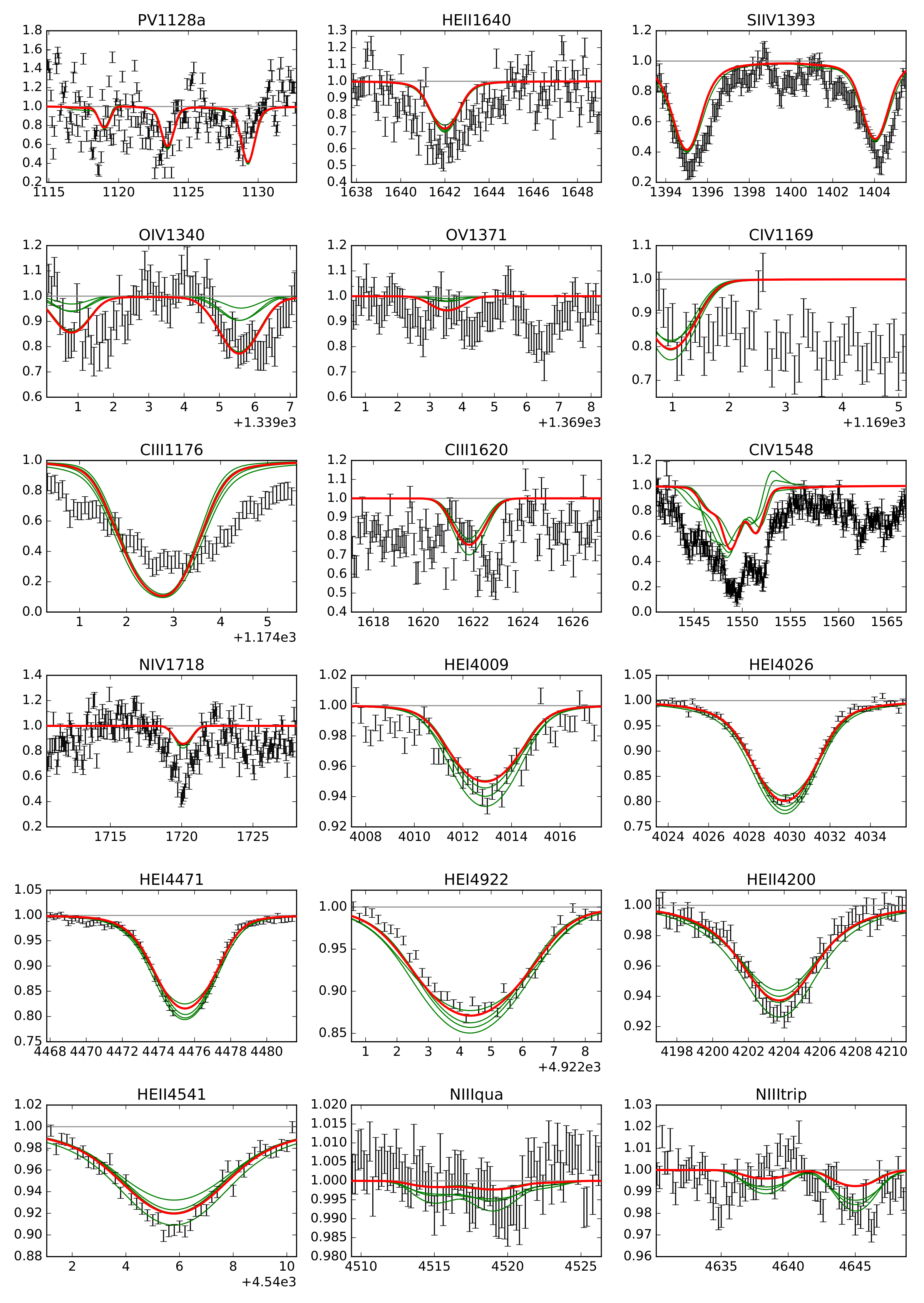}
	\includegraphics[scale=0.145]{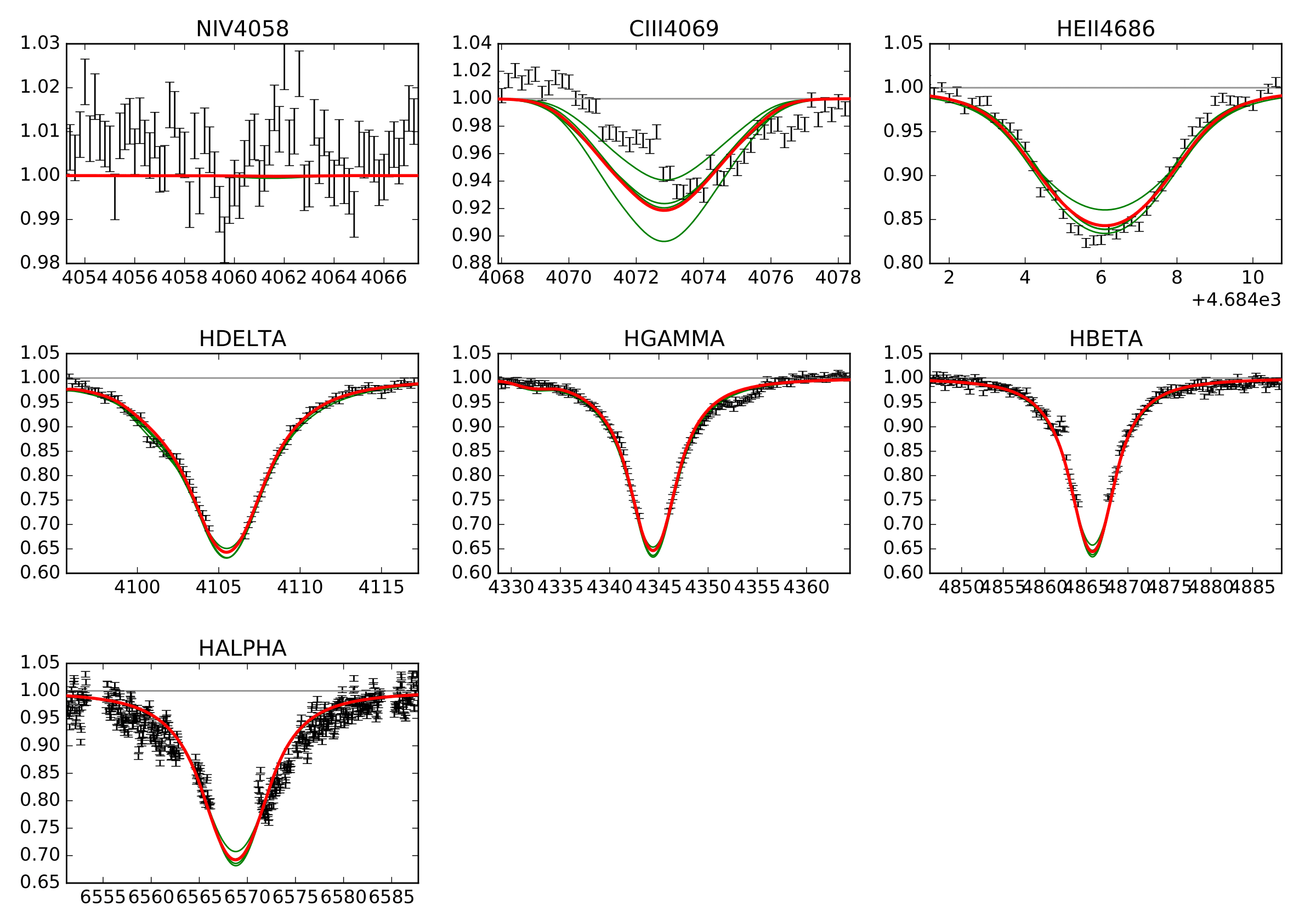}
	\caption{Best fit for VFTS280 O9V from GA with optically thick clumping.}
	\label{fig: Best-fit-VFTS280}
\end{figure}

\begin{figure}[t!]
	\centering
	\includegraphics[scale=0.3]{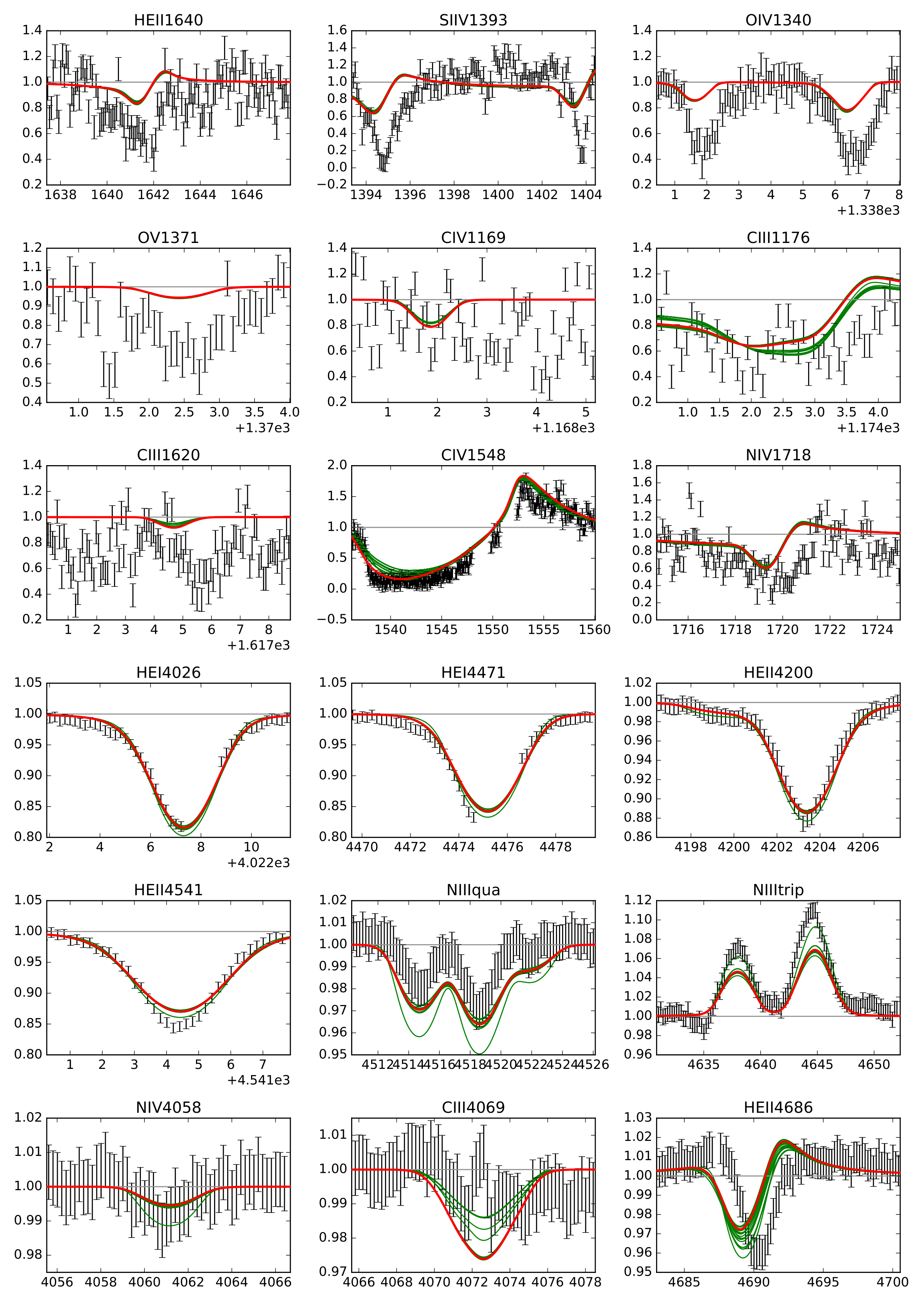}
	\includegraphics[scale=0.145]{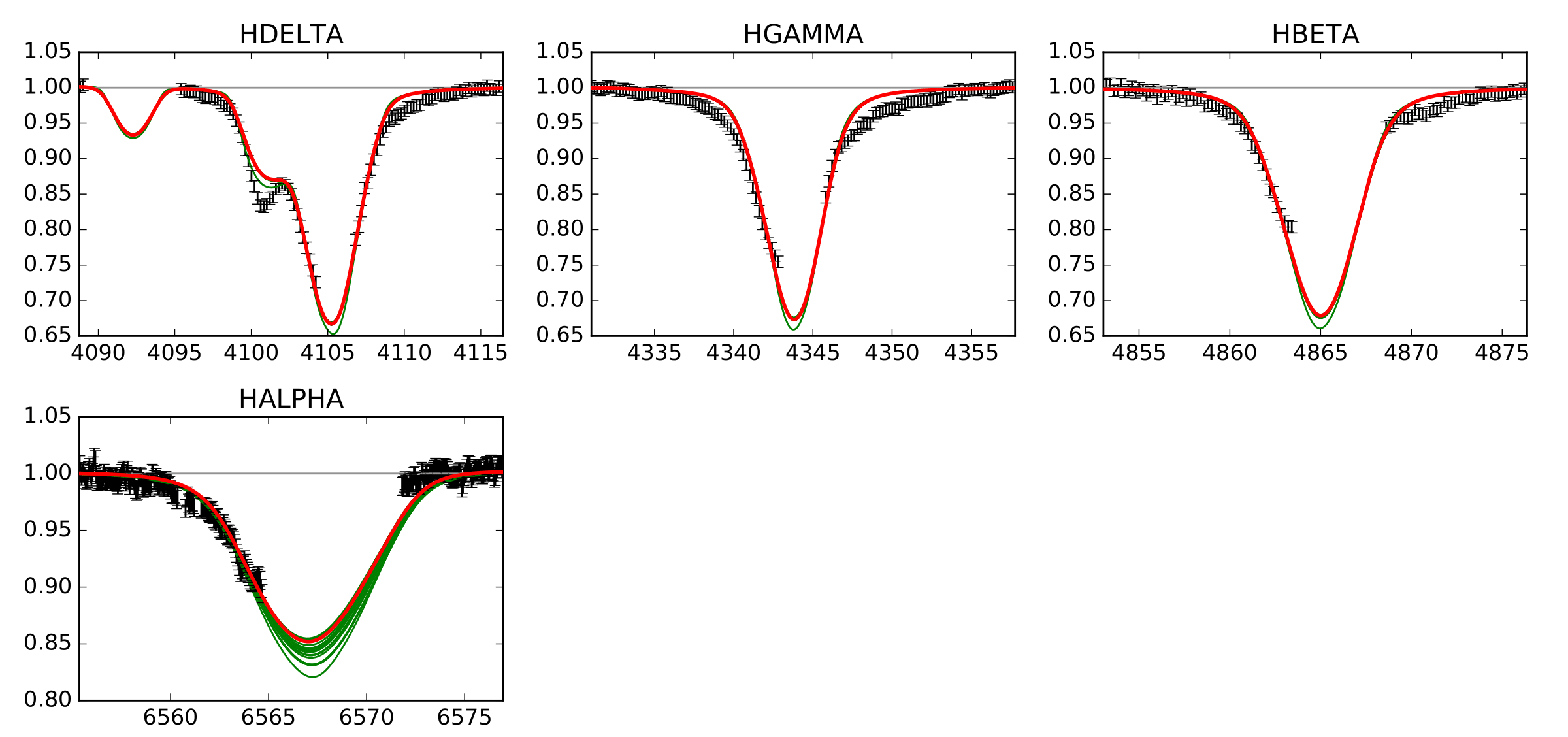}
	\caption{Best fit for VFTS440 O6-6.5II(f) from GA with optically thick clumping.}
	\label{fig: Best-fit-VFTS440}
\end{figure}

\begin{figure}[t!]
	\centering
	\includegraphics[scale=0.3]{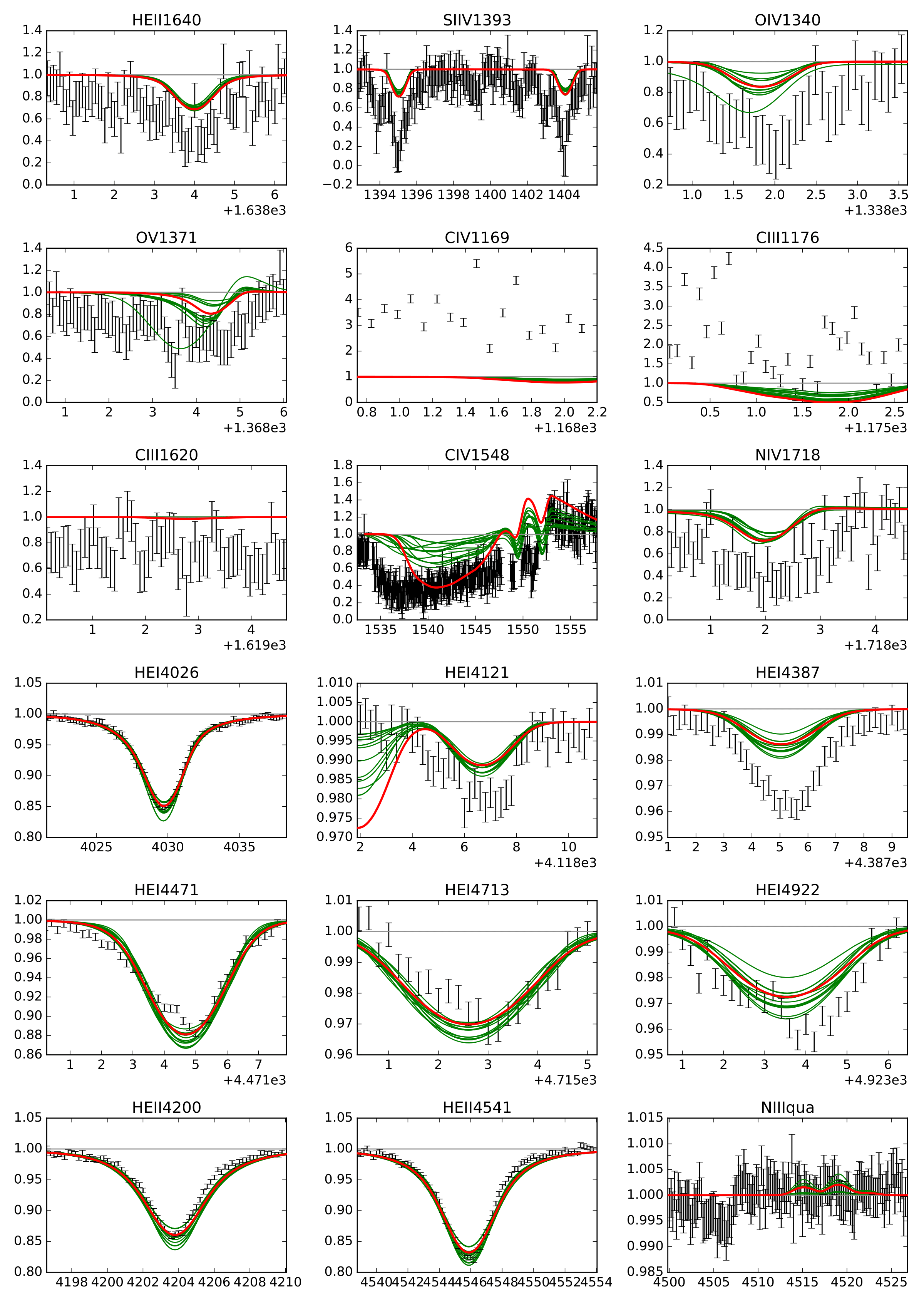}
	\includegraphics[scale=0.145]{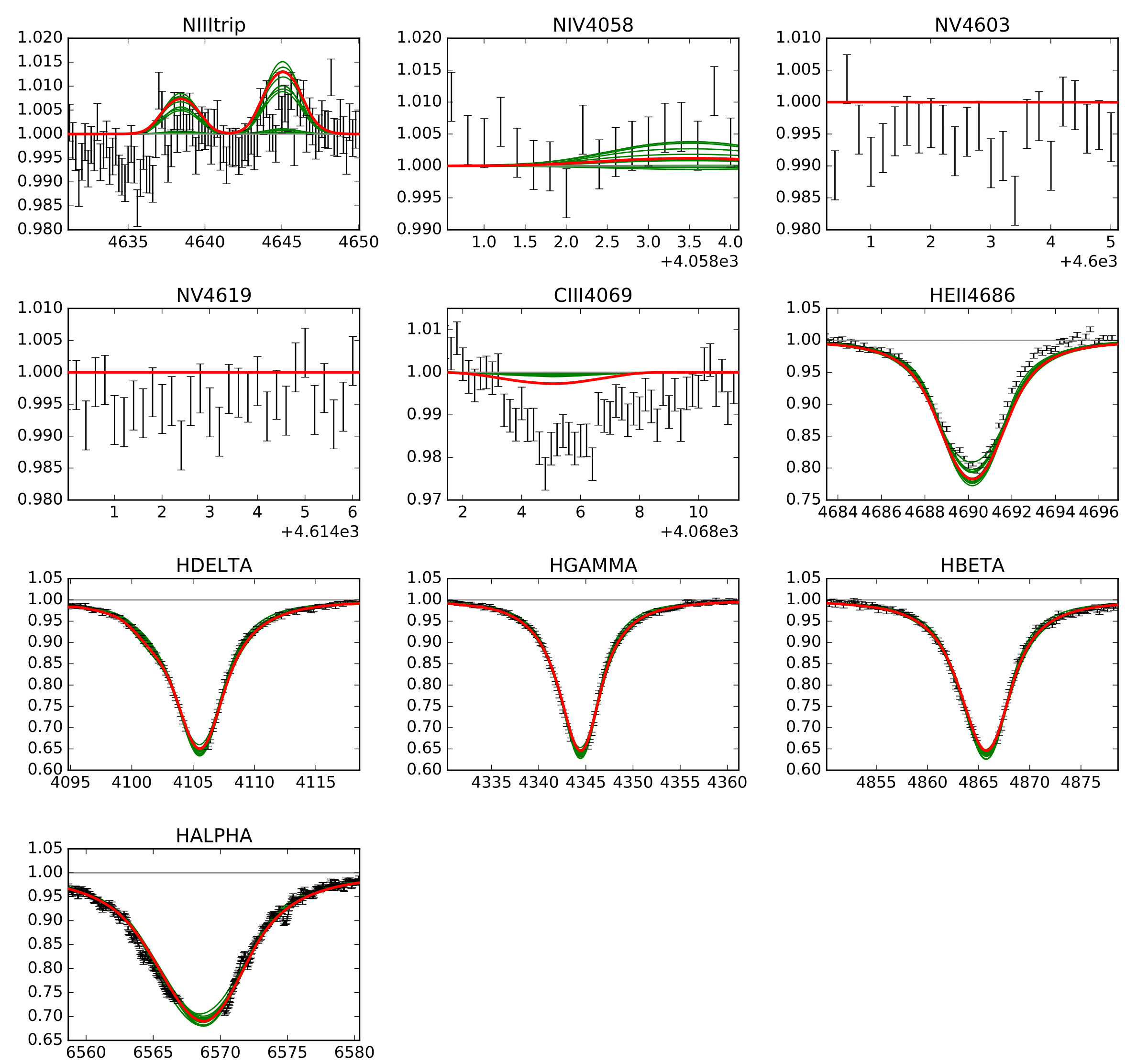}
	\caption{Best fit for VFTS096 O6V((n))((fc))z from GA with optically thick clumping.}
	\label{fig: Best-fit-VFTS096}
\end{figure}

\begin{figure}[t!]
	\centering
	\includegraphics[scale=0.3]{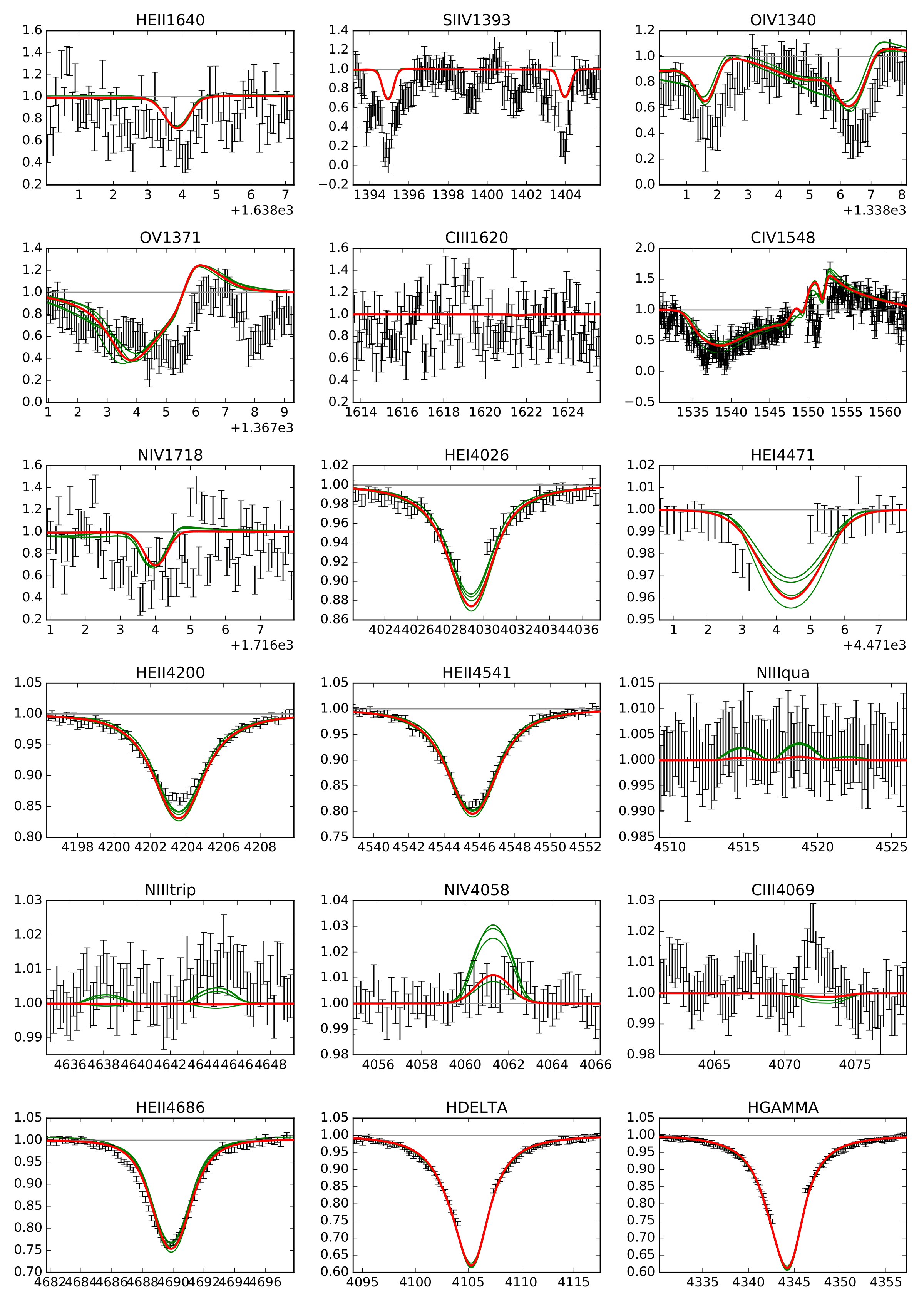}
	\includegraphics[scale=0.145]{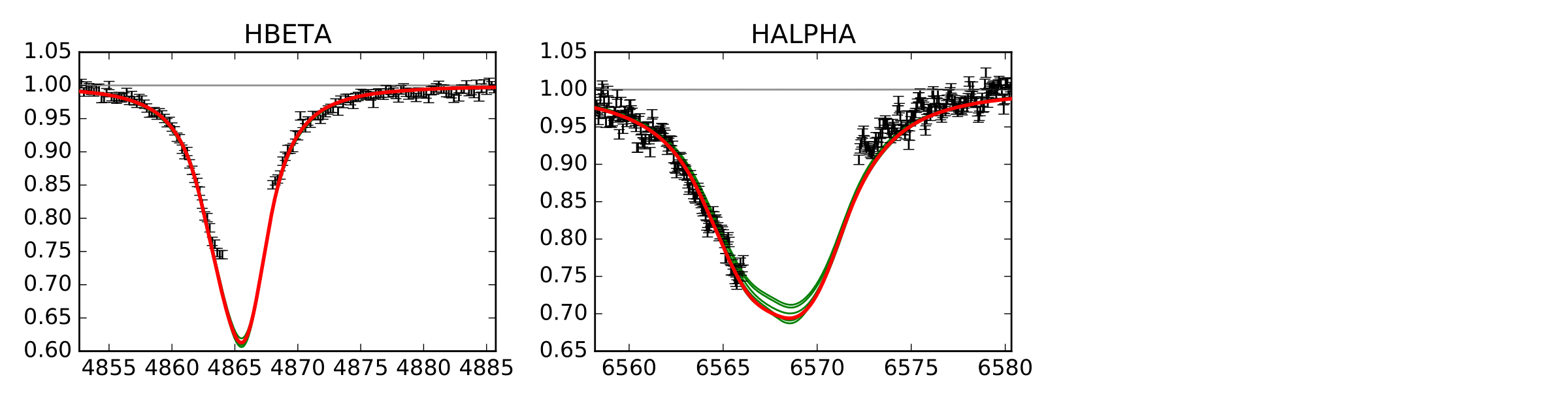}
	\caption{Best fit for VFTS586 O4V((n))((fc))z from GA with optically thick clumping.}
	\label{fig: Best-fit-VFTS586}
\end{figure}

\begin{figure}[t!]
	\centering
	\includegraphics[scale=0.3]{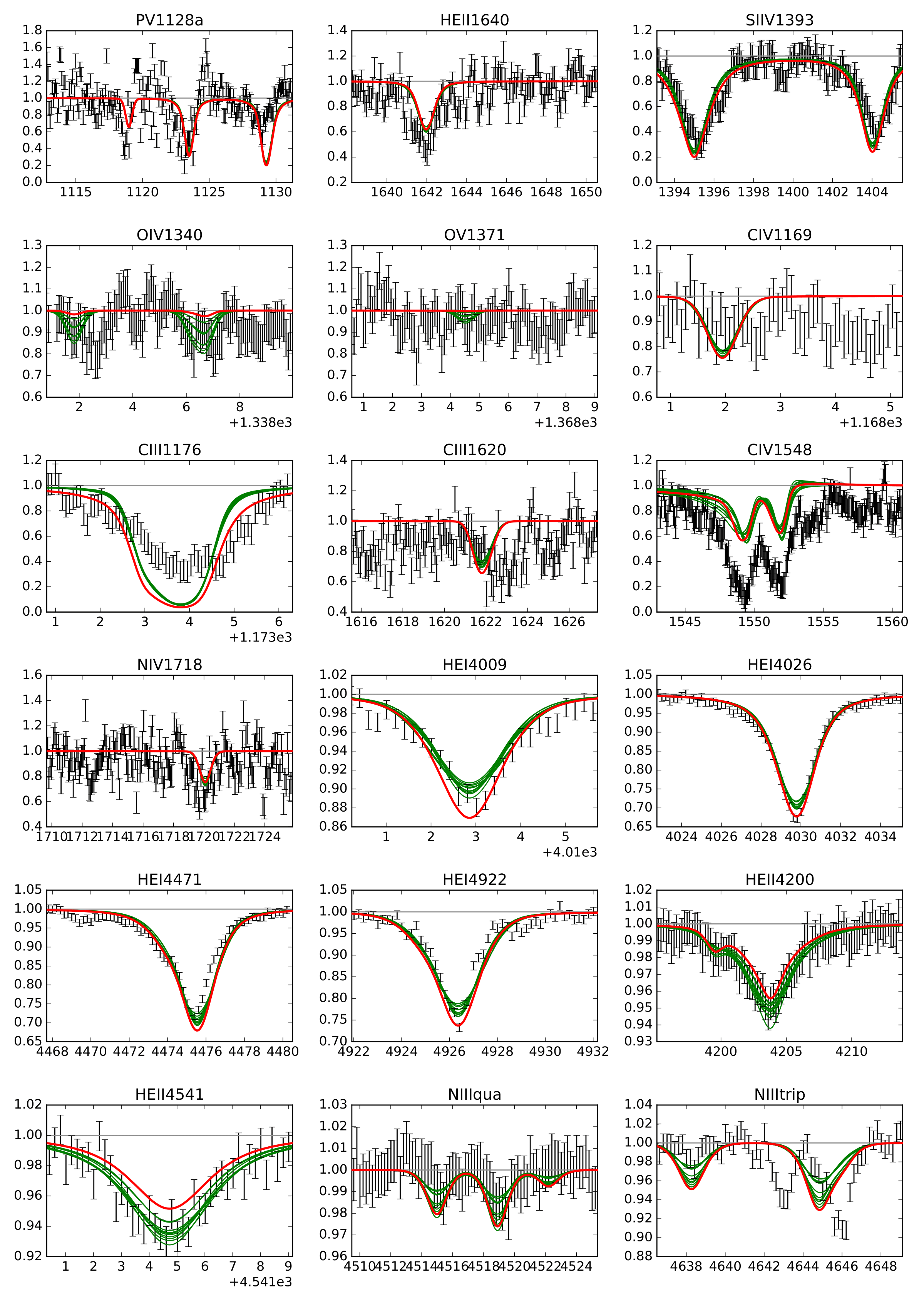}
	\includegraphics[scale=0.145]{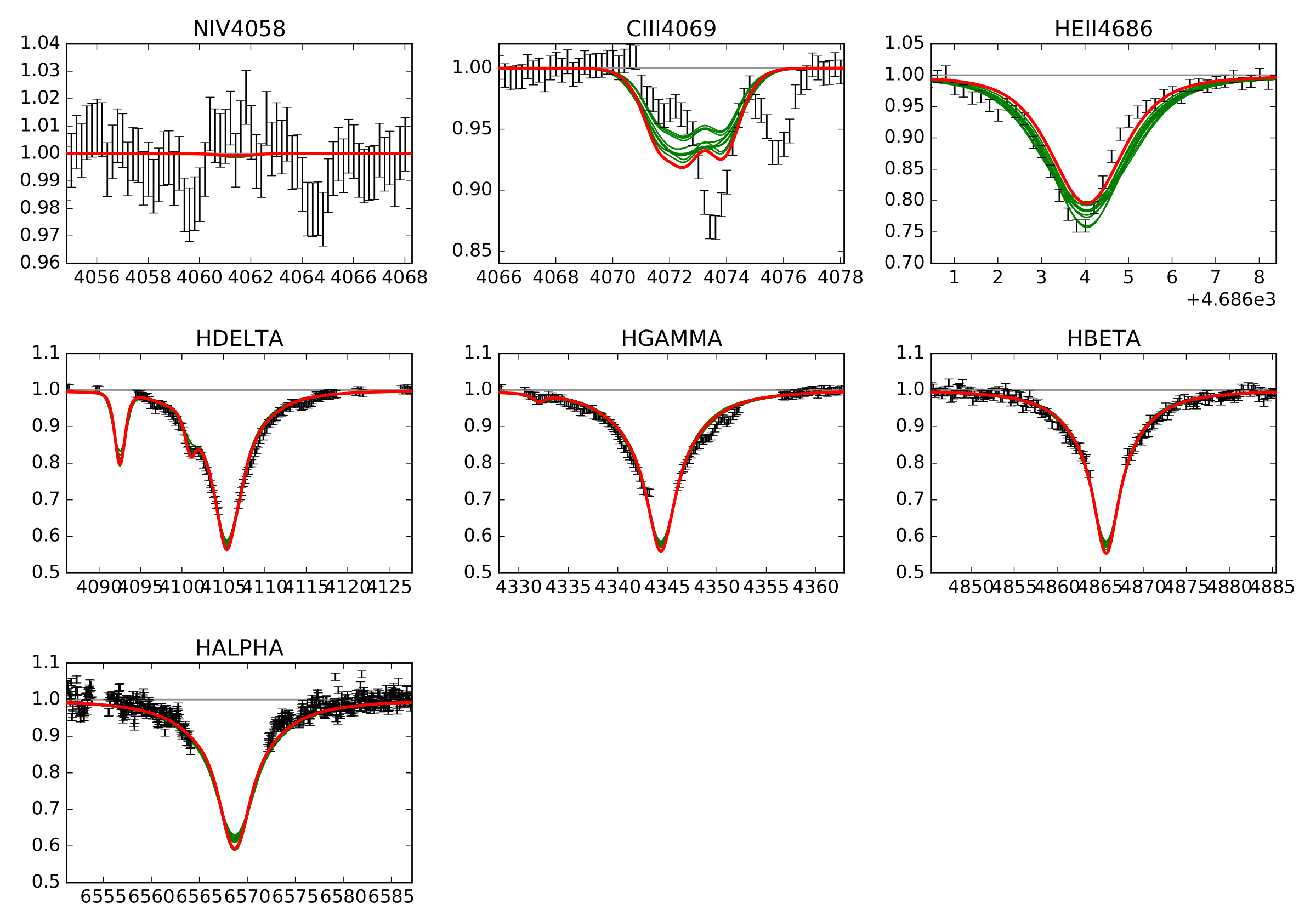}
	\caption{Best fit for VFTS627 O9.7V from GA with optically thick clumping.}
	\label{fig: Best-fit-VFTS627}
\end{figure}

\begin{figure}[t!]
	\centering
	\includegraphics[scale=0.3]{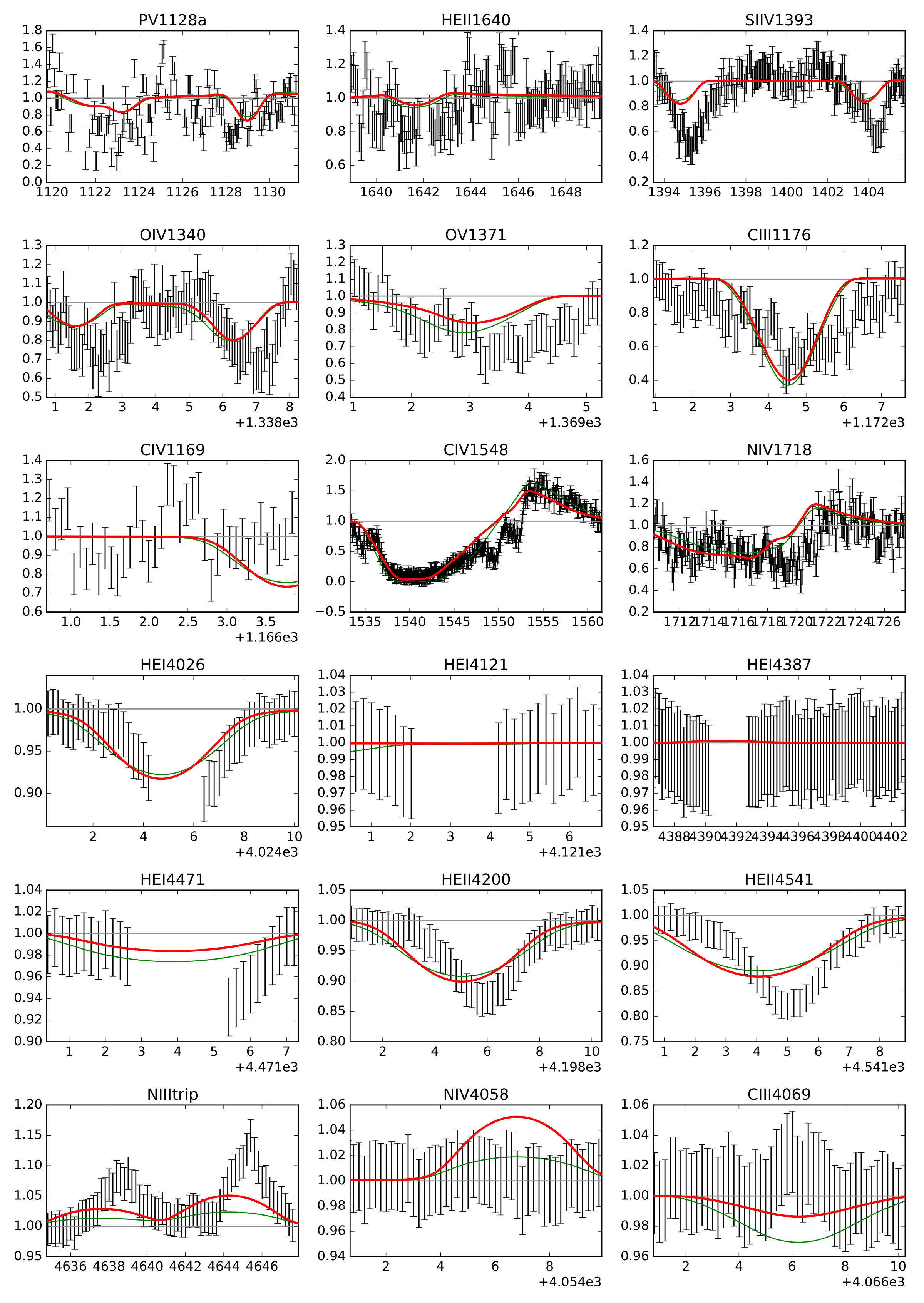}
	\includegraphics[scale=0.145]{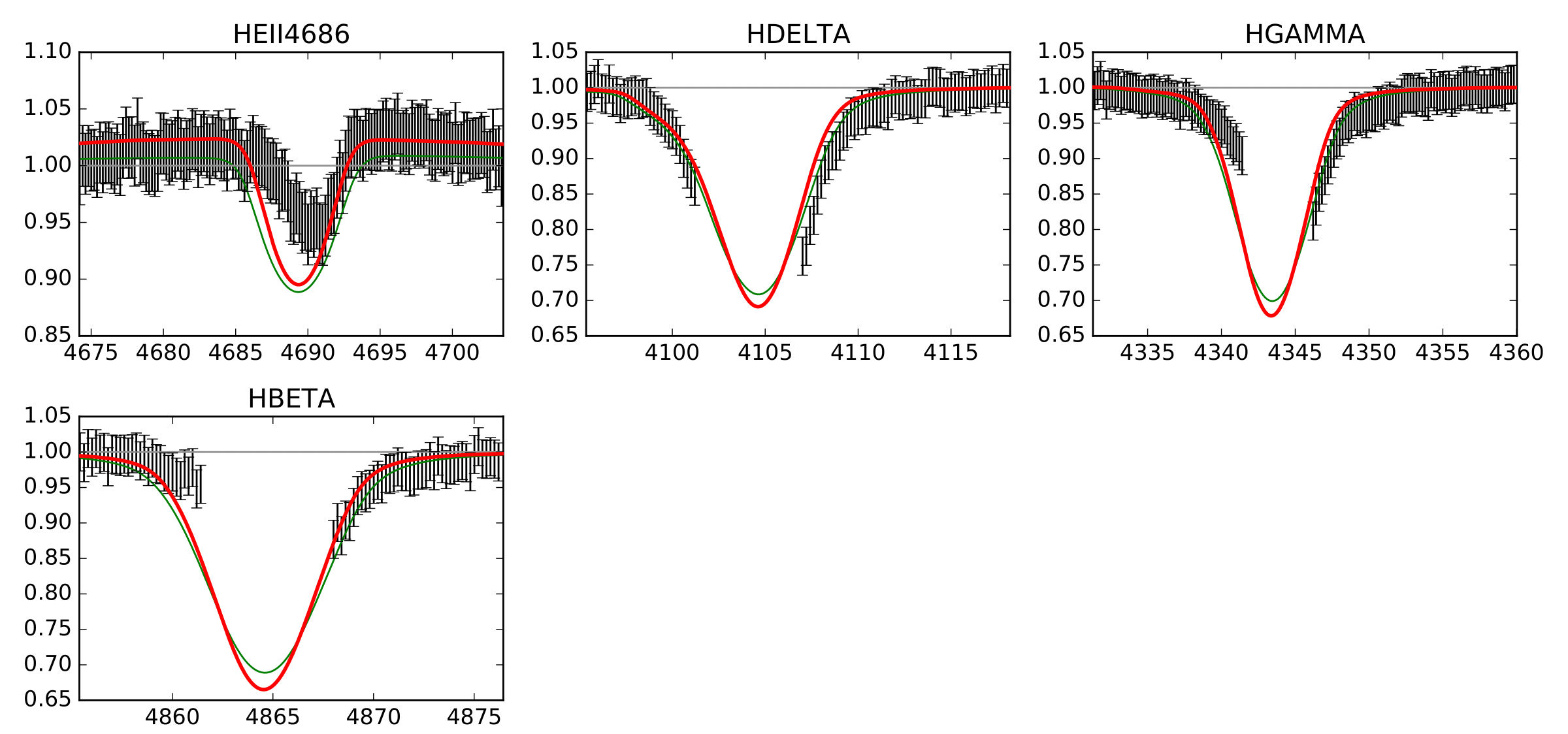}
	\caption{Best fit for VFTS422 O4III(f) from GA with optically thick clumping.}
	\label{fig: Best-fit-VFTS422}
\end{figure}

\end{appendix}

\end{document}